\documentstyle[12pt,a4,epic,eepic,epsf]{article}

\title{Quantum KZ equation with $|q|=1$ and\\
correlation functions of the XXZ model \\
in the gapless regime
\\
}
\author{
Michio Jimbo\thanks{Department of Mathematics, Faculty of Science,
                            Kyoto University, Kyoto 606, Japan.}
and Tetsuji Miwa\thanks{Research Institute for Mathematical Sciences, 
                            Kyoto University, Kyoto 606, Japan.}}

\date{January 18, 1996}

\newcommand{\End}{{\rm End}}
\renewcommand{\Im}{{\rm Im}\,}
\renewcommand{\Re}{{\rm Re}\,}
\newcommand{\Res}{{\rm Res}\,}

\newcommand{\nn}{\nonumber}
\newcommand{\omb}{\underline{\omega}}
\newcommand{\nb}{\underline{n}}
\newcommand{\eqref}[1]{(\ref{#1})}
\newcommand{\be}{\begin{equation}}
\newcommand{\en}{\end{equation}}
\newcommand{\bea}{\begin{eqnarray}}
\newcommand{\ena}{\end{eqnarray}}
\newcommand{\bean}{\begin{eqnarray*}}
\newcommand{\enan}{\end{eqnarray*}}

\newcommand{\Bbb}{\bf}
\newcommand{\Z}{{\Bbb Z}}
\newcommand{\C}{{\Bbb C}}

\newcommand{\s}{\sigma}
\renewcommand{\H}{{\cal H}}
\newcommand{\vep}{\varepsilon}

\newcommand{\bsl}{\hbox{$\scriptstyle\bar{l}\,$}\raise .9ex\hbox{$\scriptstyle*$}}

\newcommand{\brvac}[1]{{\dvac #1 \vac}}
\newcommand{\br}[1]{{\langle #1 \rangle}}
\newcommand{\lv}{\langle{\rm vac}|}
\newcommand{\rv}{|{\rm vac}\rangle}
\newcommand{\dvac}{\langle{\rm vac}|}
\newcommand{\vac}{|{\rm vac}\rangle}
\newcommand{\cR}{\check R}
\newcommand{\mt}[2]{\pmatrix{&#2\cr#1&\cr}}
\newcommand{\Ad}{{\rm Ad}\,}

\newcommand{\qed}{\hfill \fbox{}\medskip}
\newcommand{\proof}{\medskip\noindent{\it Proof.}\quad }
\def\itm#1{\begin{itemize}\item{#1}\end{itemize}}
\def\oR{\overline R}
\def\oG{\overline G}
\def\ve{\varepsilon}
\def\lb#1{\label{eqn:#1}}
\def\refeq#1{(\ref{eqn:#1})}
\def\sc[#1,#2]{{\scriptstyle{\scriptstyle#1\over\scriptstyle#2}}}
\def\con{{\rm const.}\,}
\def\Res{{\rm Res}}

\newtheorem{prop}{Proposition}[section]

\begin{document}
\maketitle
{\centerline {\it  Dedicated to the memory of Claude Itzykson}}

\begin{abstract}
An integral solution to the quantum Knizhnik-Zamolodchikov 
($q$KZ) equation with $|q|=1$ is presented. 
Upon specialization, it leads to
a conjectural formula for correlation functions of the 
XXZ model in the gapless regime.
The validity of this conjecture is verified in special cases, including 
the nearest neighbor correlator with an arbitrary coupling constant, 
and general correlators in the XXX and XY limits. 
\end{abstract}

\setcounter{section}{0}
\setcounter{equation}{0}
\def\s{\sigma}

\section{Introduction}\label{sec:1}

Consider the one-dimensional spin $1/2$ XXZ chain
\begin{equation}
H=-\frac{1}{2}\sum_{n=-\infty}^\infty
\left(\sigma^x_n\sigma^x_{n+1}+
\sigma^y_n\sigma^y_{n+1}+
\Delta \sigma^z_n\sigma^z_{n+1}\right).
\label{eqn:XXZ}
\end{equation}
In this paper we address the problem of 
describing correlation functions of \eqref{eqn:XXZ} in 
the gapless regime $|\Delta|\le 1$.
In the earlier works \cite{collin,JM}, 
the case of the anti-ferromagnetic regime $\Delta<-1$ was treated 
in the framework of representation theory of the quantum affine algebra
$U_q(\widehat{sl}_2)$.
As a result, correlation functions have been 
described by using the quantum Knizhnik-Zamolodchikov ($q$KZ) equation. 
It is this aspect that we will be concerned with in this paper. 
Before coming to the content, 
let us first recall some known results for $\Delta<-1$. 

Let $V=\C^2$, and consider the $R$ matrix 
$R(\beta)\in\End_\C(V\otimes V)$  associated with the XXZ model 
(see \refeq{RK}). 
The $q$KZ equation is the following system of linear difference equations
for an unknown function $G_n(\beta_1,\cdots,\beta_{2n})$ that 
takes values in  $V^{\otimes 2n}$:
\begin{eqnarray}
&&G_n(\beta_1,\cdots,\beta_j-2\pi i,\cdots,\beta_{2n})
\nonumber\\
&&\qquad=
R_{j\,j+1}(\beta_j-\beta_{j+1}-2\pi i)^{-1}
\cdots
R_{j\,2n}(\beta_j-\beta_{2n}-2\pi i)^{-1}
\nonumber\\
&&\qquad
\times
R_{1\,j}(\beta_1-\beta_{j})
\cdots
R_{j-1\,j}(\beta_{j-1}-\beta_{j})
G_n(\beta_1,\cdots,\beta_j,\cdots,\beta_{2n}). 
\label{eqn:qKZ}
\end{eqnarray}
Here 
$R_{ij}(\beta)\in\End_\C(V^{\otimes 2n})$
signifies the matrix acting as $R(\beta)$ on the
($i,j$)-th tensor components and as identity elsewhere. 
The correlation functions of arbitrary local operators are
obtained as the specialization
\begin{equation}
G_n(\overbrace{\beta+\pi i,\cdots,\beta+\pi i}^n,
\overbrace{\beta,\cdots,\beta}^n).
\label{eqn:spe}
\end{equation}
To be precise, in the case $\Delta<-1$, there are 
two functions $F_n^{(i)}$ ($i=0,1$) associated with the two anti-ferromagnetic
vacuum states, and 
it is their sum $G_n=F_n^{(0)}+F_n^{(1)}$ that satisfies the 
$q$KZ equation \eqref{eqn:qKZ}, as well as a 
set of relations \refeq{G1}, \refeq{G2} and \refeq{G3}.
The correlators are given by specializations of $F_n^{(i)}$ 
rather than $G_n$ itself.
In the context of representation theory, 
the functions $F^{(i)}_n$ are traces of products of certain intertwiners 
(vertex operators) taken over the integrable highest weight modules 
$V(\Lambda_i)$.
By realizing $V(\Lambda_i)$ in terms of 
bosonic free fields, an explicit integral formula was obtained 
for the functions $F^{(i)}_n$ $(i=0,1)$ and hence for
the solution $G_n$ of the $q$KZ equation. 

The argument relating correlation functions to the functions $G_n$ 
is based on the extension of the corner transfer matrix method
\cite{Bax82,JM}. 
It is applicable to the more general case of the XYZ spin chain as well 
\cite{JMN}. 
Correlation functions are related in the same way as above with 
solutions $G_n$ of the $q$KZ equation, 
this time having the elliptic $R$ matrix as coefficients. 
Unfortunately
the mathematical structure of the XYZ model is not fully 
understood yet (see \cite{FIJKMY,hwm} for a formulation of
an elliptic extension of $U_q(\widehat{sl}_2)$).
The free field realization is still unavailable 
(see however the recent development \cite{qVir,8VLuk} in this direction). 
Thus it remains an important open problem to 
construct solutions to the $q$KZ equation in the elliptic case. 

For the XXZ chain in the gapless regime $|\Delta|\le 1$, 
the corner transfer matrix fails to be well defined. 
Nevertheless this case can be viewed as a 
limiting case of the XYZ chain, 
so that the same recipe 
\eqref{eqn:spe} is expected to apply
for obtaining correlation functions. 
(Unlike the case $\Delta<-1$, the vacuum state is unique and the
distinction between $F^{(0)}$ and $F^{(1)}$ disappears.)
The problem is then to find appropriate solutions of the $q$KZ equation. 

Up to an overall scalar, the $R$ matrix 
$R(\beta)$ of the XXZ chain 
is a rational function in
\be
\zeta=e^{-\nu\beta},\quad q=-e^{\pi i \nu},
\lb{ZQ}
\en
where 
\begin{equation}
\Delta=-\cos\pi\nu=\frac{q+q^{-1}}{2}.
\label{eqn:Delta}
\end{equation}
However the nature of the solutions is quite different
depending on whether $\Delta<-1$ or $|\Delta|\le 1$.
In the case $\Delta<-1$, we have $-1<q<0$ 
and the solutions are meromorphic in $\zeta$,
typically involving infinite products of the form 
$\prod_{n=1}^\infty(1-\zeta q^{2n})$. 
On the other hand, the case $|\Delta|\le 1$ corresponds to 
$|q|=1$.
There are no analytic solutions which are
single valued in $\zeta$. 
Instead one has to look for solutions which are meromorphic in 
$\log\zeta$. 

A certain class of solutions to the $q$KZ equation with $|q|=1$ has
been studied in 
detail by Smirnov \cite{Smi92} in connection with the form factors in 
the sine-Gordon theory.
The equation relevant to the correlation functions of the XXZ model
is slightly different from Smirnov's, in particular 
the shift $-2\pi i$ in \eqref{eqn:qKZ} 
is replaced by $+2\pi i$ in his case 
(the former has `level $-4$' while the latter has 
`level $0$'; 
we will consider also the case in which 
the shift $-2\pi i$ is replaced by $-i\lambda$ where $\lambda>0$ is a 
parameter.)
In this paper we give a solution to the former, 
in the form of an integral which has a similar structure as in the case
$\Delta<-1$, 
and conjecture that its specialization \eqref{eqn:spe}
gives the correlation functions of the XXZ model with $|\Delta|\le 1$.
Our integral formula is essentially the same as the one
written down earlier by Lukyanov \cite{Luk95}.
However, in Lukyanov's case, it is given as a
generating function of the form factors of local 
operators in the sine-Gordon theory. 
Our point here is to interpret it as a formula for the 
correlation functions {\it on the lattice}. 
In general, difference equations determine the solutions 
only up to arbitrary periodic functions. 
We need to ensure that the particular solution
we present actually corresponds to correlation functions. 
As supporting evidences, we verify this statement in three special 
cases for which exact results are available:
(i) the nearest neighbour correlation $\br{\sigma^z_1\sigma^z_2}$, 
(ii) the  XXX model $\Delta=-1$ and (iii) the XY model $\Delta=0$. 
The integral formula and these verifications are the main results 
of this paper.

The text is organized as follows.
In Section 2 we formulate the $q$KZ and allied equations.
We then write down an integral formula for solutions.
In Section 3 we specialize the formula in Section 2 and 
propose that it gives correlation functions. 
In the special case $\nu=0$,
we recover the formula for the correlation functions
of the XXX model derived earlier in \cite{KIEU,Nak,JM}. 
In Section 4 we process our integral formula to reproduce the
simplest correlation function $\langle\s^z_1\s^z_2\rangle$.
This quantity can be derived by differentiating the ground state energy
of the Hamiltonian. 
In Section 5 we consider the XY limit $(\nu=1/2)$, which 
can be studied independently by using free fermions.
The correlation functions are given by the determinants of certain matrices 
whose entries are elementary functions in $\beta_j$.
Our integral formula in this case is shown to be equivalent 
to the free fermion result. 
The last
Section 6 is devoted to a discussion concerning some previous works
\cite{Smi92,Luk95,Nak} on the $q$KZ equation with $|q|=1$. 

Since most of the statements are proved by purely computational means, 
we put together technical points in the appendices in order 
to ease the reading. 
In Appendix A, a summary of Barnes' multiple gamma functions is offered. 
Appendix B contains the proof of the difference equations of Section 2. 
In Appendix C it is shown how the $n$-fold integral is reduced to an 
$(n-1)$-fold one by explicitly carrying out the integration once.
Appendix D is the derivation of the expression for
$\langle\s^z_1\s^z_2\rangle$. 
Appendix E is the evaluation of 
an integral in the case $\nu=1/2$. 
Finally, Appendix F is devoted to the free fermion theory in the XY limit.

\bigskip
\noindent{\it Acknowledgement.}\quad
We thank Sergei Lukyanov, Atsushi Nakayashiki
and Feodor Smirnov for valuable discussions.
We wish to express our sorrow at the death of Claude Itzykson.
We have always liked his nice lectures given in his fascinating voice
ever since we first met him in San Francisco, 1979. 

\setcounter{section}{1}
\setcounter{equation}{0}

\section{Integral fromula}
\subsection{The difference equations}
In this section, we formulate the system of equations we are going to 
study, including the $q$KZ equation with $|q|=1$.
We then give a particular solution in the form of an integral.
Throughout this section we fix 
parameters $\nu$ and $\lambda$ (see \refeq{ZQ})
such that $0<\nu<1$ and $\lambda>0$. 
For the convergence of the integral, we assume that 
\be
\lambda+{\pi \over\nu}>2\pi.
\en
In the application to the XXZ model, we will choose $\lambda=2\pi$.

Consider the $R$-matrix $R(\beta)\in\End_\C(V\otimes V)$ acting on the 
tensor product of $V=\C v^+\oplus\C v^-$:
\bea
&&R(\beta)(v^{\ve'_1}\otimes v^{\ve'_2})=
\sum_{\ve_1,\ve_2}R^{\ve'_1\ve'_2}_{\ve_1\ve_2}(\beta)v^{\ve_1}\otimes
v^{\ve_2},\nn\\
&&R(\beta)={1\over\kappa(\beta)}\oR(\beta).\lb{RK}
\ena
The parameter $\nu$ enters the matrix elements as follows. 
\bea
&&\oR^{++}_{++}(\beta)=\oR^{--}_{--}(\beta)=1,\nn\\
&&\oR^{+-}_{+-}(\beta)=\oR^{-+}_{-+}(\beta)=\bar b(\beta),\nn\\
&&\oR^{+-}_{-+}(\beta)=\oR^{-+}_{+-}(\beta)=\bar c(\beta),\nn\\
&&\oR^{\ve'_1\ve'_2}_{\ve_1\ve_2}(\beta)=0\quad\hbox{ in the other cases}, 
\lb{BW}
\ena
where
\[
\bar b(\beta)=
{\sinh\nu\beta\over\sinh\nu(\pi i-\beta)},
\qquad
\bar c(\beta)={\sinh\nu\pi i\over\sinh\nu(\pi i-\beta)}.
\]
The function $\kappa(\beta)$ will be specified below. 
It is chosen to ensure that 
the $R$-matrix satisfies the unitarity and the crossing symmetry relations 
\[
R_{12}(\beta)R_{21}(-\beta)={\rm id},
\qquad
R^{\vep_2'\vep_1}_{\vep_2\vep_1'}(\beta)=
R^{-\vep_1'\,\vep_2'}_{-\vep_1\,\vep_2}(\pi i-\beta).
\]

Let $n$ be a non-negative integer.
Consider a $V^{\otimes 2n}$-valued function
$G_n=G_n(\beta_1,\cdots,\beta_{2n})$, 
depending on the `spectral parameters' $\beta_1,\cdots,\beta_{2n}$.
We set 
\[
G_0=1.
\]
We study the following system of difference equations for $G_n$
involving the parameter $\lambda$:
\bea
&&G_n(\cdots,\beta_{j+1},\beta_j,\cdots)_{\cdots,\ve_{j+1},\ve_j,\cdots}\nn\\
&&=\sum_{\ve'_j,\ve'_{j+1}} R^{\ve'_j,\ve'_{j+1}}_{\ve_j,\ve_{j+1}}
(\beta_j-\beta_{j+1})
G_n(\cdots,\beta_j,\beta_{j+1},\cdots)_{\cdots,\ve'_j,\ve'_{j+1},\cdots},\lb{G1}\\
&&G_n(\beta_1,\cdots,\beta_{2n-1},\beta_{2n}-i\lambda)
_{\ve_1,\cdots,\ve_{2n}}=
G_n(\beta_{2n},\beta_1,\cdots,\beta_{2n-1})_{\ve_{2n},\ve_1,\cdots,\ve_{2n-1}},
\lb{G2}\\
&&G_n(\beta_1,\cdots,\beta_{2n})_{\ve_1,\cdots,\ve_{2n}}\bigg|_{\beta_{2n}=
\beta_{2n-1}+\pi i}
=\delta_{\ve_{2n-1}+\ve_{2n},0}G_{n-1}(\beta_1,\cdots,\beta_{2n-2})
_{\ve_1,\cdots,\ve_{2n-2}}.\nn\\\lb{G3}
\ena
In particular, 
the $q$KZ equation (\ref{eqn:qKZ}) is a consequence of \eqref{eqn:G1} 
and \eqref{eqn:G2}.
It can be shown also that \eqref{eqn:G1} and \eqref{eqn:G3} imply 
\begin{eqnarray}
&&G_n(\beta_1,\cdots,\beta_{2n})_{\ve_1,\cdots,\ve_{2n}}\bigg|_{\beta_{j+1}=
\beta_{j}+\pi i}
\nonumber\\
&&=\delta_{\ve_{j}+\ve_{j+1},0}
G_{n-1}(\beta_1,\cdots,\beta_{j-1},\beta_{j+2},\cdots,\beta_{2n})
_{\ve_1,\cdots,\ve_{j-1},\ve_{j+2},\cdots,\ve_{2n}}
\lb{G4}
\end{eqnarray}
for any $j=1,\cdots,2n-1$. 
Note that 
the equations \eqref{eqn:G1}--\eqref{eqn:G3} involve only the functions 
$G_n(\beta_1,\cdots,\beta_{2n})_{\ve_1,\cdots,\ve_{2n}}$ with 
fixed value of the `spin' $\ve_1+\cdots+\ve_{2n}$. 
Throughout this paper we will restrict ourselves to
the `spin $0$' case, 
i.e. we assume 
\[
G_n(\beta_1,\cdots,\beta_{2n})_{\ve_1,\cdots,\ve_{2n}}=0
\hbox{ unless }
\vep_1+\cdots+\vep_{2n}=0.
\]

\subsection{Auxiliary functions}
Our aim in this section is to
construct a solution to \refeq{G1}, \refeq{G2} and \refeq{G3} 
by using an $n$-fold integral. 
The formula involves certain special functions
$\kappa(\beta),\rho(\beta),\varphi(\beta),\psi(\beta)$.
Let us first give their definitions and list some of their properties.
In what follows, $S_r(x|\omega_1,\cdots,\omega_r)$ will denote
the multiple sine function (see appendix \ref{app:A} for the definition).

\itm{$\kappa(\beta)$}
\bea
&&\kappa(\beta)=-{S_2(i\beta|2\pi,\sc[\pi,\nu])S_2(\pi-i\beta|2\pi,
\sc[\pi,\nu])\over
S_2(-i\beta|2\pi,\sc[\pi,\nu])S_2(\pi+i\beta|2\pi,
\sc[\pi,\nu])}\phantom{-----}
\nn\\
&&\phantom{\kappa(\beta)}=
\exp\left\{-i\int_0^\infty \frac{dt}{t}
\frac{\sin\frac{2\beta\nu}{\pi}t\sinh(1-\nu)t}
{\sinh t\cosh\nu t}\right\},
\nn\\
&&\kappa(\beta)\kappa(-\beta)=1,\nn\\
&&\kappa(\beta)\kappa(\beta-\pi i)
=\frac{\sinh\nu\beta}{\sinh\nu(\pi i-\beta)}
=\bar{b}(\beta).\nn
\ena

\itm{$\rho(\beta)$}
\bea
&&\rho(\beta)=\sinh\sc[\pi\beta,\lambda]
{S_3(\pi-i\beta)S_3(\pi+\lambda+i\beta)
\over S_3(-i\beta)S_3(\lambda+i\beta)}
\qquad (S_3(x)=S_3(x|2\pi,\lambda,\sc[\pi,\nu])), 
\nn\\
&&\rho(\beta)=\rho(\sc[i\lambda,2])\exp\left\{\int_0^\infty{dt\over t}{\sin^2
\left((\beta-{i\lambda\over2}){\nu t\over\pi}\right)
\sinh(1-\nu)t
\over \cosh\nu t\,\sinh t\,\sinh{\lambda\nu t\over\pi}}\right\},\nn\\
&&{\rho(\beta)\over\rho(-\beta)}=\kappa(\beta),\lb{RI}\\
&&\rho(i\lambda-\beta)=\rho(\beta),\lb{RC}\\
&&\rho(\beta)={\nu\rho(\pi  i)\over i\sin\pi\nu}(\beta+\pi i)+\cdots
\hbox{ when $\beta\rightarrow-\pi i$}.\lb{A1}
\ena

\itm{$\varphi(\beta)$}
\bea
&&\varphi(\beta)={2\over S_2({\pi\over2}+i\beta|\lambda,{\pi\over\nu})
S_2({\pi\over2}-i\beta|\lambda,{\pi\over\nu})},\nn\\
&&\varphi(\beta)=\varphi(0)
\exp\left\{-2\int_0^\infty{dt\over t}
{\sin^2{\beta\nu t\over\pi}\sinh(1+\sc[\lambda-\pi,\pi]\nu)t
\over\sinh\sc[\lambda\nu t,\pi]\sinh t}\right\},
\nn\\
&&\varphi(-\beta)=\varphi(\beta),\nn\\
&&{\varphi(\beta-i\lambda)\over\varphi(\beta)}=-
{\sinh\nu(\beta-{\pi i\over2})\over
\sinh\nu(\beta+{\pi i\over2}-i\lambda)},\lb{PL}\\
&&\frac{\varphi(\beta\pm{\pi i\over\nu})}{\varphi(\beta)}
=-\frac{\sinh{\pi\over\lambda}(\beta\pm{\pi i\over2})}
{\sinh{\pi\over\lambda}(\beta\mp{\pi i\over2}\pm{\pi i\over\nu})},
\nn\\
&&\varphi(\beta)=\pm
{\sqrt{\sc[\lambda,\pi\nu]}\over iS_2(\pi|\lambda,{\pi\over\nu})
(\beta\mp{\pi i\over2})}+\cdots
\hbox{ when $\beta\rightarrow\pm{\pi i\over2}$,}\nn\\
\lb{RZ}\\
&&\rho(\beta)\rho(\beta+\pi i)\varphi(\beta+\sc[\pi i,2])
={i\over4\sinh\nu\beta}.\lb{RR}
\ena

\itm{$\psi(\beta)$}
\bea
&&\psi(\beta)=\sinh\sc[\pi\beta,\lambda]S_2(\pi+i\beta|\lambda,\sc[\pi,\nu])
S_2(\pi-i\beta|\lambda,\sc[\pi,\nu])
\nn\\
&&\psi(-\beta)=-\psi(\beta),\lb{AS}\\
&&\varphi(\beta+\sc[\pi i,2])\varphi(\beta-\sc[\pi i,2])\psi(\beta)
=\frac{1}{\sinh\nu\beta},\lb{A2}\\
&&{\psi(\beta+i\lambda)\over\psi(\beta)}=
{\sinh\nu(\beta+i\lambda-\pi i)\over\sinh\nu(\beta+\pi i)}.\lb{QL}
\ena
In addition we will use a constant $c_n$ given by
\be
\lb{NR}
c_n=(-16)^{\sc[n(n-1),2]}
\left({\pi S_2(\pi|\lambda,\sc[\pi,\nu])^2
\over\lambda\rho(\pi i)}\right)^n.
\en
In the case $\lambda=2\pi$ the formulas simplify to 
\[
\psi(\beta)=\sinh\beta,
\qquad
c_n={(-16)^{\sc[n(n-1),2]}\over\rho(\pi i)^n}.
\]

\subsection{Integral formula}
Let us present the integral formula for
$G_n(\beta_1,\cdots,\beta_{2n})_{\ve_1\cdots\ve_{2n}}$.
Given a set of indices 
$\ve_1,\cdots,\ve_{2n}\in\{+,-\}$, we define a map
$a\in\{1,\cdots,n\}\rightarrow\bar a\in\{1,\cdots,2n\}$
in such a way that
$\ve_{\bar a}=+$ and $\bar a<\bar b$  if $a<b$.
Define further a meromorphic function
\[
Q_n(\alpha|\beta)_{\ve_1\cdots\ve_{2n}}
={\prod_{j<\bar a}\sinh\nu(\alpha_a-\beta_j+{\pi i\over2})
\prod_{j>\bar a}\sinh\nu(\beta_j-\alpha_a+{\pi i\over2})
\over
\prod_{a<b}\sinh\nu(\alpha_a-\alpha_b-\pi i)}
\]
where $\alpha=(\alpha_1,\cdots,\alpha_n)$ and
$\beta=(\beta_1,\cdots,\beta_{2n})$.
After these preparations, we set
\bea
&&G_n(\beta_1,\cdots,\beta_{2n})_{\ve_1\cdots\ve_{2n}}=
c_n\prod_{j<k}\rho(\beta_j-\beta_k)
\nn\\
&&\times\prod_a\int_{C_a}{d\alpha_a\over2\pi i}
\prod_{a,j}\varphi(\alpha_a-\beta_j)\prod_{a<b}\psi(\alpha_a-\alpha_b)
Q_n(\alpha|\beta)_{\ve_1\cdots\ve_{2n}}.\lb{GF}
\ena
Clearly we have, for any $\gamma$, 
\[
G_n(\beta_1+\gamma,\cdots,\beta_{2n}+\gamma)
=
G_n(\beta_1,\cdots,\beta_{2n}).
\]

In Appendix B, we will prove that
with the appropriate choice of the integration contours $C_a$ 
$(1\le a\le n)$ as given below,
the function $G_n(\beta_1,\cdots,\beta_{2n})$ is meromorphic and satisfies
\refeq{G1},\refeq{G2} and \refeq{G3}.

In order to specify the integration contours, let us examine the poles 
of the integrand of \eqref{eqn:GF}.
The poles of $\varphi(\alpha_a-\beta_j)$ are at
\be
\alpha_a-\beta_j=\pm i(n_1\lambda+\sc[n_2,\nu]\pi+\sc[\pi,2])\quad
(n_1,n_2\ge0).
\en
The poles of $\psi(\alpha_a-\alpha_b)$ $(a<b)$ are at
\be
\alpha_a-\alpha_b=\pm i(n_1\lambda+\sc[n_2-\nu,\nu]\pi)
\quad(n_1,n_2\ge1),
\en
and the poles of ${1\over\sinh\nu(\alpha_a-\alpha_b-\pi i)}$
are at
\be
\alpha_a-\alpha_b=\sc[n+\nu,\nu]\pi i\quad(n\in\Z).
\en
Since $\psi(\alpha_a-\alpha_b)$ has zeros at
\be
\alpha_a-\alpha_b=\pm i(n_1\lambda+\sc[n_2,\nu]\pi+\pi)\quad(n_1,n_2\ge0).
\en
the poles of $\sc[\psi(\alpha_a-\alpha_b),\sinh\nu(\alpha_a-\alpha_b-\pi i)]$
are at
\be
\alpha_a-\alpha_b=i(n_1\lambda+\sc[n_2-\nu,\nu]\pi)
\quad(n_1,n_2\ge1),
\en
or
\be
\alpha_a-\alpha_b=-i(n_1\lambda+\sc[n_2-\nu,\nu]\pi)\quad(n_1\ge0,n_2\ge1).
\en
Therefore, the poles in the variable $\alpha_a$ of the integrand
are contained in the set
\bea	
&&\{\beta_j\pm i(n_1\lambda+\sc[n_2,\nu]\pi+\sc[\pi,2])
\quad(1\le j\le2n;n_1,n_2\ge0);\nn\\
&&\alpha_b+i(n_1\lambda+\sc[n_2-\nu,\nu]\pi)
\quad(a<b\le n;n_1,n_2\ge1);\nn\\
&&\alpha_b-i(n_1\lambda+\sc[n_2-\nu,\nu]\pi)
\quad(a<b\le n;n_1\ge0,n_2\ge1);\nn\\
&&\alpha_b+i(n_1\lambda+\sc[n_2-\nu,\nu]\pi)
\quad(1\le b<a;n_1\ge0,n_2\ge1);\nn\\
&&\alpha_b-i(n_1\lambda+\sc[n_2-\nu,\nu]\pi)
\quad(1\le b<a;n_1,n_2\ge1)\}.\nn
\ena

We choose the contour $C_a$ for $\alpha_a$ by the following rule: 
\bea
&&
\hbox{$\alpha_a$ lies on the real line for $|\alpha_a|\gg 0$,}
\lb{C1}\\
&&
\hbox{$\beta_j+\sc[\pi i,2]$ $(1\le j\le 2n)$,
$\alpha_b+i(\lambda+\sc[1-\nu,\nu]\pi)$ $(a<b\le n)$,}
\nonumber\\
&&\qquad \hbox{$\alpha_b+i\sc[1-\nu,\nu]\pi$ $(1\le b<a)$
are above $C_a$,}
\lb{C2}\\
&&
\hbox{$\beta_j-\sc[\pi i,2]$ $(1\le j\le 2n)$,
$\alpha_b-i\sc[1-\nu,\nu]\pi$ $(a<b\le n)$,}
\nonumber\\
&&\qquad 
\hbox{$\alpha_b-i(\lambda+\sc[1-\nu,\nu]\pi)$ $(1\le b<a)$
are below $C_a$.}
\lb{C3}
\ena
Note that we can choose $C_a$ to be the same contour $C$
for all $a$ such that $\beta_j+\sc[\pi i,2]$ $(1\le j\le 2n)$
are above $C$ and $\beta_j-\sc[\pi i,2]$ $(1\le j\le 2n)$
are below $C$.

Let check the convergence of the integral.
Recall that the periods of the double sine function
$S_2$ used in $\varphi$ and $\psi$ are such that
$\omega_1=\lambda>0$ and $\omega_2=\sc[\pi,\nu]>\pi$.
In the proof below, we use ``\con'' to mean different constants
which appear in the estimates. From \refeq{ES}, we have
\bea
|\varphi(\alpha_a-\beta_j)|&\le&\con
e^{\sc[\pi(\pi-\omega_1-\omega_2),\omega_1\omega_2]|\alpha_a|},\nn\\
\left|{\psi(\alpha_a-\alpha_b)\over\sinh\nu(\alpha_a-\alpha_b-\pi i)}\right|
&\le&\con e^{\sc[2\pi(\omega_2-\pi),\omega_1\omega_2](|\alpha_a|+|\alpha_b|)},
\nn\\
|\sinh\nu(\alpha_a-\beta_j\pm\sc[\pi i,2])|
&\le&\con e^{\sc[\pi,\omega_2]|\alpha_a|}.\nn
\ena
Collecting these estimates we see that
\[
\left|\prod_{a,j}\varphi(\alpha_a-\beta_j)\prod_{a<b}\psi(\alpha_a-\alpha_b)
Q_n(\alpha|\beta)_{\ve_1\cdots\ve_{2n}}\right|
\le\con e^{\sc[\pi(2\pi-\omega_1-\omega_2),\omega_1\omega_2]\sum_a|\alpha_a|}.
\]
Since we have assumed
that $2\pi-\omega_1-\omega_2=2\pi-\lambda-\sc[\pi,\nu]<0$,
the integral is convergent.

\subsection{One-time integration}
In the case of interest $\lambda=2\pi$, 
the $n$-fold integral for 
$G_n(\beta_1,\cdots,\beta_{2n})$ can be reduced to an $(n-1)$-fold integral 
by carrying out the integration once. 
The result is stated as follows.

\begin{eqnarray}
&&G_n(\beta_1,\cdots,\beta_{2n})_{\vep_1,\cdots,\vep_{2n}}
=\tilde{c}_n
\prod_{1\le j<k\le 2n}\rho(\beta_j-\beta_k)
\nonumber\\
&&\times
\frac{\pi}{\nu}\frac{1}{e^{\sum\beta_j/2}\sum e^{-\beta_j}}
\sum_{l=1}^{n}(-1)^{\bar{l}+1}
\int\!\!\cdots\!\!\int \prod_{k\neq l}d\alpha_k D(\alpha_1,\cdots,
\alpha_{l-1},\alpha_{l+1},
\cdots,\alpha_n)
\nonumber\\
&&\times
\prod_{k\neq l}\Bigl[\prod_{j=1}^{2n}\varphi(\alpha_k-\beta_j)
\prod_{j<\bar{k}}\sinh\nu\bigl(\alpha_k-\beta_j+\sc[\pi i,2]\bigr)
\prod_{j>\bar{k}}\sinh\nu\bigl(-\alpha_k+\beta_j+\sc[\pi i,2]\bigr)\Bigr]
\nonumber\\
&&\times
\frac{\sinh\nu\left(\sum_{k\neq l}\alpha_k+
\frac{1}{2}\beta_{\bar{l}}-\frac{1}{2}\sum_{j\neq\bar{l}}\beta_j
+\pi i(\bar{l}-2l+\frac{1}{2})\right)}
{\prod_{r<s,r,s\neq l}\sinh\nu(\alpha_r-\alpha_s-\pi i)}.
\label{eqn:once}
\end{eqnarray}
Here we have set 
\[
\tilde{c}_n
=2^{3n(n-1)/2}\bigl(-\pi i \rho(\pi i)\bigr)^{-n},
\]
and 
\[
D(x_1,\cdots,x_{n-1})
=\det\left(e^{-(n-2k-1)x_j}\right)_{1\le j,k\le n-1}.
\]
As before, the numbers $\bar{1}<\cdots<\bar{n}$ are determined by
\[
\{\bar{l}\mid  1\le l\le n\}=\{j\mid 1\le j\le 2n, \vep_j=+\}.
\]
The integration is taken along a path going from 
$-\infty$ to $+\infty$ in such a way 
that $-\pi/2<{\rm Im}(\alpha_k-\beta_j)<\pi/2$ for all $k,j$. 
In the above, we assume that $0<\nu<1/2$
for the convergence of the integral.
It should be possible to treat also the case $1/2\le \nu<1$
by introducing a suitable regularization as in \cite{Smi92}, 
but we do not go into this question here. 

The derivation of \eqref{eqn:once} will be given in Appendix C.

\setcounter{section}{2}
\setcounter{equation}{0}

\section{Correlation functions}\label{sec:3}
\def\itm#1{\begin{itemize}\item{#1}\end{itemize}}
\def\oR{\overline R}
\def\oG{\overline G}
\def\ve{\varepsilon}
\def\sc[#1,#2]{{\scriptstyle{\scriptstyle#1\over\scriptstyle#2}}}

We now proceed to the description of correlation functions of the 
XXZ model. 
From now on, we assume that $\lambda=2\pi$. 

First let us set up the notation. 	
Let $E_{\vep\vep'}$ denote the $2\times 2$ matrix with 
 $1$ at the $(\vep,\vep')$-th place and $0$ elsewhere. 
Thus the Pauli spin operators read 
\[
\sigma^x=E_{+-}+E_{-+},
\quad
\sigma^y=-iE_{+-}+iE_{-+},
\quad
\sigma^z=E_{++}-E_{--}.
\]
In the tensor product $\cdots \otimes V_j \otimes V_{j+1}\otimes\cdots $ 
of $V_j\simeq\C^2$, we let 
$\sigma^\alpha_j$, $E^{(j)}_{\vep\vep'}$ denote respectively
the operators acting as $\sigma^\alpha$ or $E_{\vep\vep'}$ 
on the $j$-th component and as identity elsewhere. 
	
By a local operator we mean an element of the algebra generated by
$\sigma^\alpha_j$'s. 
Any local operator is a linear combination of 
operators of the form 
$O=E^{(r)}_{\ve'_r\ve_r}E^{(r+1)}_{\ve'_{r+1}\ve_{r+1}}\cdots 
E^{(s)}_{\ve'_s\ve_s}$ ($r\le s$). 
The correlation function of $O$ is its expected value with respect to 
the ground state eigenvector of the XXZ Hamiltonian. 
We conjecture that it is given by the following special value of $G_n$
($n=s-r+1$):
\begin{equation}
\langle
E^{(r)}_{\ve'_r\ve_r}\cdots E^{(s)}_{\ve'_s\ve_s}
\rangle{\buildrel{\rm def}\over=}
G_n(\overbrace{\beta+\pi i,\cdots,\beta+\pi i}^n,
\overbrace{\beta,\cdots,\beta}^n)_{-\ve'_r,\cdots,-\ve'_s,\ve_s,\cdots,\ve_r}.
\label{eqn:SC}
\end{equation}
We shall consider a slightly more general object
\begin{equation}
G_n(\beta_r+\pi i,\cdots,\beta_s+\pi i,\beta_s,\cdots,\beta_r)
_{-\ve'_r,\cdots,-\ve'_s,\ve_s,\cdots,\ve_r}.
\label{eqn:GC}
\end{equation}
This corresponds to introducing spectral parameters $\beta_j$ 
as inhomogeneity of the model
(in the terminology of \cite{BaxRS}, the corresponding 
model is `$Z$-invariant').
We shall denote \eqref{eqn:GC} by 
\be
\langle
E^{(r)}_{\ve'_r\ve_r}\cdots E^{(s)}_{\ve'_s\ve_s}
\rangle(\beta_r,\cdots,\beta_s).
\lb{GC1}
\en
To see the specialization \eqref{eqn:GC} is well-defined,
we note the following. In the general formula \refeq{GF}
the contour $C$ for integration is such that
$\beta_j+i(2\pi n_1+\sc[\pi,\nu]n_2+\sc[\pi,2])$
$(n_1,n_2\ge0)$
(resp.,
$\beta_j-i(2\pi n_1+\sc[\pi,\nu]n_2+\sc[\pi,2])$
$(n_1,n_2\ge0)$)
is above (resp., below) $C$. 
When $\beta_j=\beta_k+\pi i$, the contour $C$ is 
pinched by $\beta_j-\sc[\pi i,2]$ and $\beta_k+\sc[\pi i,2]$. 
However, $Q_n(\alpha|\beta)$ 
has a zero at $\alpha_a=\beta_j-\sc[\pi i,2]$
for $\bar a>j$,
and also at $\alpha_a=\beta_k+\sc[\pi i,2]$ for $\bar a<k$.
Therefore, if $j<k$, there is no pinching by poles of the total integrand.

In order for \eqref{eqn:SC} to make sense as a correlator, we must check 
the following property:
\begin{prop}\label{prop:3.1}
\begin{eqnarray}
&&\br{E^{(r)}_{\ve_r'\ve_r}\cdots E^{(s-1)}_{\ve_{s-1}'\ve_{s-1}}
(E^{(s)}_{++}+E^{(s)}_{--})}(\beta_r,\cdots,\beta_{s-1},\beta_s)
\nonumber\\
&&\qquad =
\br{E^{(r)}_{\ve_r'\ve_r}\cdots E^{(s-1)}_{\ve_{s-1}'\ve_{s-1}}}
(\beta_r,\cdots,\beta_{s-1}),
\label{eqn:Res1}
\\
&&\br{(E^{(r)}_{++}+E^{(r)}_{--})E^{(r+1)}_{\ve_{r+1}'\vep_{r+1}}
\cdots E^{(s)}_{\ve_{s}'\ve_{s}}}
(\beta_r,\beta_{r+1},\cdots,\beta_s)
\nonumber\\
&&\qquad=
\br{E^{(r+1)}_{\ve_{r+1}'\ve_{r+1}}\cdots E^{(s)}_{\ve_{s}'\ve_{s}}}
(\beta_{r+1},\cdots,\beta_s).
\label{eqn:Res2}
\end{eqnarray}
\end{prop}

\proof 
Let us take $\beta_j=\beta_{j+1}+\pi i$ in \eqref{eqn:G1}. 
Since
\[
R(\pi i)=
\pmatrix{ 0 & & &   \cr
            &1&1&   \cr
            &1&1&   \cr
            & & & 0 \cr}
\]
we find from \eqref{eqn:G4} that
\begin{eqnarray*}
&&G_{n-1}(\beta_1,\cdots,\beta_{j-1},\beta_{j+2},\cdots,\beta_{2n})
_{\ve_{1},\cdots,\ve_{j-1},\ve_{j+2},\cdots,\ve_{2n}}\\
&&=
\sum_{\ve}G_n(\beta_1,\cdots,\beta_j,\beta_{j+1},\cdots,
\beta_{2n})_{\ve_{1},\cdots,\ve,-\ve,\cdots,\ve_{2n}}
\Bigl|_{\beta_j=\beta_{j+1}+\pi i}.
\end{eqnarray*}
Eq.\eqref{eqn:Res1} is a direct consequence of this. 
The above equation together with \eqref{eqn:G2} imply 
(since $\lambda=2\pi$) that 
\[
G_{n-1}(\beta_2,\cdots,\beta_{2n-1})
_{\ve_{2},\cdots,\ve_{2n-1}}
=
\sum_{\ve}G_n(\beta_1,\cdots,\beta_{2n})_{-\ve,\ve_2,\cdots,\ve_{2n-1},\ve}
\Bigl|_{\beta_{2n}=\beta_{1}+\pi i}.
\]
From this follows \eqref{eqn:Res2}.\qed

We now write down the integral formula for \eqref{eqn:GC}.
Set $n=s-r+1$, and define 
$\bar{1},\cdots,\bar{n}$ ($r\le \bar{1}<\cdots<\bar{n}\le s+n=r+2n-1$)
by the following rule:
\[
\{\bar{1},\cdots,\bar{n}\}
=\{j\mid r\le j\le s,\,\vep_j'=-\}
\cup
\{ j^*\mid  r\le j\le s,\,\vep_j=+\}
\]
where $j^*=2s+1-j$.
We have then the following expression for \eqref{eqn:GC}.
\begin{eqnarray}
&&\langle E^{(r)}_{\ve'_r\ve_r}\cdots E^{(s)}_{\ve'_s\ve_s}
\rangle(\beta_r,\cdots,\beta_s)
\nonumber\\
&&=
\prod_{r\le j<k\le s}\frac{\sinh(\beta_j-\beta_k)}{\sinh\nu(\beta_j-\beta_k)}
\int\!\cdots\!\int\prod_{l=1}^n\frac{d\alpha_l}{2\pi}
\prod_{1\le l<l'\le n}\frac{\sinh(\alpha_l-\alpha_{l'})}
{\sinh\nu(\alpha_l-\alpha_{l'}-\pi i)}
\nonumber\\
&&\times 
\prod_{r\le \bar{l}\le s}
\Bigl[\prod_{j=r}^s\frac{i}{\sinh(\alpha_l-\beta_j+i0)}
\prod_{r\le j<\bar{l}}\sinh\nu(\alpha_l-\beta_j)
\prod_{\bar{l}< j\le s}\sinh\nu(-\alpha_l+\beta_j+\pi i)\Bigr]
\nonumber\\
&&\times 
\prod_{s+1\le \bar{l}\le s+n}
\Bigl[\prod_{j=r}^s\frac{-i}{\sinh(\alpha_l-\beta_j-i0)}
\prod_{r\le j<\bsl}
\sinh\nu(-\alpha_l+\beta_j)
\prod_{\bsl< j\le s}\sinh\nu(\alpha_l-\beta_j+\pi i)\Bigr].
\nonumber\\
&&\label{eqn:CORR}
\end{eqnarray}
Here the symbol $\alpha_l-\beta_j+i0$ 
(resp. $\alpha_l-\beta_j-i0$) indicates that the contour 
for $\alpha_l$ runs above (resp. below) $\beta_j$. 

Let us put the formula 
in a form closer to the known result for $\nu=0$, 
taking $r=1, s=n$. 
We choose the integration contour
$C_+$ for $\alpha_a$ $(1\le\bar a\le n)$
and $C_-$ for $\alpha_a$ $(n+1\le\bar a\le 2n)$
 in such a way that
$\beta_j+\pi i$ 
(resp., $\beta_j$)
$(1\le j\le n)$ are above (resp., below) $C_+$
and $\beta_j$ (resp., $\beta_j-\pi i$)
$(1\le j\le n)$ are above (resp., below) $C_-$.
(Here the contours are directed from $-\infty$ to $\infty$, 
as opposed to the $C^{\pm}$ in page 122-3 of ref.\cite{JM}.)

Set
$A'=\{j\mid \ve'_j=-\}$ and $A=\{j\mid \ve_j=+\}$.
We suppose that $\sharp(A')+\sharp(A)=n$ since otherwise
$\langle
E^{(1)}_{\ve'_1\ve_1}\cdots E^{(n)}_{\ve'_n\ve_n}\rangle
(\beta_1,\cdots,\beta_n)=0$.
We define a mapping
\be
a\in\{1,\cdots,n\}=A_+\sqcup A_-\rightarrow
\bar a\in \{1,\cdots,n\}
\en
by the condition 
that (i) $\{\bar a\mid a\in A_+\}=A'$, $\{\bar a\mid a\in A_-\}=A$;
(ii) if $a,b\in A_+$ and $a<b$ then $\bar a<\bar b$;
(iii) if $a,b\in A_-$ and $a<b$ then $\bar a>\bar b$.
In other words, the $+$'s in the sequence
$-\ve'_1,\cdots,-\ve'_n$, $\ve_n,\cdots,\ve_1$
are
$-\ve'_{\bar1},\cdots,-\ve'_{{\bar s}'}$, $\ve_{\bar s},\cdots,\ve_{\bar1}$
where $s=\sharp(A_-)=n-s'$.
Then we have
\bea
&&G_n(\beta_1+\sc[\pi i,2],\cdots,\beta_n+\sc[\pi i,2],
\beta_n-\sc[\pi i,2],\cdots,\beta_1-\sc[\pi i,2])
_{-\ve'_1,\cdots,-\ve'_n,\ve_n,\cdots,\ve_1}\nn\\
%&&=(-1)^{\sum_{a\in A_-}\bar a+n\sharp(A_+)+\sharp(A_-) +\sc[n(n-1),2]}
%\prod_{1\le j<k\le n}{\sinh(\beta_j-\beta_k)\over\sinh\nu(\beta_j-\beta_k)}
%\nn\\
%&&\times\prod_{a\in A_+}\int_{C_+}{d\alpha_a\over2\pi i}
%\prod_{a\in A_-}\int_{C_-}{d\alpha_a\over2\pi i}
%\prod_{1\le a<b\le n}{\sinh(\alpha_a-\alpha_b)\over
%\sinh\nu(\alpha_a-\alpha_b-\pi i)}\nn\\
%&&\times
%{\prod_{\{{a\in A_+\atop j<\bar a}\}\cup\{{a\in A_-\atop j<\bar a}\}}
%\sinh\nu(\alpha_a-\beta_j)
%\over
%\prod_{1\le a,j\le n}\sinh(\alpha_a-\beta_j)}
%\nn\\
%&&\times\prod_{a\in A_+\atop j>\bar a}\sinh\nu(\beta_j-\alpha_a+\pi i)
%\prod_{a\in A_-\atop j>\bar a}\sinh\nu(\alpha_a-\beta_j+\pi i)\nn\\
&&=(-1)^{\sum_{a\in A_+}\bar a+\sum_{a\in A_-}\bar a+\sharp(A_-)
+\sc[n(n-1),2]}
\prod_{1\le j<k\le n}{\sinh(\beta_j-\beta_k)\over\sinh\nu(\beta_j-\beta_k)}
\nn\\
&&\times\prod_{a\in A_+}\int_{C_+}
{d\alpha_a\over2\pi i\sinh\nu(\alpha_a-\beta_{\bar a})}
\prod_{a\in A_-}\int_{C_-}
{d\alpha_a\over2\pi i\sinh\nu(\alpha_a-\beta_{\bar a})}\nn\\
&&\times\prod_{1\le a<b\le n}{\sinh(\alpha_a-\alpha_b)\over
\sinh\nu(\alpha_a-\alpha_b-\pi i)}
\prod_{1\le a,j\le n}{\sinh\nu(\alpha_a-\beta_j)\over\sinh(\alpha_a-\beta_j)}
\nn\\
&&\times\prod_{a\in A_+\atop j>\bar a}
{\sinh\nu(\beta_j-\alpha_a+\pi i)\over\sinh\nu(\beta_j-\alpha_a)}
\prod_{a\in A_-\atop j>\bar a}
{\sinh\nu(\alpha_a-\beta_j+\pi i)\over\sinh\nu(\alpha_a-\beta_j)}.
\ena
In the limit $\nu=0$ we recover the integral
formula for the XXX correlation functions (\cite{KIEU,Nak,JM}).

\setcounter{section}{3}
\setcounter{equation}{0}
\section{Nearest neighbour correlator}\label{sec:4}

In this section we examine the simplest cases of the general formula 
for the correlators proposed in the previous section.

First consider the case $G_1(\beta_1,\beta_2)$. 
Taking $n=1$ in the formula \eqref{eqn:once}, 
we immediately find the following.

\begin{prop}
\[
G_1(\beta_1,\beta_2)_{-+}=
G_1(\beta_1,\beta_2)_{+-}=
\frac{1}{2\nu}\frac{\rho(\beta_1-\beta_2)}{\rho(\pi i)}
\frac{\sinh\frac{\nu}{2}(\beta_1-\beta_2-\pi i)}
{\sinh\frac{1}{2}(\beta_1-\beta_2-\pi i)}.
\]
In particular, by setting $\beta_1=\beta_2+\pi i$, we have
\begin{equation}
\br{E_{++}^{(1)}}=\br{E_{--}^{(1)}}=\sc[1,2].
\label{eqn:1pt}
\end{equation}
\end{prop}

Next let us take $n=2$ in \eqref{eqn:once}.

\begin{prop} Assuming $0<\nu<1/2$, 
we have
\begin{eqnarray}
&&G_2(\beta_1,\cdots,\beta_4)_{++--}
=\frac{\prod_{j<k}\rho(\beta_j-\beta_k)}
{\rho(0)^2\rho(\pi i)^4}
\frac{e^{-\sum_{j=1}^4\beta_j/2}}{2\pi\nu^2\sum_{j=1}^4e^{-\beta_j}}
\label{eqn:generic}\\
&&\times
\int d\alpha e^\alpha \,
\prod_{j=1}^4\varphi(\alpha-\beta_j)\,
\sinh\nu\bigl(-\alpha+\beta_3+\sc[\pi i,2]\bigr)
\sinh\nu\bigl(-\alpha+\beta_4+\sc[\pi i,2]\bigr)
\nonumber\\
&&
\times 
\Bigl[
\sinh\nu\bigl(-\alpha+\beta_2+\sc[\pi i,2]\bigr)
\sinh\nu\bigl(
\alpha+\sc[\beta_2-\beta_1-\beta_3-\beta_4,2]-\sc[3\pi i,2]\bigr)
\nonumber\\
&&
-\sinh\nu\bigl(\alpha-\beta_1+\sc[\pi i,2]\bigr)
\sinh\nu\bigl(\alpha+
\sc[\beta_1-\beta_2-\beta_3-\beta_4,2]-\sc[\pi i,2]\bigr)
\Bigr].
\nonumber
\end{eqnarray}
The integral is taken along a path from $-\infty$ to $+\infty$ such that 
$-\sc[\pi,2]<\Im(\alpha-\beta_j)<\sc[\pi,2]$ for all $j$. 
\end{prop}
Here we have used the relation $\rho(0)\rho(\pi i)=-\sc[1,4\sqrt{\nu}]$ which 
follows from \eqref{eqn:RZ}, \eqref{eqn:RR} 
and $S_2(\pi|2\pi,\sc[\pi,\nu])=\sqrt{2}$.

Upon specialization $(\beta_1,\cdots,\beta_4)=
(\beta+\pi i,\beta+\pi i,\beta,\beta)$, 
this integral can be processed further. 
After a chain of steps detailed in Appendix \ref{app:D}, 
we obtain the following result.

\begin{prop}\label{prop:p1}We have
\begin{eqnarray}
\br{E^{(1)}_{--}E^{(2)}_{--}}&=&
G_2(\beta+\pi i,\beta+\pi i,\beta, \beta)_{++--}
\nonumber\\
&=&\frac{1}{\pi^2\sin\pi\nu}
\frac{d}{d\nu}
\left(\sin\pi\nu
\int_0^\infty \frac{\sinh(1-\nu)t}{\sinh t \cosh\nu t}dt
\right)
+\frac{1}{2}.
\label{eqn:ans}
\end{eqnarray}
\end{prop}
We note that, 
since both sides are holomorphic with respect to $\nu$ for 
$0<\Re\,\nu<1$, \eqref{eqn:ans} is valid without 
the restriction $0<\nu<\frac{1}{2}$.

We now compare the formulas \eqref{eqn:1pt}, \eqref{eqn:ans} 
with known answers.
For this purpose let us quote 
from \cite{Bax82} the results concerning the XXZ model 
which are relevant to the following discussion.

The XXZ model for a periodic chain of circumference $N$ 
is given by the Hamiltonian 
\begin{equation}
H=-\frac{1}{2}\sum_{j=1}^N
\left(\sigma^x_j\sigma^x_{j+1}
+\sigma^y_j\sigma^y_{j+1}
+\Delta\sigma^z_j\sigma^z_{j+1}
\right). 
\label{eqn:Hamil}
\end{equation}
This Hamiltonian \eqref{eqn:Hamil} is associated with 
the six vertex model with the Boltzmann weights (\cite{Bax82}, eq.(8.8,9))
\begin{equation}
a=\sin\frac{\mu-w}{2},
\quad 
b=\sin\frac{\mu+w}{2},
\quad
c=\sin \mu.
\label{eqn:Boltz}
\end{equation}
Denoting by $T(w)$ the transfer matrix of the periodic system with 
$N$ columns, 
we have 
\begin{equation}
\frac{d}{dw}\log T(w)\Bigl|_{w=-\mu}
=-\frac{1}{2\sin\mu}\left(H+\frac{N}{2}\cos\mu\right)
\label{eqn:TH}
\end{equation}
where $\Delta$ is related to $\mu$ via 
\begin{equation}
\Delta=-\cos\mu.
\label{eqn:del}
\end{equation}
The gapless regime $|\Delta|\le 1$ corresponds to $\mu$ being real.

For a local operator $O$, let 
$\brvac{O}$ denote its 
ground state average (in the limit $N\rightarrow\infty$).
Since in the gapless regime the vacuum $\vac$ is invariant
under the $+\leftrightarrow-$ symmetry, one must have 
\[
\brvac{\sigma^z_1}=\brvac{\left(E^{(1)}_{++}-E^{(1)}_{--}\right)}=0.
\]
Along with $1=\brvac{\left(E^{(1)}_{++}+E^{(1)}_{--}\right)}$, this means
\begin{equation}
\brvac{E^{(1)}_{++}}=\brvac{E^{(1)}_{--}}=\frac{1}{2}.
\label{eqn:1pt1}
\end{equation}
Our formula \eqref{eqn:1pt} is consistent with this.

In the limit $N\rightarrow\infty$, 
the free energy per site $f$ is given by (\cite{Bax82},eq.(8.8.17))
\begin{equation}
-\frac{f}{kT}=\log a +\int_{-\infty}^\infty
\frac{\sinh(\mu+w)x\sinh(\pi-\mu)x}{2x\sinh\pi x\cosh\mu x}dx.
\label{eqn:free}
\end{equation}
It follows from the relation \eqref{eqn:TH} that 
the ground state energy per site $e_0$ of the XXZ chain is 
\begin{equation}
e_0=-\frac{1}{2}\cos\mu
-2\sin\mu \frac{\partial}{\partial w}\left(-\frac{f}{kT}\right)\Bigl|_{w=-\mu}. 
\label{eqn:e0}
\end{equation}

Differentiating $e_0$ with respect to $\Delta$, 
we can obtain the nearest neighbour correlator for the $\sigma^z$ operators:
\[
\brvac{\sigma^z_1\sigma^z_2}
=-2\frac{d e_0}{d\Delta}
=-\frac{2}{\sin\mu}\frac{d e_0}{d\mu}.
\]
Inserting \eqref{eqn:free} into \eqref{eqn:e0}, we find 
the following expression for this quantity: 
\begin{equation}
\brvac{\sigma^z_1\sigma^z_2}
=1+\frac{4}{\sin\mu}
\frac{d}{d\mu}
\left(\sin\mu
\int_0^\infty \frac{\sinh(\pi-\mu)x}{\sinh\pi x\cosh\mu x}dx
\right).
\label{eqn:s1s2}
\end{equation}
On the other hand, in view of \eqref{eqn:1pt1} we have
\[
\brvac{\sigma^z_1\sigma^z_2}
=\brvac{(1-2E_{--}^{(1)})(1-2E_{--}^{(2)})}
=4\brvac{E_{--}^{(1)}E_{--}^{(2)}}-1.
\]
Therefore, the formula
\eqref{eqn:ans} agrees with \eqref{eqn:s1s2} with the identification
$\mu=\pi\nu$.

\setcounter{section}{4}
\setcounter{equation}{0}

\section{The XY limit}\label{sec:5}

In this section we study the integral formula for the correlation
functions \eqref{eqn:CORR} at a special value of the parameter $\nu=1/2$.
This is the case where the XXZ chain reduces to the XY chain $\Delta=0$. 
It is well-known that the XY chain 
is equivalent to the two-dimensional Ising model. 
To be more precise, the XXZ model with $\Delta=0$ corresponds to
the critical Ising model. In this case, diagonalizing the transfer matrix 
in terms of free fermions, one can 
calculate the correlation functions directly 
in the presence of arbitrary  spectral parameters.
The diagonalization is worked out
in Appendix \ref{app:F}.
Here we show that the formulas thus obtained 
give the same result as the integral formula \eqref{eqn:CORR}.

We shall consider the function
\[
\br{E^{(r)}_{\vep_r'\vep_r}E^{(r+1)}_{\vep_{r+1}'\vep_{r+1}}
\cdots E^{(s)}_{\vep_s'\vep_s}}(\beta_r,\beta_{r+1},\cdots,\beta_s)
\]
given by \refeq{GC1}.
In order to simplify the presentation, 
we shall take all $\beta_j$'s to be real
throughout this section. 
(This is a matter of convenience and not actually a restriction.
The final formulas are valid as meromorphic functions in $\beta_j$'s.)

A special feature about $\nu=\sc[1,2]$ is that \eqref{eqn:CORR} 
becomes a determinant. 
\begin{prop}
If $\nu=\sc[1,2]$, we have
\begin{equation}
\br{E^{(r)}_{\vep_r'\vep_r}\cdots E^{(s)}_{\vep_s'\vep_s}}
(\beta_r,\cdots,\beta_s) 
=
\prod_{r\le j<k\le s}2i\cosh\sc[1,2](\beta_j-\beta_k)
\times
\det\left(I_{k\bar{k'}}\right)_{1\le k,k'\le n}.
\label{eqn:EE}
\end{equation}
Here $n=s-r+1$, the $I_{kl}$ are given for $r\le l\le s$ by
\begin{eqnarray}
I_{k l}
&=&
2i^{2r-s+1-l}\int
\frac{d\alpha}{2\pi}\,
e^{(k-\frac{n+1}{2})\alpha}
\nonumber\\
&&\times
\prod_{j=r}^{l}\frac{1}{2\cosh\frac{1}{2}(\alpha-\beta_j)}
\prod_{j=l}^{s}\frac{1}{2\sinh\frac{1}{2}(\alpha-\beta_j+i0)}\,,
\label{eqn:Iki}
\end{eqnarray}
and for $s+1\le l\le s+n$ 
\[
I_{kl}=\overline{I_{kl^*}}
\]
with the bar denoting the complex conjugate.
In \eqref{eqn:Iki}, 
the integration contour is a line above the 
real axis, as indicated by the symbol $+i0$.
\end{prop}

\proof
Specializing the formula \eqref{eqn:CORR}  to  $\nu=1/2$ we find
\begin{eqnarray*}
&&\prod_{r\le j<k\le s}2\cosh\frac{\beta_j-\beta_k}{2}
\int\!\cdots\!\int \prod_{l=1}^n\frac{d\alpha_l}{2\pi} 
\prod_{l<l'}2i\sinh\frac{1}{2}(\alpha_l-\alpha_{l'})
\\
&&\times
\prod_{r\le \bar{l}\le s}
\Bigl[2i^{n+s-\bar{l}}
\prod_{j=r}^{\bar{l}}\frac{1}{2\cosh\frac{1}{2}(\alpha_l-\beta_j)}
\prod_{j=\bar{l}}^{s}\frac{1}{2\sinh\frac{1}{2}(\alpha_l-\beta_j+i0)}
\Bigr]
\\
&&\times
\prod_{s+1\le \bar{l}\le s+n}
\Bigl[2i^{-r-1+\bsl}
\prod_{j=r}^{\bsl}\frac{1}{2\cosh\frac{1}{2}(\alpha_l-\beta_j)}
\prod_{j=\bsl}^{s}\frac{1}{2\sinh\frac{1}{2}(\alpha_l-\beta_j-i0)}
\Bigr].
\end{eqnarray*}
Inserting
\[
\prod_{l<l'}2i\sinh\frac{1}{2}(\alpha_l-\alpha_{l'})
=
i^{-n(n-1)/2}e^{-\frac{n+1}{2}(\alpha_1+\cdots+\alpha_n)}\,
\det\left(e^{k\alpha_{k'}}\right)_{1\le k,k'\le n}
\]
we obtain the right hand side of \eqref{eqn:EE}.
\qed

In Appendix \ref{app:E} we evaluate the integral \eqref{eqn:Iki} explicitly
(see \eqref{eqn:Iki1}). 

We now proceed to the calculation of correlation functions of the 
fermion operators 
\begin{eqnarray*}
\psi_m^*&=&\cdots \sigma^z_{m-2}\sigma^z_{m-1}\sigma^+_m,
\\
\psi_m&=&\cdots \sigma^z_{m-2}\sigma^z_{m-1}\sigma^-_m.
\end{eqnarray*}
We shall consider only monomials consisting 
of an {\it even} number of such operators.
They are local operators in the sense of Section 3.
For instance
\[
\psi_m\psi^*_l
=
\cases{ 
\sigma^-_{m}\sigma^z_{m+1}\cdots\sigma^z_{l-1}\sigma^+_l & ($m<l$),\cr
E^{(m)}_{--} & ($m=l$),\cr
\sigma^+_{l}\sigma^z_{l+1}\cdots\sigma^z_{m-1}\sigma^-_m & ($m>l$).\cr
}.
\]
Clearly the function $\br{O}$ for a monomial $O$ is $0$ (see \refeq{GC1})
unless it consists of the same number of $\psi$'s and $\psi^*$'s.

The following two propositions will be proved in Appendix \ref{app:E}.
\begin{prop}\label{prop:wick}
\begin{equation}
\br{\psi_{m_1}\cdots\psi_{m_k}\psi^*_{l_k}\cdots\psi^*_{l_1}}
=\det\left(\br{\psi_{m_j}\psi^*_{l_i}}\right)_{1\le j,i\le k}.
\label{eqn:wick}
\end{equation}
\end{prop}

\begin{prop}\label{prop:two}
\begin{eqnarray}
\br{\psi_m\psi^*_l}&=&(-1)^{m+l}\br{\psi^*_m\psi_l}
\nonumber\\
&=&-\frac{i^{l-m+1}}{\pi}\left(B_mB_l\right)^{1/2}
\sum_{j=m}^l\beta_j\frac{\prod_{i=m+1}^{l-1}(B_j+B_i)}
{\prod_{i=m\atop i\neq j}^l(B_j-B_i)}
\qquad (m<l),
\nonumber\\
&&\label{eqn:psipsi1}
\\
&=&\frac{1}{2}
\qquad (m=l)
\label{eqn:psipsi2}.
\end{eqnarray}
\end{prop}

The formulas \eqref{eqn:psipsi1}, \eqref{eqn:psipsi2} give the 
same results 
for the corresponding quantities \eqref{eqn:CF} 
$\dvac \psi_m\psi^*_l\vac$ obtained directly by 
diagonalizing the Hamiltonian (see Appendix \ref{app:F}).
In general, the multiple correlators of the fermions are given by applying 
Wick's theorem. Since $\brvac{\psi_m\psi_l}=\brvac{\psi^*_m\psi^*_l}=0$, 
the result is given as a determinant in the same way as \eqref{eqn:wick}.
Any local operator (i.e., a finite linear combination of monomials in 
$\sigma^\alpha_j$'s) can also be written as a linear combination of monomials 
of the fermions. 
Therefore, we can state

\begin{prop}
For an arbitrary local operator $O$, $\br{O}=\dvac O\vac$ holds. 
\end{prop}

\setcounter{section}{5}
\setcounter{equation}{0}
\section{Discussions}

Before concluding the paper, let us 
touch upon previous works on the $q$KZ equation with $|q|=1$. 
In \cite{Smir} Smirnov introduced and solved a
system of difference equations for the form factors
of local operators in the sine-Gordon theory.
His equations are the same as \refeq{G1} and \refeq{G2} in this paper
except that $S=-R$ is used (see page 29 in \cite{Smir}) and that 
$\lambda=-2\pi$ instead of $\lambda>0$ (the case 
relevant to the correlation function is $\lambda=2\pi$).
There is a significant difference between \refeq{G3} in this paper
and the equation (16) (page 11) in Smirnov's.
The latter requires that the solution has a simple pole at the point
$\beta_{2n}=\beta_{2n-1}+\pi i$, while the former requires that the
solution is regular there. The physical origin of this difference
is that in Smirnov's case the poles are the annihilation
poles of the form factors while in our case \refeq{G3} is the
normalization of the correlation functions (see Proposition \ref{prop:3.1}).

There are other mathematical differences between Smirnov's
formula and ours. The number of integration is $n$-fold in our formula
in contrast to the $(n-1)$-fold integrals in Smirnov's.
Since the integration can be carried out once (see \eqref{eqn:once}), 
this difference
is rather superficial. 
The significant difference is that in Smirnov's formula 
the $(n-1)$-fold integral reduces to the determinant of an
$(n-1)\times(n-1)$ matrix with entries given by integrals with respect to
a single variable. This is not the case in our formula (except for 
$\nu=\sc[1,2]$ ---the case of the XY model). 
This lack of determinantal structure
is already noted by Nakayashiki in \cite{Nak},
where the special case $\nu=0$ was studied.

In a recent paper \cite{Count}, Smirnov has constructed an affluent family
of solutions that corresponds to a family of local operators in the sine-Gordon
theory. In our case, the structure of the total space of solutions is 
absolutely unknown. 
In this connection, 
let us mention an open problem: 
to show that our integrals satisfy 
$G_n(\beta_1,\cdots,\beta_{2n})_{\ve_1,\cdots,\ve_{2n}}
=G_n(\beta_1,\cdots,\beta_{2n})_{-\ve_1,\cdots,-\ve_{2n}}$. 
A direct verification seems difficult, and 
we suspect that it should follow 
from the uniqueness of the solution satisfying certain 
analyticity and asymptotics.

In fact, the form factors and the correlation functions are closely related.
As was discussed in \cite{DFJMN,JM}, in the  regime $\Delta<-1$,
the former are represented by the type II vertex operators and the latter
by the type I vertex operators. The type I vertex operators generate a family
of solutions to the form factor equations, and vice versa.
In the sine-Gordon theory, this viewpoint was explored by Lukyanov in
\cite{Luk95}. Lukyanov has introduced the appropriate commutation relations
for the vertex operators, and has given a bosonization of the 
sine-Gordon theory with a cut-off parameter. 
Though we have not checked the details, 
it seems likely that the integral formula for $G_n$ in 
this paper is derivable from Lukyanov's bosonization after taking the 
cut-off parameter to infinity. 

In the approach of this paper, the role of the quantum affine algebra
$U_q(\widehat{sl}_2)$ is unclear. In the $\Delta<-1$ regime,
the free energy, the excitation spectrum and the correlation functions
depend on the spectral parameters $\beta_j$ through $\zeta_j=e^{-\nu\beta_j}$.
In the gapless regime, these quantities are single-valued only in $\beta_j$
and the period $\sc[2\pi i,\nu]$ is lost. 
What is the implication of this fact in the representation theory? 
This is an interesting question to be asked.

\appendix
\setcounter{equation}{0}
\section{Multiple gamma functions}\label{app:A}

The multiple gamma and sine functions were introduced by 
Barnes\cite{Barn1,Barn2}, 
Shintani\cite{Shin} and Kurokawa\cite{Kuro}. 
Here we follow the notation of \cite{Kuro}. 
In what follows we fix an $r$-tuple of complex numbers 
$\omb=(\omega_1,\cdots,\omega_r)$. 
For simplicity we shall assume that $\Re\omega_i>0$. 
We set $\nb\cdot\omb=n_1\omega_1+\cdots+n_r\omega_r$ 
($\nb=(n_1,\cdots,n_r)$), 
$|\omb|=\omega_1+\cdots+\omega_r$. 

The multiple gamma and associated functions are defined as follows. 

\begin{description}
\item[Multiple Hurwitz zeta function]
\begin{equation}
\zeta_r(s,x|\omb)=\sum_{n_1,\cdots,n_r\ge 0}(\nb\cdot\omb+x)^{-s}
\label{eqn:Hzeta}
\end{equation}
\item[Multiple gamma function]
\begin{equation}
\Gamma_r(x|\omb)=\exp\bigl(\zeta_r'(0,x|\omb)\bigr)
\qquad ('=\frac{\partial}{\partial s})
\label{eqn:gam}
\end{equation}
\item[Multiple digamma function]
\begin{equation}
\psi_r(x|\omb)=\frac{d}{dx}\log \Gamma_r(x|\omb)
\label{eqn:digam}
\end{equation}
\item[Multiple sine function]
\begin{equation}
S_r(x|\omb)=\Gamma_r(x|\omb)^{-1}
\Gamma_r(|\omb|-x|\omb)^{(-1)^r}
\label{eqn:sin}
\end{equation}
\item[Multiple Bernoulli polynomials]
\begin{equation}
\frac{t^re^{xt}}{\prod_{i=1}^r(e^{\omega_it}-1)}
=\sum_{n=0}^\infty\frac{t^n}{n!}B_{r,n}(x|\omb)
\label{eqn:Bern}
\end{equation}
\end{description}

When $r=1$, they are related with the ordinary gamma and other functions 
via
\begin{eqnarray*}
&&\zeta_1(s,x|\omega_1)=\omega_1^{-s}\zeta(s,\frac{x}{\omega_1}),
\\
&&\Gamma_1(x|\omega_1)=\omega_1^{\frac{x}{\omega_1}-\frac{1}{2}}
\frac{\Gamma\bigl(\frac{x}{\omega_1}\bigr)}{\sqrt{2\pi}},
\\
&&\psi_1(x|\omega_1)=\frac{1}{\omega_1}\Bigl(\psi\bigl(\frac{x}{\omega_1}\bigr)
+\log\omega_1\Bigr),
\\
&&S_1(x|\omega_1)
=2\sin\bigl(\frac{\pi x}{\omega_1}\bigr).
\end{eqnarray*}

We list here the basic properties of these functions. 

\begin{description}
\item[Difference equations]
Set $\omb(i)=(\omega_1,\cdots,\omega_{i-1},\omega_{i+1},\cdots,\omega_r)$. 
\begin{eqnarray}
&&
\zeta_r(s,x+\omega_i|\omb)-\zeta_r(s,x|\omb)=\zeta_{r-1}(s,x|\omb(i)),
\label{eqn:diffzeta}\\
&&
\frac{\Gamma_r(x+\omega_i|\omb)}{\Gamma_r(x|\omb)}
=\frac{1}{\Gamma_{r-1}(x|\omb(i))},
\label{eqn:diffgam}\\
&&
\frac{S_r(x+\omega_i|\omb)}{S_r(x|\omb)}
=\frac{1}{S_{r-1}(x|\omb(i))},
\label{eqn:diffsin}\\
&&
B_{r,n}(x+\omega_i|\omb)-B_{r,n}(x|\omb)=nB_{r-1,n-1}(x|\omb(i)).
\label{eqn:diffBer}
\end{eqnarray}
\item[Analyticity]
As a function of $s$, 
$\zeta_r(s,x|\omb)$ is continued meromorphically on the whole complex
plane and is holomorphic except for simple poles at $s=1,\cdots,r$.
We have 
\begin{eqnarray*}
&&\zeta_r(n,x|\omb)=\frac{(-1)^n}{(n-1)!}\psi^{(n-1)}_r(x|\omb)
\qquad (n>r),
\\
&&\zeta_r(-n,x|\omb)=(-1)^r\frac{n!}{(n+r)!}B_{r,n+r}(x|\omb)
\qquad (n\ge 0),
\\
&&\lim_{s\rightarrow n}(s-n)\zeta_r(s,x|\omb)
=(-1)^{n-r}\frac{B_{r,r-n}(x|\omb)}{(n-1)!(r-n)!}
\qquad (r\ge n\ge 1).
\end{eqnarray*}

$\Gamma_r(x|\omb)^{-1}$ is an entire function of $x$. 
$\Gamma_r(x|\omb)$ is meromorphic with poles at 
$x=\nb\cdot\omb$ ($n_1,\cdots,n_r\le 0$). 

$S_r(x|\omb)$ is entire in $x$ when $r$ is odd, and 
is meromorphic when $r$ is even. Its zeroes and poles are given by 
\begin{eqnarray*}
\hbox{r:odd }&&\hbox{ zeroes at }x=\nb\cdot\omb 
\qquad (n_1,\cdots,n_r\ge 1 \hbox{ or } n_1,\cdots,n_r\le 0 ), 
\\
\hbox{r:even }&&\hbox{ zeroes at }x=\nb\cdot\omb 
\qquad (n_1,\cdots,n_r\le 0 ), 
\\
&&\hbox{ poles at }x=\nb\cdot\omb 
\qquad (n_1,\cdots,n_r\ge 1).
\end{eqnarray*}
All zeroes and poles are simple if $\nb\cdot\omb$'s do not overlap. 

\item[Integral representations]
If $\Re x>0$ then 
\begin{eqnarray*}
\zeta_r(s,x|\omb)&=&
-\Gamma(1-s)\int_C
\frac{e^{-xt}(-t)^{s-1}}{\prod_{i=1}^r(1-e^{-\omega_it})}\frac{dt}{2\pi i},
\\
\log\Gamma_r(x|\omb)&=&
\gamma\frac{(-1)^r}{r!}B_{r,r}(x|\omb)
+\int_C
\frac{e^{-xt}\log(-t)}{\prod_{i=1}^r(1-e^{-\omega_it})}\frac{dt}{2\pi i t},
\\
\psi_r(x|\omb)&=&
\gamma\frac{(-1)^r}{r!}B_{r,r}'(x|\omb)
-\int_C
\frac{e^{-xt}\log(-t)}{\prod_{i=1}^r(1-e^{-\omega_it})}\frac{dt}{2\pi i},
\end{eqnarray*}
where $\gamma=$Euler's constant and $\Gamma(x)$ denotes the ordinary 
gamma function. 
The contour $C$ is shown in Figure \ref{fig:af1}.

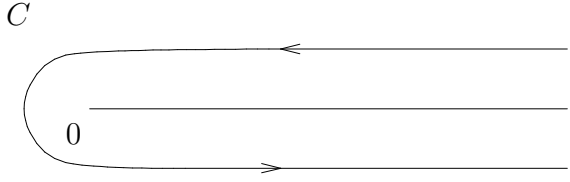
\begin{figure}[htb]
\begin{center}
\setlength{\unitlength}{0.0125in}
\begin{picture}(235,86)(0,-10)
\drawline(115,0)(235,0)
\drawline(75,0)(115,0)
\drawline(107.000,-2.000)(115.000,0.000)(107.000,2.000)
\drawline(235,50)(115,50)
\drawline(123.000,52.000)(115.000,50.000)(123.000,48.000)
\drawline(35,25)(235,25)
\drawline(115,50)
	(110.068,49.998)
	(105.273,49.990)
	(100.615,49.978)
	(96.094,49.961)
	(91.709,49.939)
	(87.461,49.912)
	(83.350,49.880)
	(79.375,49.844)
	(75.537,49.802)
	(71.836,49.756)
	(68.271,49.705)
	(64.844,49.648)
	(61.553,49.587)
	(58.398,49.521)
	(55.381,49.451)
	(52.500,49.375)
	(49.756,49.294)
	(47.148,49.209)
	(42.344,49.023)
	(38.086,48.818)
	(34.375,48.594)
	(31.211,48.350)
	(28.594,48.086)
	(25.000,47.500)
	(20.312,45.781)
	(16.250,43.125)
	(12.812,39.531)
	(10.000,35.000)
	(8.906,32.500)
	(8.125,30.000)
	(7.656,27.500)
	(7.500,25.000)
	(7.656,22.500)
	(8.125,20.000)
	(8.906,17.500)
	(10.000,15.000)
	(12.812,10.469)
	(16.250,6.875)
	(20.312,4.219)
	(25.000,2.500)
	(27.969,1.914)
	(31.875,1.406)
	(36.719,0.977)
	(39.492,0.791)
	(42.500,0.625)
	(45.742,0.479)
	(49.219,0.352)
	(52.930,0.244)
	(56.875,0.156)
	(61.055,0.088)
	(65.469,0.039)
	(70.117,0.010)
	(75.000,0.000)

\put(0,60){\makebox(0,0)[lb]{\raisebox{0pt}[0pt][0pt]{\shortstack[l]{{\twlrm $C$}}}}}
\put(25,10){\makebox(0,0)[lb]{\raisebox{0pt}[0pt][0pt]{\shortstack[l]{{\twlrm $0$}}}}}
\end{picture}
\vskip 2cm
\caption{The contour $C$.}
\label{fig:af1}
\end{center}
\end{figure}

\item[Asymptotic expansion]
Assume $\omega_1,\cdots,\omega_r>0$. Then for any $N\ge 1$ we have 
\begin{eqnarray}
\log\Gamma_r(z|\omb)
&=&
(-1)^{r-1}\sum_{k=0}^r
\frac{B_{r,r-k}(0)}{(r-k)!}\frac{z^k}{k!}
\Bigl(\log z+\gamma-\sum_{j=1}^k\frac{1}{j}\Bigr)
+\gamma\zeta_r(0,z)
\nonumber\\
&&+\sum_{n=1}^N (-1)^{n-r}\frac{(n-1)!}{(n+r)!}B_{r,r+n}(0)z^{-n}
+o(z^{-N})
\label{eqn:asymp}
\end{eqnarray}
as $z\rightarrow\infty$ in the angular domain $|Arg(z-x)|\le \pi-\epsilon$, 
where $x>0$ and $0<\epsilon<\pi$. 
\end{description}

The case $r=2$ is of special interest to us. 
In this case the following formulas hold. 
\begin{eqnarray}
&&\log S_2(x|\omb)
=\int_C \frac{\sinh\left(x-\frac{\omega_1+\omega_2}{2}\right)t}
{2\sinh\frac{\omega_1t}{2}\sinh\frac{\omega_2t}{2}}
\log(-t)\frac{dt}{2\pi i t},
\qquad (0<\Re x<\omega_1+\omega_2)
\nonumber\\
&&\frac{S_2(x+\omega_1|\omb)}{S_2(x|\omb)}
=\frac{1}{2\sin\displaystyle\frac{\pi x}{\omega_2}},
\\
&&S_2(x|\omb)S_2(-x|\omb)
=-4\sin\frac{\pi x}{\omega_1}\sin\frac{\pi x}{\omega_2},
\\
&&S_2(x|\omb)=\frac{2\pi}{\sqrt{\omega_1\omega_2}}\,x+O(x^2)
\qquad (x\longrightarrow 0),
\\
&&S_2(\omega_1|\omb)=\sqrt{\frac{\omega_2}{\omega_1}},
\qquad
S_2\Bigl(\frac{\omega_1}{2}|\omb\Bigr)=\sqrt{2},
\qquad
S_2\Bigl(\frac{\omega_1+\omega_2}{2}|\omb\Bigr)=1.
\nonumber
\end{eqnarray}
In addition, as $x\rightarrow \infty$ ($\pm\Im x>0$), we have 
\begin{eqnarray}
&&\log S_2(x|\omb)=
\pm \pi i\left(\frac{x^2}{2\omega_1\omega_2}
-\frac{\omega_1+\omega_2}{2\omega_1\omega_2}x
-\frac{1}{12}\bigl(\frac{\omega_1}{\omega_2}+\frac{\omega_2}{\omega_1}+3
\bigr)\right) + o(1),
\nonumber\\
&&\log S_2(a+x|\omb)S_2(a-x|\omb)=
\pm\pi i  \frac{(2a-\omega_1-\omega_2)}{\omega_1\omega_2}x
+o(1).
\label{eqn:ES}
\end{eqnarray}

\setcounter{equation}{0}
\section{Proof of difference equations for $G_n$}\label{app:C}

\def\oR{\overline R}
\def\oG{\overline G}
\def\ve{\varepsilon}
\def\sc[#1,#2]{{\scriptstyle{\scriptstyle#1\over\scriptstyle#2}}}

We prove here that the integral formula \eqref{eqn:GF} possesses the 
required properties \eqref{eqn:G1}, \eqref{eqn:G2}, \eqref{eqn:G3}.

\par\smallskip\noindent{\it Proof of \refeq{G1}.}\quad
Let $\oG_n=G_n/\rho_n$, $\rho_n=\prod_{j<k}\rho(\beta_j-\beta_k)$.
Because of \refeq{RI}, \refeq{G1} reduces to th same equation for
$\oG_n$ wherein $R$ is replaced by
$\oR$.

There are four cases to consider: 
$(\ve_j,\ve_{j+1})=(-,-)$, $(+,-)$, $(-,+)$ and $(+,+)$.
\itm{Case $(\ve_j,\ve_{j+1})=(-,-)$}
We are to show that
\be
\oG_n(\cdots,\beta_{j+1},\beta_j,\cdots)_{\cdots--\cdots}
=\oG_n(\cdots,\beta_j,\beta_{j+1},\cdots)_{\cdots--\cdots}.
\en
This is obvious because the integrand of $\oG_n$ is symmetric with respect to
$\beta_j$ and $\beta_{j+1}$ if $(\ve_j,\ve_{j+1})=(-,-)$.
\itm{Case $(\ve_j,\ve_{j+1})=(-,+)$}
Suppose that $\bar a=j+1$, and set $\alpha=\alpha_a$.
Comparing the integrands of 
$\oG_n(\cdots,\beta_{j+1},\beta_j,\cdots)_{\cdots+-\cdots}$
and
$\oG_n(\cdots,\beta_j,\beta_{j+1},\cdots)_{\cdots\pm\mp\cdots}$,
we see that the desired equality follows from
\begin{eqnarray*}
&&\bar b(\beta_j-\beta_{j+1})\sinh\nu(\alpha-\beta_j+\sc[\pi i,2])
+\bar c(\beta_j-\beta_{j+1})\sinh\nu(\beta_{j+1}-\alpha+\sc[\pi i,2])
\\
&&=\sinh\nu(\beta_j-\alpha+\sc[\pi i,2]).
\end{eqnarray*}

\itm{Case $(\ve_j,\ve_{j+1})=(+,-)$}
This is similar to the case $(\ve_j,\ve_{j+1})=(-,+)$.

\itm{Case $(\ve_j,\ve_{j+1})=(+,+)$}
We are to show that
\be
\oG_n(\cdots,\beta_{j+1},\beta_j,\cdots)_{\cdots++\cdots}
=\oG_n(\cdots,\beta_j,\beta_{j+1},\cdots)_{\cdots++\cdots}.
\lb{PP}
\en
Suppose that $\bar a=j$, and set $\alpha=\alpha_a$ and $\alpha'=\alpha_{a+1}$.
Apart from the factors
that are symmetric with respect to $\beta_j$
and $\beta_{j+1}$ and antisymmetric with respect to $\alpha$ and $\alpha'$,
the integrand of the LHS of \refeq{PP} contains
\be
{\sinh\nu(\alpha'-\beta_{j+1}+\sc[\pi i,2])
\sinh\nu(\beta_j-\alpha+\sc[\pi i,2])
\over
\sinh\nu(\alpha-\alpha'-\pi i)}.
\en
Antisymmetrizing it with respect to the variables $\alpha$ and $\alpha'$, we
obtain an expression that is symmetric with respect to $\beta_j$ and $\beta_{j+1}$. Therefore, we have \refeq{PP}.
\par\smallskip\noindent{\it Proof of \refeq{G2}.}\quad
Because of \refeq{RC}, the equality \refeq{G2} is equivalent to
\be
\oG_n(\beta_1,\cdots,\beta_{2n-1},\beta_{2n}-i\lambda)_{\ve_1\cdots\ve_{2n}}
=\oG_n(\beta_{2n},\beta_1,\cdots,\beta_{2n-1})
_{\ve_{2n}\ve_1\cdots\ve_{2n-1}}.
\en
If $\ve_{2n}=-$, then $\bar n\not=2n$.
In this case, the integrands of the LHS and the RHS coincide because 
of \refeq{PL}:
\be
\varphi(\alpha_a-\beta_{2n}+i\lambda)\sinh
\nu(\beta_{2n}-\alpha_a+\sc[\pi i,2]-i\lambda)
=\varphi(\alpha_a-\beta_{2n})\sinh
\nu(\alpha_a-\beta_{2n}+\sc[\pi i,2]).
\en
If $\ve_{2n}=+$, then $\bar n=2n$.
We make the following change of integration variables:
\bea
&&\alpha_n\rightarrow\alpha_n-i\lambda\hbox{ in the LHS},\nn\\
\noalign\\
&&\cases{\alpha_1\rightarrow\alpha_n;\cr
\alpha_2\rightarrow\alpha_1;\cr
\cdots\cr
\alpha_n\rightarrow\alpha_{n-1}&in the RHS.\cr}\nn
\ena
Then the integrands become the same by virtue of 
 \refeq{PL},
\be
\varphi(\alpha_n-i\lambda-\beta_j)
\sinh\nu(\alpha_n-i\lambda-\beta_j+\sc[\pi i,2])
=\varphi(\alpha_n-\beta_j)\sinh\nu(\beta_j-\alpha_n+\sc[\pi i,2])\nn
\en
and \refeq{AS},\refeq{QL},
\be
{\psi(\alpha_a-\alpha_n+i\lambda)
\over\sinh\nu(\alpha_a-\alpha_n+i\lambda-\pi i)}
={\psi(\alpha_n-\alpha_a)
\over\sinh\nu(\alpha_n-\alpha_a-\pi i)}.
\en

We must also check that the contours for the LHS and the RHS
are the same.
Consider the contour $\tilde C_a$ corresponding to $\alpha_a$
except for the case when $\ve_{2n}=+$ and $a=n$.
(We use $C_a$ and $\tilde C_a$ to distinguish the contours before and after
the change of variables.)
The condition \refeq{C1},\refeq{C2} and \refeq{C3} for $j\not=2n$ are unchanged
for either $\ve_{2n}=+$ or $\ve_{2n}=-$, and for both LHS and RHS.
As for the case $\ve_{2n}=-$ and $j=2n$, the condition is that,
in the LHS,
\be
\beta_{2n}-i\lambda\pm i(n_1\lambda+\sc[n_2,\nu]\pi
+\sc[\pi,2])\quad(n_1,n_2\ge0)
\lb{LH}
\en
are above (below) $\tilde C_a$; in the RHS,
\be
\beta_{2n}\pm i(n_1\lambda+\sc[n_2,\nu]\pi
+\sc[\pi,2])\quad(n_1,n_2\ge0)
\lb{RH}
\en
are above (below) $\tilde C_a$. They are not the same, but
not contradictory, i.e., no points are required to be in opposite sides of
a contour at the same time. Because we know that the integrands are the same,
it means that those points that appear only in either \refeq{LH} or \refeq{RH},
are actually not poles. Therefore, the difference between
\refeq{LH} and \refeq{RH} makes no difference in the integrals.
As for the case $\ve_{2n}=+$ and $j=2n$, the conditions \refeq{C2} and
\refeq{C3} are unchanged for $\alpha_a$ $(a\not=n)$.

If $\ve_{2n}=+$, the contour for $\alpha_n$ is such that for $j\not=2n$,
in the LHS
\[
\beta_j+i\lambda\pm i(n_1\lambda+\sc[n_2,\nu]\pi+\sc[\pi,2])
\quad(n_1,n_2\ge0)
\]
are above (below) the contour for $\alpha_n$; in the RHS
\[
\beta_j\pm i(n_1\lambda+\sc[n_2,\nu]\pi+\sc[\pi,2])
\quad(n_1,n_2\ge0)
\]
are above (below) the contour for $\alpha_n$.
These two conditions are not contradictory
in the same sense as above.
For $j=2n$, the conditions \refeq{C2} and \refeq{C3} are unchanged.

If $\ve_{2n}=+$, the mutual position of $\alpha_a$ and $\alpha_n$ changes
from the original one because of the change of variables.
However, the resulting positions of $\alpha_a$ and $\alpha_n$
in the LHS and the RHS are identical. Therefore, the intagrals are the same.

\par\smallskip\noindent{\it Proof of \refeq{G3}.}\quad
The factor $\rho(\beta_{2n-1}-\beta_{2n})$ has a zero at
$\beta_{2n}=\beta_{2n-1}+\pi i$: we see from \refeq{RZ} that
\be
\rho(\beta_{2n-1}-\beta_{2n})=
\frac{\nu i\rho(\pi i)}{\sin\pi\nu}
(\beta_{2n}-\beta_{2n-1}-\pi i)+\cdots.
\en

The integral may have a pole at $\beta_{2n}=\beta_{2n-1}+\pi i$ because 
the contour $C_a$ is pinched by the pole of $\varphi(\alpha_a-\beta_{2n-1})$
at $\alpha_a=\beta_{2n-1}+\sc[\pi i,2]$ and that of
$\varphi(\alpha_a-\beta_{2n})$ at $\alpha_a=\beta_{2n}-\sc[\pi i,2]$.
We will check if this is indeed a pole, and, if so, compute the residue.

Let us consider the four cases separately.

\itm{Case $(\ve_{2n-1},\ve_{2n})=(-,-)$}
The pole of $\varphi(\alpha_a-\beta_{2n-1})$ at
$\alpha_a=\beta_{2n-1}+\sc[\pi i,2]$ is cancelled
by the zero of $\sinh\nu(\beta_{2n-1}-\alpha_a+\sc[\pi i,2])$.
Therefore, there is no pinching in this case.

\itm{Case $(\ve_{2n-1},\ve_{2n})=(+,+)$}
If $\bar a\not=2n-1,2n$, for the same reason, there is no pinching of $C_a$.
Consider the integrals $I_i$ $(i=1,2,3)$ corresponding to
the following contours

\begin{figure}[htb]
\begin{center}
\setlength{\unitlength}{0.0125in}
\begin{picture}(334,357)(0,-10)
\drawline(290,25)(165,25)
\drawline(290,45)(165,45)
\drawline(290,155)(165,155)
\drawline(145,25)(20,25)
\drawline(145,45)(20,45)
\drawline(145,155)(20,155)
\drawline(145,45)(165,45)
\drawline(157.000,43.000)(165.000,45.000)(157.000,47.000)
\drawline(145,25)(165,25)
\drawline(157.000,23.000)(165.000,25.000)(157.000,27.000)
\drawline(160,285)(180,305)
\drawline(175.757,297.929)(180.000,305.000)(172.929,300.757)
\drawline(140,300)(160,320)
\drawline(155.757,312.929)(160.000,320.000)(152.929,315.757)
\drawline(145,155)(165,155)
\drawline(157.000,153.000)(165.000,155.000)(157.000,157.000)
\drawline(145,185)(165,205)
\drawline(160.757,197.929)(165.000,205.000)(157.929,200.757)
\drawline(300,225)	(295.759,225.223)
	(291.667,225.432)
	(287.720,225.627)
	(283.915,225.807)
	(280.247,225.972)
	(276.714,226.123)
	(273.312,226.260)
	(270.037,226.383)
	(266.885,226.491)
	(263.852,226.584)
	(260.936,226.663)
	(258.132,226.728)
	(255.436,226.779)
	(252.846,226.815)
	(247.966,226.843)
	(243.461,226.815)
	(239.304,226.728)
	(235.464,226.584)
	(231.912,226.383)
	(228.619,226.123)
	(225.555,225.807)
	(222.692,225.432)
	(220.000,225.000)

\drawline(220,225)	(215.979,224.166)
	(211.465,222.973)
	(208.973,222.224)
	(206.297,221.363)
	(203.417,220.383)
	(200.313,219.276)
	(196.965,218.037)
	(193.352,216.656)
	(189.455,215.127)
	(185.254,213.442)
	(180.728,211.594)
	(175.857,209.576)
	(173.286,208.501)
	(170.621,207.380)
	(167.860,206.214)
	(165.000,205.000)

\drawline(10,165)	(14.241,164.777)
	(18.333,164.568)
	(22.280,164.373)
	(26.085,164.193)
	(29.753,164.028)
	(33.286,163.877)
	(36.688,163.740)
	(39.963,163.617)
	(43.115,163.509)
	(46.148,163.416)
	(49.064,163.337)
	(51.868,163.272)
	(54.564,163.221)
	(57.154,163.185)
	(62.034,163.156)
	(66.539,163.185)
	(70.696,163.272)
	(74.536,163.416)
	(78.088,163.617)
	(81.381,163.877)
	(84.445,164.193)
	(87.308,164.568)
	(90.000,165.000)

\drawline(90,165)	(94.021,165.834)
	(98.535,167.027)
	(101.027,167.776)
	(103.703,168.637)
	(106.583,169.617)
	(109.687,170.724)
	(113.035,171.963)
	(116.648,173.344)
	(120.545,174.873)
	(124.746,176.558)
	(129.272,178.406)
	(134.143,180.424)
	(136.714,181.499)
	(139.379,182.620)
	(142.140,183.786)
	(145.000,185.000)

\drawline(295,340)	(290.759,340.223)
	(286.667,340.432)
	(282.720,340.627)
	(278.915,340.807)
	(275.247,340.972)
	(271.714,341.123)
	(268.312,341.260)
	(265.037,341.383)
	(261.885,341.491)
	(258.852,341.584)
	(255.936,341.663)
	(253.132,341.728)
	(250.436,341.779)
	(247.846,341.815)
	(242.966,341.843)
	(238.461,341.815)
	(234.304,341.728)
	(230.464,341.584)
	(226.912,341.383)
	(223.619,341.123)
	(220.555,340.807)
	(217.692,340.432)
	(215.000,340.000)

\drawline(215,340)	(210.979,339.166)
	(206.465,337.973)
	(203.973,337.224)
	(201.297,336.363)
	(198.417,335.383)
	(195.313,334.276)
	(191.965,333.037)
	(188.352,331.656)
	(184.455,330.127)
	(180.254,328.442)
	(175.728,326.594)
	(170.857,324.576)
	(168.286,323.501)
	(165.621,322.380)
	(162.860,321.214)
	(160.000,320.000)

\drawline(315,325)	(310.759,325.223)
	(306.667,325.432)
	(302.720,325.627)
	(298.915,325.807)
	(295.247,325.972)
	(291.714,326.123)
	(288.312,326.260)
	(285.037,326.383)
	(281.885,326.491)
	(278.852,326.584)
	(275.936,326.663)
	(273.132,326.728)
	(270.436,326.779)
	(267.846,326.815)
	(262.966,326.843)
	(258.461,326.815)
	(254.304,326.728)
	(250.464,326.584)
	(246.912,326.383)
	(243.619,326.123)
	(240.555,325.807)
	(237.692,325.432)
	(235.000,325.000)

\drawline(235,325)	(230.979,324.166)
	(226.465,322.973)
	(223.973,322.224)
	(221.297,321.363)
	(218.417,320.383)
	(215.313,319.276)
	(211.965,318.037)
	(208.352,316.656)
	(204.455,315.127)
	(200.254,313.442)
	(195.728,311.594)
	(190.857,309.576)
	(188.286,308.501)
	(185.621,307.380)
	(182.860,306.214)
	(180.000,305.000)

\drawline(25,265)	(29.241,264.777)
	(33.333,264.568)
	(37.280,264.373)
	(41.085,264.193)
	(44.753,264.028)
	(48.286,263.877)
	(51.688,263.740)
	(54.963,263.617)
	(58.115,263.509)
	(61.148,263.416)
	(64.064,263.337)
	(66.868,263.272)
	(69.564,263.221)
	(72.154,263.185)
	(77.034,263.157)
	(81.539,263.185)
	(85.696,263.272)
	(89.536,263.416)
	(93.088,263.617)
	(96.381,263.877)
	(99.445,264.193)
	(102.308,264.568)
	(105.000,265.000)

\drawline(105,265)	(109.021,265.834)
	(113.535,267.027)
	(116.027,267.776)
	(118.703,268.637)
	(121.583,269.617)
	(124.687,270.724)
	(128.035,271.963)
	(131.648,273.344)
	(135.545,274.873)
	(139.746,276.558)
	(144.272,278.406)
	(149.143,280.424)
	(151.714,281.499)
	(154.379,282.620)
	(157.140,283.786)
	(160.000,285.000)

\drawline(5,280)	(9.241,279.777)
	(13.333,279.568)
	(17.280,279.373)
	(21.085,279.193)
	(24.753,279.028)
	(28.286,278.877)
	(31.688,278.740)
	(34.963,278.617)
	(38.115,278.509)
	(41.148,278.416)
	(44.064,278.337)
	(46.868,278.272)
	(49.564,278.221)
	(52.154,278.185)
	(57.034,278.157)
	(61.539,278.185)
	(65.696,278.272)
	(69.536,278.416)
	(73.088,278.617)
	(76.381,278.877)
	(79.445,279.193)
	(82.308,279.568)
	(85.000,280.000)

\drawline(85,280)	(89.021,280.834)
	(93.535,282.027)
	(96.027,282.776)
	(98.703,283.637)
	(101.583,284.617)
	(104.687,285.724)
	(108.035,286.963)
	(111.648,288.344)
	(115.545,289.873)
	(119.746,291.558)
	(124.272,293.406)
	(129.143,295.424)
	(131.714,296.499)
	(134.379,297.620)
	(137.140,298.786)
	(140.000,300.000)

\put(10,0){\makebox(0,0)[lb]{\raisebox{0pt}[0pt][0pt]{\shortstack[l]{{\twlrm $C_n$}}}}}
\put(10,55){\makebox(0,0)[lb]{\raisebox{0pt}[0pt][0pt]{\shortstack[l]{{\twlrm $C_{n-1}$}}}}}
\put(15,130){\makebox(0,0)[lb]{\raisebox{0pt}[0pt][0pt]{\shortstack[l]{{\twlrm $C_n$}}}}}
\put(0,180){\makebox(0,0)[lb]{\raisebox{0pt}[0pt][0pt]{\shortstack[l]{{\twlrm $C_{n-1}$}}}}}
\put(15,245){\makebox(0,0)[lb]{\raisebox{0pt}[0pt][0pt]{\shortstack[l]{{\twlrm $C_n$}}}}}
\put(5,285){\makebox(0,0)[lb]{\raisebox{0pt}[0pt][0pt]{\shortstack[l]{{\twlrm $C_{n-1}$}}}}}
\put(195,80){\makebox(0,0)[lb]{\raisebox{0pt}[0pt][0pt]{\shortstack[l]{{\twlrm $\beta_{2n}-\sc[\pi i,2]$}}}}}
\put(300,50){\makebox(0,0)[lb]{\raisebox{0pt}[0pt][0pt]{\shortstack[l]{{\twlrm $I_3$}}}}}
\put(300,165){\makebox(0,0)[lb]{\raisebox{0pt}[0pt][0pt]{\shortstack[l]{{\twlrm $I_2$}}}}}
\put(300,280){\makebox(0,0)[lb]{\raisebox{0pt}[0pt][0pt]{\shortstack[l]{{\twlrm $I_1$}}}}}
\put(75,80){\makebox(0,0)[lb]{\raisebox{0pt}[0pt][0pt]{\shortstack[l]{{\twlrm $\beta_{2n-1}+\sc[\pi i,2]$}}}}}
\put(75,200){\makebox(0,0)[lb]{\raisebox{0pt}[0pt][0pt]{\shortstack[l]{{\twlrm $\beta_{2n-1}+\sc[\pi i,2]$}}}}}
\put(195,200){\makebox(0,0)[lb]{\raisebox{0pt}[0pt][0pt]{\shortstack[l]{{\twlrm $\beta_{2n}-\sc[\pi i,2]$}}}}}
\put(190,320){\makebox(0,0)[lb]{\raisebox{0pt}[0pt][0pt]{\shortstack[l]{{\twlrm $\beta_{2n}-\sc[\pi i,2]$}}}}}
\put(70,320){\makebox(0,0)[lb]{\raisebox{0pt}[0pt][0pt]{\shortstack[l]{{\twlrm $\beta_{2n-1}+\sc[\pi i,2]$}}}}}
\put(205,65){\makebox(0,0)[lb]{\raisebox{0pt}[0pt][0pt]{\shortstack[l]{{\twlrm $\times$}}}}}
\put(85,65){\makebox(0,0)[lb]{\raisebox{0pt}[0pt][0pt]{\shortstack[l]{{\twlrm $\times$}}}}}
\put(85,185){\makebox(0,0)[lb]{\raisebox{0pt}[0pt][0pt]{\shortstack[l]{{\twlrm $\times$}}}}}
\put(205,185){\makebox(0,0)[lb]{\raisebox{0pt}[0pt][0pt]{\shortstack[l]{{\twlrm $\times$}}}}}
\put(205,305){\makebox(0,0)[lb]{\raisebox{0pt}[0pt][0pt]{\shortstack[l]{{\twlrm $\times$}}}}}
\put(80,305){\makebox(0,0)[lb]{\raisebox{0pt}[0pt][0pt]{\shortstack[l]{{\twlrm $\times$}}}}}
\end{picture}
\vskip 2cm
\caption{The contours for $I_1$, $I_2$ and $I_3$.}
\label{fig:af3}
\end{center}
\end{figure}
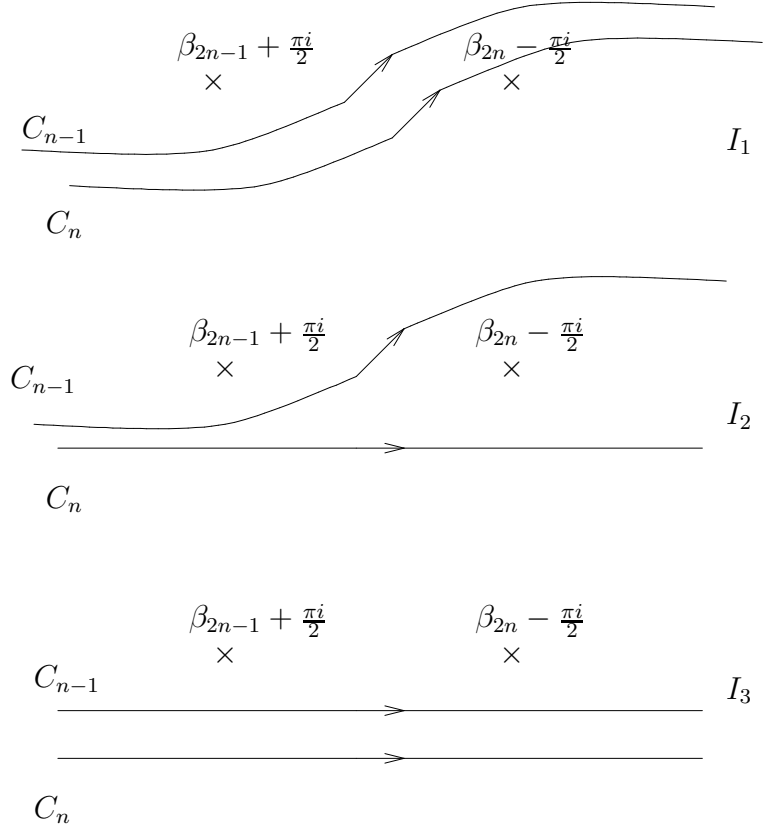

The integral $I_3$ has no pinching at $\beta_{2n}=\beta_{2n-1}+\pi i$.
Let us show that $I_1-I_2$ and $I_2-I_3$ are regular at
$\beta_{2n}=\beta_{2n-1}+\pi i$.
After integration with respect to
$\alpha_1,\cdots,\alpha_{n-2}$, the integral reads as
\bea
&&\int{d\alpha_{n-1}\over2\pi i}
\int{d\alpha_n\over2\pi i}
A(\alpha_{n-1},\alpha_n)
\prod_{a=n-1,n\atop j=2n-1,2n}\varphi(\alpha_a-\beta_j)\nn\\
&&\times\psi(\alpha_{n-1}-\alpha_n)
{\sinh\nu(\alpha_n-\beta_{2n-1}+{\pi i\over2})
\sinh\nu(\beta_{2n}-\alpha_{n-1}+{\pi i\over2})
\over\sinh\nu(\alpha_{n-1}-\alpha_n-\pi i)}.\nn
\ena
Here, $A(\alpha_{n-1},\alpha_n)$ is holomorphic and symmetric with respect
to $\alpha_{n-1}$ and $\alpha_n$. Since $\psi(\beta)=-\psi(-\beta)$,
we can antisymmetrize the last factor and obtain
\bea
&&B(\alpha_{n-1},\alpha_n;\beta_{2n-1},\beta_{2n})
={1\over2}\Biggl\{
{\sinh\nu(\alpha_n-\beta_{2n-1}+{\pi i\over2})
\sinh\nu(\beta_{2n}-\alpha_{n-1}+{\pi i\over2})
\over\sinh\nu(\alpha_{n-1}-\alpha_n-\pi i)}
\nn\\
&&-{\sinh\nu(\alpha_{n-1}-\beta_{2n-1}+{\pi i\over2})
\sinh\nu(\beta_{2n}-\alpha_n+{\pi i\over2})
\over\sinh\nu(\alpha_n-\alpha_{n-1}-\pi i)}
\Biggr\}.\nn
\ena
The integral $I_1-I_2$ is equal to the integral over the contour

\eject

\begin{figure}[htb]
\begin{center}
\setlength{\unitlength}{0.0125in}
\begin{picture}(324,102)(0,-10)
\drawline(135,45)(155,65)
\drawline(150.757,57.929)(155.000,65.000)(147.929,60.757)
\drawline(200,20)	(196.739,21.437)
	(193.961,22.762)
	(189.683,25.194)
	(185.000,30.000)

\drawline(185,30)	(183.877,32.666)
	(183.093,35.731)
	(182.639,39.062)
	(182.506,42.524)
	(182.685,45.983)
	(183.166,49.305)
	(183.941,52.355)
	(185.000,55.000)

\drawline(185,55)	(186.431,57.480)
	(188.286,60.040)
	(193.032,64.897)
	(195.803,66.939)
	(198.762,68.555)
	(201.847,69.618)
	(205.000,70.000)

\drawline(205,70)	(208.153,69.618)
	(211.238,68.555)
	(214.197,66.939)
	(216.968,64.897)
	(221.714,60.040)
	(223.569,57.480)
	(225.000,55.000)

\drawline(225,55)	(226.376,51.844)
	(227.561,48.226)
	(228.453,44.293)
	(228.952,40.190)
	(228.958,36.063)
	(228.368,32.059)
	(227.082,28.322)
	(225.000,25.000)

\drawline(225,25)	(221.327,21.814)
	(219.029,20.742)
	(216.326,19.997)
	(213.145,19.565)
	(209.413,19.432)
	(205.055,19.581)
	(202.620,19.758)
	(200.000,20.000)

\drawline(208.160,21.186)(200.000,20.000)(207.759,17.206)
\drawline(290,85)	(285.759,85.223)
	(281.667,85.432)
	(277.720,85.627)
	(273.915,85.807)
	(270.247,85.972)
	(266.714,86.123)
	(263.312,86.260)
	(260.037,86.383)
	(256.885,86.491)
	(253.852,86.584)
	(250.936,86.663)
	(248.132,86.728)
	(245.436,86.779)
	(242.846,86.815)
	(237.966,86.843)
	(233.461,86.815)
	(229.304,86.728)
	(225.464,86.584)
	(221.912,86.383)
	(218.619,86.123)
	(215.555,85.807)
	(212.692,85.432)
	(210.000,85.000)

\drawline(210,85)	(205.979,84.166)
	(201.465,82.973)
	(198.973,82.224)
	(196.297,81.363)
	(193.417,80.383)
	(190.313,79.276)
	(186.965,78.037)
	(183.352,76.656)
	(179.455,75.127)
	(175.254,73.442)
	(170.728,71.594)
	(165.857,69.576)
	(163.286,68.501)
	(160.621,67.380)
	(157.860,66.214)
	(155.000,65.000)

\drawline(0,25)	(4.241,24.777)
	(8.333,24.568)
	(12.280,24.373)
	(16.085,24.193)
	(19.753,24.028)
	(23.286,23.877)
	(26.688,23.740)
	(29.963,23.617)
	(33.115,23.509)
	(36.148,23.416)
	(39.064,23.337)
	(41.868,23.272)
	(44.564,23.221)
	(47.154,23.185)
	(52.034,23.157)
	(56.539,23.185)
	(60.696,23.272)
	(64.536,23.416)
	(68.088,23.617)
	(71.381,23.877)
	(74.445,24.193)
	(77.308,24.568)
	(80.000,25.000)

\drawline(80,25)	(84.021,25.834)
	(88.535,27.027)
	(91.027,27.776)
	(93.703,28.637)
	(96.583,29.617)
	(99.687,30.724)
	(103.035,31.963)
	(106.648,33.344)
	(110.545,34.873)
	(114.746,36.558)
	(119.272,38.406)
	(124.143,40.424)
	(126.714,41.499)
	(129.379,42.620)
	(132.140,43.786)
	(135.000,45.000)

\put(230,60){\makebox(0,0)[lb]{\raisebox{0pt}[0pt][0pt]{\shortstack[l]{{\twlrm $\beta_{2n}-\sc[\pi i,2]$}}}}}
\put(185,0){\makebox(0,0)[lb]{\raisebox{0pt}[0pt][0pt]{\shortstack[l]{{\twlrm $C_n$}}}}}
\put(0,30){\makebox(0,0)[lb]{\raisebox{0pt}[0pt][0pt]{\shortstack[l]{{\twlrm $C_{n-1}$}}}}}
\put(65,65){\makebox(0,0)[lb]{\raisebox{0pt}[0pt][0pt]{\shortstack[l]{{\twlrm $\beta_{2n-1}+\sc[\pi i,2]$}}}}}
\put(200,50){\makebox(0,0)[lb]{\raisebox{0pt}[0pt][0pt]{\shortstack[l]{{\twlrm $\times$}}}}}
\put(75,50){\makebox(0,0)[lb]{\raisebox{0pt}[0pt][0pt]{\shortstack[l]{{\twlrm $\times$}}}}}
\end{picture}
\vskip 2cm
\caption{The contour for $I_1-I_2$.}
\label{fig:af4}
\end{center}
\end{figure}
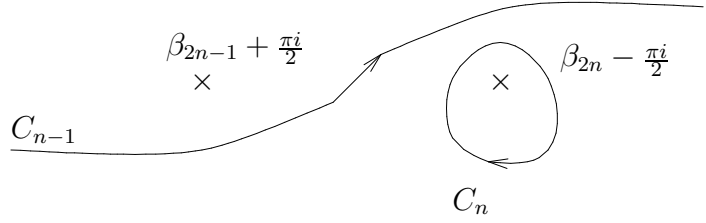

Taking the residue at $\alpha_n=\beta_{2n}-\sc[\pi i,2]$
(the minus sign in front of $\Res$ below
comes from the clockwise orientation of the
integration contour), we get
\bea
&&\int{d\alpha_{n-1}\over{2\pi i}}\left\{-\hbox{\rm Res}
_{\alpha_n=\beta_{2n}-\sc[\pi i,2]}
\varphi(\alpha_n-\beta_{2n})\right\}
A(\alpha_{n-1},\beta_{2n}-\sc[\pi i,2])\nn\\
&&\times\varphi(\alpha_{n-1}-\beta_{2n-1})
\varphi(\alpha_{n-1}-\beta_{2n})
\varphi(\beta_{2n}-\beta_{2n-1}-\sc[\pi i,2])\nn\\
&&\times\psi(\alpha_{n-1}-\beta_{2n}+\sc[\pi i,2])
B(\alpha_{n-1},\beta_{2n}-\sc[\pi i,2];\beta_{2n-1},\beta_{2n}).\nn
\ena
The integrand has no pole at $\alpha_{n-1}=\beta_{2n}-\sc[\pi i,2]$
because $\psi(\alpha_{n-1}-\beta_{2n}+\sc[\pi i,2])$ vanishes.
The integral has no pole at $\beta_{2n}=\beta_{2n-1}+\pi i$ because
$B(\alpha_{n-1},\beta_{2n}-\sc[\pi i,2];\beta_{2n-1},\beta_{2n})$
vanishes. Therefore, the integral is regular at
$\beta_{2n}=\beta_{2n-1}+\pi i$. By a similar argement
we can show that $I_2-I_3$ is regular at $\beta_{2n}=\beta_{2n-1}+\pi i$.
\itm{Case $(\ve_{2n-1},\ve_{2n})=(-,+)$}
Taking into account the zero of the factor $\rho(\beta_{2n-1}-\beta_{2n})$
and the pole of the residue $-\Res_{\alpha_n=\beta_{2n}-\sc[\pi i,2]}$,
both at $\beta_{2n}=\beta_{2n-1}+\pi i$, we have
\bea
&&{c_n\over c_{n-1}}
\Biggl\{
\rho(\beta_{2n-1}-\beta_{2n})\varphi(\alpha_n-\beta_{2n-1})
\Bigl(-\Res_{\alpha_n=\beta_{2n}-\sc[\pi i,2]}
\varphi(\alpha_n-\beta_{2n})\Bigr)\nn\\
&&\times\sinh\nu(\alpha_n-\beta_{2n-1}+\sc[\pi i,2])
\nn\\
&&\times 
\prod_{1\le j\le 2n-2}
\rho(\beta_j-\beta_{2n-1})
\rho(\beta_j-\beta_{2n})\varphi(\alpha_n-\beta_j)
\sinh\nu(\alpha_n-\beta_j+\sc[\pi i,2])
\nn\\
&&\times\prod_{1\le a\le n-1}\varphi(\alpha_a-\beta_{2n-1})
\varphi(\alpha_a-\beta_{2n})\psi(\alpha_a-\alpha_n)
\nn\\
&&\times
{\sinh\nu(\beta_{2n-1}-\alpha_a+\sc[\pi i,2])
\sinh\nu(\beta_{2n}-\alpha_a+\sc[\pi i,2])
\over
\sinh\nu(\alpha_a-\alpha_n-\pi i)}\Biggr\}
\Bigg|_{\alpha_n=\beta_{2n}-\sc[\pi i,2]
\atop
\beta_{2n}=\beta_{2n-1}+\pi i}=1.
\nn
\ena
Using \refeq{A1}, \refeq{RZ}, \refeq{RR}, \refeq{A2} and $c_0=1$,
we obtain \refeq{NR}.

The case $(\ve_{2n-1},\ve_{2n})=(+,-)$ is similar.

\setcounter{equation}{0}
\section{One-time integration}
	
In this appendix we show how to reduce 
the $n$-fold integral for 
$G(\beta_1,\cdots,\beta_{2n})$ 
to an $(n-1)$-fold integral \eqref{eqn:once}.
We shall follow the method suggested earlier to us by F. Smirnov. 
A similar calculation has been published in Nakayashiki's paper
\cite{Nak} where the limiting case $\nu\rightarrow 0$ was discussed. 
Since our working is entirely similar to the one in \cite{Nak}, we 
shall only indicate the necessary steps, omitting further details.
In the sequel we set $\beta=(\beta_1,\cdots,\beta_{2n}),
\vep=(\vep_1,\cdots,\vep_{2n})$.
We restrict to $\lambda=2\pi$, so that 
$\psi(\beta)=\sinh \beta$ in the general formula \eqref{eqn:GF}.

\begin{description}
\item[Step 1] Using
\[
\prod_{r>s}2\sinh(\alpha_r-\alpha_s)
=\det\bigl(e^{-(n-2l+1)\alpha_k}\bigr)_{1\le k,l\le n}, 
\]
we rewrite the main part of $G(\beta)_\vep$ as  
\begin{equation}
J(\beta)_\vep
=
\int\!\!\cdots\!\!\int\prod_{k=1}^nd\alpha_k
\prod_{k,j}\varphi(\alpha_k-\beta_j)
\det\bigl(e^{-(n-2l+1)\alpha_k}\bigr)_{1\le k,l\le n}
Q(\alpha|\beta)_\vep,
\end{equation}
where 
\[
Q(\alpha|\beta)_\vep=
\frac{\prod_{j<\bar{k}}\sinh\nu\Bigl(\alpha_k-\beta_j+\frac{\pi i}{2}\Bigr)
\prod_{j>\bar{k}}\sinh\nu\Bigl(-\alpha_k+\beta_j+\frac{\pi i}{2}\Bigr)}
{\prod_{r<s}\sinh\nu(\alpha_r-\alpha_s-\pi i)}.
\]
\item[Step 2] 
In the first column of the determinant, substitute 
$e^{-(n-1)\alpha_k}$ by the right hand side of the identity
\[
e^{-(n-1)\alpha_k}
=\frac{i^{n+1}2^{2n-1}}{e^{\sum_j\beta_j/2}\sum_je^{-\beta_j}}
\left(F_{+}(\alpha_k)-F_{-}(\alpha_k)
+\sum_{l=2}^n c_l(\beta)e^{-(n-2l+1)\alpha_k}
\right).
\]
Here
\[
F_{+}(\alpha)=\prod_{j=1}^{2n}\sinh\frac{1}{2}
\Bigl(\alpha-\beta_j+\frac{\pi i}{2}\Bigr),
\quad 
F_{-}(\alpha)=(-1)^n\prod_{j=1}^{2n}\sinh\frac{1}{2}
\Bigl(\alpha-\beta_j-\frac{\pi i}{2}\Bigr)
\]
and $c_l(\beta)$ denotes some function of $\beta_j$'s.
Terms containing $c_l(\beta)$ vanish in the determinant.

\item[Step 3] Expand the determinant at the first column to obtain 
\begin{eqnarray*}
J(\beta)_\vep&=&
\frac{i^{n+1}2^{2n-1}}{e^{\sum_j\beta_j/2}\sum_je^{-\beta_j}}
\sum_{l=1}^n(-1)^{l-1}J_{l,\vep},
\\
J_{l,\vep}&=& 
\int\!\!\cdots\!\!\int\prod_{k=1}^nd\alpha_k
\prod_{k,j}\varphi(\alpha_k-\beta_j)
\left(F_+(\alpha_l)-F_-(\alpha_l)\right)D_l(\alpha)
Q(\alpha|\beta)_\vep,
\end{eqnarray*}
where for brevity we set 
$D_l(\alpha)=D(\alpha_1,\cdots,\alpha_{l-1},\alpha_{l+1},\cdots,\alpha_n)$.

\item[Step 4] 
Next we carry out the integral over $\alpha_l$. 
For each $l$, set
\[
Q_l(\alpha|\beta)_\vep=
\frac{\prod_{j(<\bar{l})}\sinh\nu\Bigl(\alpha_l-\beta_j+\frac{\pi i}{2}\Bigr)
\prod_{j(>\bar{l})}\sinh\nu\Bigl(-\alpha_l+\beta_j+\frac{\pi i}{2}\Bigr)}
{\prod_{r(<l)}\sinh\nu(\alpha_r-\alpha_l-\pi i)
\prod_{r(>l)}\sinh\nu(\alpha_l-\alpha_r-\pi i)}.
\]
Consider the integrals 
\begin{eqnarray*}
K_{l,\vep}^{(\pm)}&=&\pm \int_{C_\pm} H^{(\pm)}_{l,\vep} d\alpha_l,
\\
H^{(\pm)}_{l,\vep}&=&
F_{\pm}(\alpha_l) 
\prod_{j=1}^{2n}\varphi(\alpha_l-\beta_j)\,Q_l(\alpha|\beta)_\vep, 
\end{eqnarray*}
taken along the contours 
$C_{\pm}=\sum_{i=1}^4C_{\pm,i}$, respectively, 
which are shown in Figure \ref{fig:af2}. 
(To fix the idea we draw the figure assuming 
the $\beta_j$ are all real, but the necessary 
modification in the general situation should be obvious.)

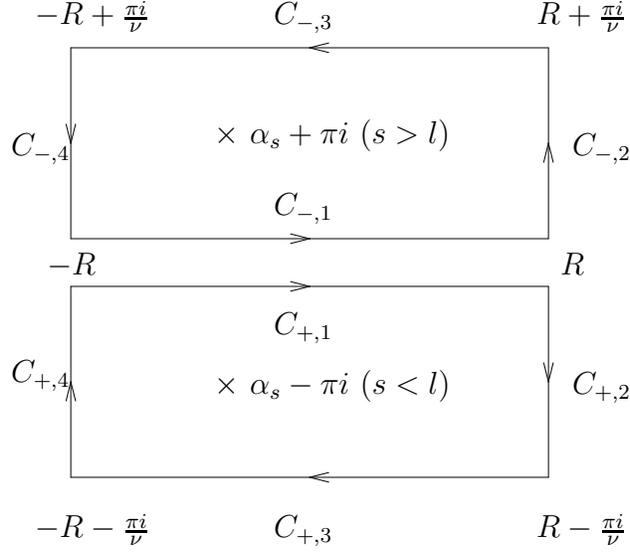
\begin{figure}[htb]
\begin{center}
\setlength{\unitlength}{0.0125in}
\begin{picture}(307,243)(0,-10)
\drawline(225,205)(125,205)
\drawline(133.000,207.000)(125.000,205.000)(133.000,203.000)
\drawline(25,205)(25,165)
\drawline(23.000,173.000)(25.000,165.000)(27.000,173.000)
\drawline(25,125)(125,125)
\drawline(117.000,123.000)(125.000,125.000)(117.000,127.000)
\drawline(225,125)(225,165)
\drawline(227.000,157.000)(225.000,165.000)(223.000,157.000)
\drawline(125,205)(25,205)
\drawline(25,165)(25,125)
\drawline(125,125)(225,125)
\drawline(225,165)(225,205)
\drawline(25,105)(125,105)
\drawline(117.000,103.000)(125.000,105.000)(117.000,107.000)
\drawline(225,105)(225,65)
\drawline(223.000,73.000)(225.000,65.000)(227.000,73.000)
\drawline(225,25)(125,25)
\drawline(133.000,27.000)(125.000,25.000)(133.000,23.000)
\drawline(25,25)(25,65)
\drawline(27.000,57.000)(25.000,65.000)(23.000,57.000)
\drawline(125,105)(225,105)
\drawline(225,65)(225,25)
\drawline(125,25)(25,25)
\drawline(25,65)(25,105)
\put(100,60){\makebox(0,0)[lb]{\raisebox{0pt}[0pt][0pt]{\shortstack[l]{{\twlrm \hbox{$\alpha_s-\pi i$ $(s<l)$}}}}}}
\put(85,60){\makebox(0,0)[lb]{\raisebox{0pt}[0pt][0pt]{\shortstack[l]{{\twlrm $\times$}}}}}
\put(15,110){\makebox(0,0)[lb]{\raisebox{0pt}[0pt][0pt]{\shortstack[l]{{\twlrm $-R$}}}}}
\put(10,215){\makebox(0,0)[lb]{\raisebox{0pt}[0pt][0pt]{\shortstack[l]{{\twlrm $-R+\sc[\pi i,\nu]$}}}}}
\put(10,0){\makebox(0,0)[lb]{\raisebox{0pt}[0pt][0pt]{\shortstack[l]{{\twlrm $-R-\sc[\pi i,\nu]$}}}}}
\put(220,0){\makebox(0,0)[lb]{\raisebox{0pt}[0pt][0pt]{\shortstack[l]{{\twlrm $R-\sc[\pi i,\nu]$}}}}}
\put(110,0){\makebox(0,0)[lb]{\raisebox{0pt}[0pt][0pt]{\shortstack[l]{{\twlrm $C_{+,3}$}}}}}
\put(0,65){\makebox(0,0)[lb]{\raisebox{0pt}[0pt][0pt]{\shortstack[l]{{\twlrm $C_{+,4}$}}}}}
\put(110,85){\makebox(0,0)[lb]{\raisebox{0pt}[0pt][0pt]{\shortstack[l]{{\twlrm $C_{+,1}$}}}}}
\put(235,60){\makebox(0,0)[lb]{\raisebox{0pt}[0pt][0pt]{\shortstack[l]{{\twlrm $C_{+,2}$}}}}}
\put(235,160){\makebox(0,0)[lb]{\raisebox{0pt}[0pt][0pt]{\shortstack[l]{{\twlrm $C_{-,2}$}}}}}
\put(110,135){\makebox(0,0)[lb]{\raisebox{0pt}[0pt][0pt]{\shortstack[l]{{\twlrm $C_{-,1}$}}}}}
\put(0,160){\makebox(0,0)[lb]{\raisebox{0pt}[0pt][0pt]{\shortstack[l]{{\twlrm $C_{-,4}$}}}}}
\put(100,165){\makebox(0,0)[lb]{\raisebox{0pt}[0pt][0pt]{\shortstack[l]{{\twlrm \hbox{$\alpha_s+\pi i$ $(s>l)$}}}}}}
\put(85,165){\makebox(0,0)[lb]{\raisebox{0pt}[0pt][0pt]{\shortstack[l]{{\twlrm $\times$}}}}}
\put(110,215){\makebox(0,0)[lb]{\raisebox{0pt}[0pt][0pt]{\shortstack[l]{{\twlrm $C_{-,3}$}}}}}
\put(230,110){\makebox(0,0)[lb]{\raisebox{0pt}[0pt][0pt]{\shortstack[l]{{\twlrm $R$}}}}}
\put(220,215){\makebox(0,0)[lb]{\raisebox{0pt}[0pt][0pt]{\shortstack[l]{{\twlrm $R+\sc[\pi i,\nu]$}}}}}
\end{picture}
\vskip 2cm
\caption{The contour $C_\pm$.}
\label{fig:af2}
\end{center}
\end{figure}

It can be verified that 
inside the contours 
the only poles of the integrand $H^{(\pm)}_{l,\vep}$ 
are $\alpha_l=\alpha_s\mp \pi i$ for $s<l$ or $s>l$ respectively. 
Collecting the residues we obtain 
\begin{eqnarray*}
&&K_{l,\vep}^{(+)}+K_{l,\vep}^{(-)}
= -2\pi i (\sum_{s(<l)}M_{l,s}^{(+)}+
\sum_{s(>l)}M_{l,s}^{(-)}),
\\
&&M_{l,s}^{(\pm)}=\Res_{\alpha_l=\alpha_s\mp \pi i}
F_{\pm}(\alpha_l)\prod_j\varphi(\alpha_l-\beta_j)\,
Q_l(\alpha|\beta)_\vep d\alpha_l.
\end{eqnarray*}
One can show that, 
upon integration by the other variables and summing over $l$, 
these terms cancel with each other.
More precisely, set
\[
Q_{l,s}'(\alpha|\beta)_\vep=
\frac{\prod_{j(<\bar{l})}\sinh\nu\Bigl(\alpha_l-\beta_j+\frac{\pi i}{2}\Bigr)
\prod_{j(>\bar{l})}\sinh\nu\Bigl(-\alpha_l+\beta_j+\frac{\pi i}{2}\Bigr)}
{\prod_{r(<l)\atop r\neq s}\sinh\nu(\alpha_r-\alpha_l-\pi i)
\prod_{r(>l)\atop r\neq s}\sinh\nu(\alpha_l-\alpha_r-\pi i)}.
\]
Then we have, for any pair $r<s$,
\[
\int d\alpha_r M^{(+)}_{sr}D_s(\alpha)Q'_{rs}(\alpha|\beta)_\vep
+
(-1)^{r-s}\int d\alpha_s M^{(-)}_{rs}D_r(\alpha)Q'_{sr}(\alpha|\beta)_\vep
=0.
\]
This can be seen by changing the variable $\alpha_r\rightarrow 
\alpha_s+\pi i$.
\item[Step 5] 
From the transformation properties of $\varphi(\beta)$ it follows that 
\[
H^{(\pm)}_{l,\vep}
\Bigl|_{\alpha_l\rightarrow \alpha_l\mp \pi i/\nu}
=
H^{(\mp)}_{l,\vep}
\]
which implies that the integrals corresponding to 
$C_{\pm,3}$ give the same results as for $C_{\mp,1}$. 

As $R\rightarrow\infty$, 
the integrals along $C_{\pm,2,4}$ are calculated from the following
asymptotics of the integrand 
as $\alpha_l\rightarrow\pm\infty$:
\begin{eqnarray*}
&&\varphi(\alpha_l-\beta_j)
\sim 2\exp\left(\mp\frac{1+\nu}{2}(\alpha_l-\beta_j)\right),
\\
&&F_\sigma(\alpha_l)\prod_{j=1}^{2n}\varphi(\alpha_l-\beta_j)
\sim i^n\exp\left(\mp\nu(n\alpha_l -\frac{1}{2}\sum_{j=1}^{2n}\beta_j)
\right),
\\
&&Q_l(\alpha|\beta)_\vep
\sim
\frac{(-1)^{l+\bar{l}+1}}{2^n}\exp\bigl(\pm \nu(n\alpha_l+A_l)\bigr)
\times(\pm 1)^n
\end{eqnarray*}
with
\[
A_l=\sum_{k\neq l}\alpha_k-\sum_{j\neq \bar{l}}\beta_j
+\pi i (\bar{l}-2l+\frac{1}{2}).
\]
\end{description}

From the last two steps we find that 
\[
\int_{-R}^R  (H^{(+)}_{l,\vep}-H^{(-)}_{l,\vep})d\alpha_l
=\frac{\pi i}{\nu}\frac{(-1)^{l+\bar{l}+1}i^n}{2^n}
\left(e^{\nu\tilde{A}_l}-e^{-\nu\tilde{A}_l}\right)
+{\cal R},
\]
where
\[
\tilde{A}_l=A_l+\frac{1}{2}\sum_{j=1}^{2n}\beta_j
\]
and $\cal R$ signifies a term which 
vanishes when $R\rightarrow\infty$. 
Hence we arrive at the result \eqref{eqn:once}
stated in the beginning.

\setcounter{equation}{0}
\section{Derivation of the nearest neighbour correlator}\label{app:D}

Here we derive the formula \eqref{eqn:ans}.
We start from \eqref{eqn:generic} and consider the specialization
\[
G=G(\beta+\pi i,\beta+\pi i,\beta, \beta)_{++--}.
\]
\begin{prop}
\begin{equation}
G+\frac{1}{2}
=\frac{1}{2\pi\nu^2}
\int_{-\infty}^\infty
\frac{e^\alpha}{\cosh^2\alpha}
\frac{\varphi'(\alpha)}{\varphi(\alpha)}
\left(1-\cos\pi\nu\cosh2\nu\alpha\right) d\alpha.
\label{eqn:Isp}
\end{equation}
\end{prop}
\proof
First let $(\beta_1,\cdots,\beta_4)=
(\beta+\pi i,\beta+\pi i,\beta+\vep,\beta+\vep)$.
Then \eqref{eqn:generic} becomes 
\begin{eqnarray*}
&&G(\beta+\pi i,\beta+\pi i,\beta+\vep,\beta+\vep)_{++--}
=
\frac{-1}{4\pi\nu^2}\frac{e^{-\beta-\vep}}{1-e^{-\vep}}
\frac{\rho(\pi i-\vep)^4}{\rho(\pi i)^4}
\\
&&
\times\int d\alpha\, e^\alpha 
\varphi(\alpha-\beta-\pi i)^2\varphi(\alpha-\beta-\vep)^2
\sinh^2\nu\bigl(\alpha-\beta-\vep-\frac{\pi i}{2}\bigr)
\\
&&
\times
\Bigl[\sinh\nu\bigl(\alpha-\beta-\frac{3\pi i}{2}\bigr)
\sinh\nu\bigl(\alpha-\beta-\vep-\frac{3\pi i}{2}\bigr)
\\
&&+\sinh\nu\bigl(\alpha-\beta-\frac{\pi i}{2}\bigr)
\sinh\nu\bigl(\alpha-\beta-\vep-\frac{\pi i}{2}\bigr)\Bigr].
\end{eqnarray*}
Since $\alpha=\beta+\vep+\pi i/2$ is not a pole, the contour 
can be taken as $\pi/2<\Im(\alpha-\beta)<3\pi/2$. 
Changing the variable $\alpha\rightarrow\alpha+\beta+\vep+\pi i$ and 
using 
$\varphi(\alpha+\pi i)\varphi(\alpha)=
-i/\cosh\alpha\sinh\nu(\alpha+\pi i/2)$, we find
\begin{eqnarray*}
&&G=\lim_{\varepsilon\rightarrow0}\frac{1}{4\pi\nu^2}\frac{1}{1-e^{-\vep}}
\int_{-\infty}^\infty d\alpha\, e^\alpha 
\frac{\varphi(\alpha+\vep)^2}{\varphi(\alpha)^2}
\frac{-1}{\cosh^2\alpha}
\\
&&
\times
\Bigl[\sinh\nu\bigl(\alpha+\frac{\pi i}{2}+\vep\bigr)
\sinh\nu\bigl(\alpha+\frac{\pi i}{2}\bigr)
+\sinh\nu\bigl(\alpha-\frac{\pi i}{2}+\vep\bigr)
\sinh\nu\bigl(\alpha-\frac{\pi i}{2}\bigr)\Bigr].
\end{eqnarray*}
Noting
\[
0=
\int_{-\infty}^\infty d\alpha\,
\frac{e^\alpha}{\cosh^2\alpha}
\Bigl[\sinh^2\nu\bigl(\alpha+\frac{\pi i}{2}\bigr)
+\sinh^2\nu\bigl(\alpha-\frac{\pi i}{2}\bigr)\Bigr]
\]
and letting $\vep\rightarrow 0$, we obtain
\begin{eqnarray*}
G&=&\frac{1}{2\pi\nu^2}\frac{\partial}{\partial\vep}
\biggl(
\int_{-\infty}^\infty d\alpha\, e^\alpha 
\frac{\varphi(\alpha+\vep)^2}{\varphi(\alpha)^2}
\frac{-1}{\cosh^2\alpha}
\\
&&\times
\Re\sinh\nu\bigl(\alpha+\frac{\pi i}{2}+\vep\bigr)
\sinh\nu\bigl(\alpha+\frac{\pi i}{2}\bigr)
\biggr)\biggl|_{\vep=0}
\\
&=&
I_1+I_2,
\end{eqnarray*}
where $I_1$ is given by the right hand side of \eqref{eqn:Isp}
and 
\[
I_2=\frac{-1}{4\pi\nu}
\int_{-\infty}^\infty d\alpha\,
\frac{e^\alpha}{\cosh^2\alpha}\, 
\Re\sinh2\nu\bigl(\alpha+\frac{\pi i}{2}\bigr).
\]
Using
\begin{equation}
\int_{-\infty}^\infty d\alpha\frac{e^{\lambda\alpha}}{\cosh^2\alpha}
=\frac{\pi\lambda}{\sin\displaystyle\frac{\pi\lambda}{2}}
\label{eqn:coshint}
\end{equation}
we find
\[
I_2=-\frac{1}{2},
\]
thereby completing the proof of the lemma.
\qed

\begin{prop}
\begin{eqnarray}
G&=&J_1+J_2-\frac{1}{2},
\label{eqn:J1J2}\\
J_1&=&
\frac{1}{\pi^2}
\int_0^\infty tdt\, \frac{\sinh(1+\nu)t}{\sinh t}
\left(\frac{1}{\sin\pi\nu}
\Im\frac{1}{\cosh\nu(t+\pi i)}
+\frac{\sinh\nu t}{\cosh^2\nu t}\right),
\nonumber\\
J_2&=&
\frac{1}{\pi\sin\pi\nu}
\int_0^\infty dt\, \frac{\sinh(1+\nu)t}{\sinh t}
\left(\Re\frac{1}{\cosh\nu(t+\pi i)}
-\frac{\cos\pi\nu}{\cosh\nu t}\right).\nonumber
\end{eqnarray}
\end{prop}

\proof
Substituting the integral formula for 
$\log\varphi(\alpha)$ into $I_1$ above, we obtain
\begin{eqnarray*}
G+\frac{1}{2}&=&
\frac{1}{\pi^2\nu}
\int_0^\infty dt\, \frac{\sinh(1+\nu)t}{\sinh t\sinh2\nu t}
\\
&&\times
\int_{-\infty}^\infty d\alpha\,
\frac{e^\alpha}{\cosh^2\alpha}\sin\left(\frac{2\nu t}{\pi}\alpha\right)
\left(\cosh2\nu\alpha\cos\pi\nu-1\right).
\end{eqnarray*}
By integrating over $\alpha$ using \eqref{eqn:coshint}, 
the right hand side becomes
\[
\frac{1}{\pi^2}\int_0^\infty dt\,
\frac{\sinh(1+\nu)t}{\sinh t\sinh 2\nu t}
\left(
\cos\pi\nu\Bigl(\frac{t-\pi i}{\cosh\nu(t-\pi i)}
+\frac{t+\pi i}{\cosh\nu(t+\pi i)}\Bigr)
-\frac{2t}{\cosh\nu t}
\right).
\]
After a little algebra we obtain \eqref{eqn:J1J2}.\qed

\begin{prop} We have
\begin{equation}
G-{1\over2}=J_1+J_2-1=\frac{1}{\pi^2\sin\pi\nu}
\frac{\partial}{\partial\nu}\left(\sin\pi\nu
\int_0^\infty dt\, \frac{\sinh(1-\nu)t}{\sinh t\cosh\nu t}\right).
\label{eqn:K}
\end{equation}
\end{prop}

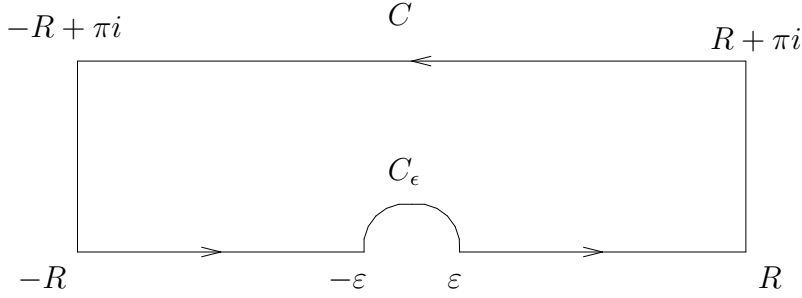
\begin{figure}[htb]
\begin{center}
\setlength{\unitlength}{0.0125in}
\begin{picture}(335,136)(0,-10)
\drawline(170.000,35.000)(164.558,34.931)(159.395,33.210)
	(155.000,30.000)(151.790,25.605)(150.069,20.442)
	(150.000,15.000)
\drawline(190.000,15.000)(189.931,20.442)(188.210,25.605)
	(185.000,30.000)(180.605,33.210)(175.442,34.931)
	(170.000,35.000)
\drawline(250,15)(310,15)
\drawline(90,15)(150,15)
\drawline(82.000,13.000)(90.000,15.000)(82.000,17.000)
\drawline(90,15)(30,15)
\drawline(310,95)(310,15)
\drawline(30,95)(30,15)
\drawline(30,95)(170,95)
\drawline(242.000,13.000)(250.000,15.000)(242.000,17.000)
\drawline(250,15)(190,15)
\drawline(178.000,97.000)(170.000,95.000)(178.000,93.000)
\drawline(170,95)(310,95)
\put(185,0){\makebox(0,0)[lb]{\raisebox{0pt}[0pt][0pt]{\shortstack[l]{{\twlrm $\varepsilon$}}}}}
\put(135,0){\makebox(0,0)[lb]{\raisebox{0pt}[0pt][0pt]{\shortstack[l]{{\twlrm $-\varepsilon$}}}}}
\put(315,0){\makebox(0,0)[lb]{\raisebox{0pt}[0pt][0pt]{\shortstack[l]{{\twlrm $R$}}}}}
\put(5,0){\makebox(0,0)[lb]{\raisebox{0pt}[0pt][0pt]{\shortstack[l]{{\twlrm $-R$}}}}}
\put(295,100){\makebox(0,0)[lb]{\raisebox{0pt}[0pt][0pt]{\shortstack[l]{{\twlrm $R+\pi i$}}}}}
\put(0,105){\makebox(0,0)[lb]{\raisebox{0pt}[0pt][0pt]{\shortstack[l]{{\twlrm $-R+\pi i$}}}}}
\put(160,110){\makebox(0,0)[lb]{\raisebox{0pt}[0pt][0pt]{\shortstack[l]{{\twlrm $C$}}}}}
\put(160,45){\makebox(0,0)[lb]{\raisebox{0pt}[0pt][0pt]{\shortstack[l]{{\twlrm $C_\epsilon$}}}}}
\end{picture}
\vskip 2cm
\caption{The contour $C$ and the semi-circle $C_\varepsilon$.}
\label{fig:af5}
\end{center}
\end{figure}

\proof
Consider the integral 
\[
\int_C dt\,\frac{\sinh(1+\nu)(t-\pi i)}{\sinh(t-\pi i)}
\frac{t-\pi i}{\cosh \nu t}=0,
\]
where the contour $C$ is as shown in Figure \ref{fig:af5}.

Taking the imaginary part, we obtain 
\begin{eqnarray*}
&&\int_{-R}^Rtdt\,
\frac{\sinh(1+\nu)t}{\sinh t}\,
\Im\left(\frac{1}{\cosh\nu(t+\pi i)}\right)
\\
&&\quad=
\Im
\left(\int_{-R}^{-\varepsilon}+\int_{C_\varepsilon}+\int^R_\varepsilon\right)
dt\left(\frac{\sinh(1+\nu)(t-\pi i)}{\sinh(t-\pi i)}
\frac{t-\pi i}{\cosh \nu t}\right)\nn\\
&&\quad+
\int_0^\pi dt \,
\Re\biggl(
\frac{\sinh(1+\nu)(R+ti-\pi i)}{\sinh(R+ti-\pi i)}
\frac{R+ti-\pi i}{\cosh\nu(R+ti)}
\\
&&
\quad -\frac{\sinh(1+\nu)(-R+ti-\pi i)}{\sinh(-R+ti-\pi i)}
\frac{-R+ti-\pi i}{\cosh\nu(-R+ti)}
\biggr).
\end{eqnarray*}
As $\varepsilon\rightarrow0$, the integral over the semi-circle
$C_\varepsilon$ gives $\pi^2\sin\pi\nu$, and
as $R\rightarrow \infty$, the last term behaves like
\[
4\pi R\cos\pi \nu-2\pi^2\sin\pi\nu
+O(R^{-1}).
\]
Since
\begin{eqnarray*}
&&\Im\!\left(\frac{\sinh(1+\nu)(t-\pi i)}{\sinh(t-\pi i)}
\frac{t-\pi i}{\cosh \nu t}\right)
\\
&&\quad \qquad=
-{t\cosh(1+\nu)t\sin\pi\nu\over\sinh t\cosh\nu t}
-{\pi\sinh(1+\nu)t\cos\pi\nu\over\sinh t\cosh\nu t}
\end{eqnarray*}
and
\[
{\sinh(1+\nu)t\sinh\nu t\over\sinh t\cosh^2\nu t}
-{\cosh(1+\nu)t\over\sinh t\cosh^2\nu t}
=-{\cosh t\over\sinh t\cosh{\nu t}},
\]
we have
\begin{eqnarray*}
&&
\int_{-R}^Rtdt\,\frac{\sinh(1+\nu)t}{\sinh t}
\left(\frac{\sinh\nu t}{\cosh^2\nu t}
+\frac{1}{\sin\pi\nu}\Im\frac{1}{\cosh\nu(t+\pi i)}\right)
\\
&&
\quad =-\int_{-R}^Rtdt\,\frac{\cosh t}{\sinh t\cosh^2\nu t}
-\pi\cot\pi\nu \int_{-R}^R\frac{\sinh(1+\nu)t}{\sinh t\cosh\nu t}dt
\\
&&\qquad +4\pi R\cot\pi\nu-\pi^2+O(R^{-1})
\end{eqnarray*}
Noting that
\begin{eqnarray*}
&&\lim_{R\rightarrow\infty}
\left(4\pi R\cot\pi\nu-\pi\cot\pi\nu\int_{-R}^Rdt\,
{\sinh(1+\nu)t\over\sinh t\cosh\nu t}\right)
\\
&&\qquad
=2\pi\cot\pi\nu\int_0^\infty dt\,{\sinh(1-\nu)t\over\sinh t\cosh\nu t},
\end{eqnarray*}
we have
\[
J_1=-{1\over2}-{1\over\pi^2}\int_0^\infty tdt\,
{\cosh t\over\sinh t\cosh^2\nu t}
+{\cot\pi\nu\over\pi}\int_0^\infty dt\,
{\sinh(1-\nu)t\over\sinh t\cosh\nu t}.
\]
By a similar calculation, starting from
\[
\int_C dt\,\frac{\sinh(1+\nu)(t-\pi i)}{\sinh(t-\pi i)}
\frac{1}{\cosh \nu t}=0,
\]
we obtain $J_2={3\over2}$. Collecting these terms, we obtain (\ref{eqn:K}).
\qed

\setcounter{equation}{0}
\section{Integrals related to the case $\nu=1/2$}\label{app:E}

Here we supply proofs to the formulas presented in section \ref{sec:5}.
We retain the notation there.

First let us evaluate the integrals $I_{ki}$ \eqref{eqn:Iki}
and $J_{ki}$ \eqref{eqn:Jki}.
Set $B_j=e^{\beta_j}$, and define
\begin{eqnarray}
F_{ji}&=&\frac{1}{\prod_{k=r\atop k\neq j}^i(B_j-B_k)
\prod_{k=i}^s(B_j+B_k)}
\qquad (r\le j\le i\le s),
\nonumber\\
G_{ji}&=&\frac{(-1)^n}{\prod_{k=r}^i(B_j+B_k)
\prod_{k=i\atop k\neq j}^s(B_j-B_k)}
\qquad (r\le i\le j\le s).
\label{eqn:FG}
\end{eqnarray}

\begin{figure}[htb]
\begin{center}
\setlength{\unitlength}{0.0125in}
\begin{picture}(313,280)(0,-10)
\drawline(153,115)(233,115)
\drawline(233,150)(153,150)
\drawline(161.000,152.000)(153.000,150.000)(161.000,148.000)
\drawline(113,130)(313,130)
\drawline(153,150)	(148.852,150.294)
	(144.991,150.549)
	(141.405,150.764)
	(138.077,150.941)
	(134.993,151.078)
	(132.139,151.176)
	(129.500,151.234)
	(127.061,151.254)
	(122.725,151.176)
	(119.014,150.941)
	(115.812,150.549)
	(113.000,150.000)

\drawline(113,150)	(108.161,148.741)
	(105.313,147.843)
	(102.387,146.744)
	(99.538,145.427)
	(96.920,143.875)
	(93.000,140.000)

\drawline(93,140)	(91.625,136.544)
	(91.167,132.500)
	(91.625,128.456)
	(93.000,125.000)

\drawline(93,125)	(96.920,121.125)
	(99.538,119.573)
	(102.387,118.256)
	(105.313,117.157)
	(108.161,116.259)
	(113.000,115.000)

\drawline(113,115)	(115.812,114.451)
	(119.014,114.059)
	(122.725,113.824)
	(127.061,113.746)
	(129.500,113.766)
	(132.139,113.824)
	(134.993,113.922)
	(138.077,114.059)
	(141.405,114.236)
	(144.991,114.451)
	(148.852,114.706)
	(153.000,115.000)

\drawline(145.167,112.423)(153.000,115.000)(144.876,116.413)
\drawline(113,265)	(116.452,264.865)
	(119.782,264.722)
	(122.994,264.573)
	(126.090,264.415)
	(129.074,264.250)
	(131.949,264.075)
	(134.717,263.892)
	(137.381,263.699)
	(139.945,263.496)
	(142.411,263.282)
	(147.063,262.822)
	(151.360,262.314)
	(155.326,261.755)
	(158.985,261.141)
	(162.361,260.469)
	(165.477,259.734)
	(168.358,258.933)
	(171.026,258.063)
	(173.507,257.120)
	(178.000,255.000)

\drawline(178,255)	(180.958,253.235)
	(184.046,251.083)
	(187.230,248.588)
	(190.481,245.795)
	(193.765,242.747)
	(197.051,239.489)
	(200.307,236.064)
	(203.501,232.517)
	(206.601,228.890)
	(209.576,225.230)
	(212.393,221.578)
	(215.021,217.980)
	(217.427,214.479)
	(219.580,211.119)
	(221.449,207.945)
	(223.000,205.000)

\drawline(223,205)	(224.630,201.257)
	(226.093,196.942)
	(226.770,194.519)
	(227.417,191.893)
	(228.037,189.043)
	(228.634,185.949)
	(229.210,182.591)
	(229.771,178.949)
	(230.319,175.002)
	(230.858,170.731)
	(231.392,166.116)
	(231.925,161.135)
	(232.192,158.502)
	(232.459,155.770)
	(232.729,152.937)
	(233.000,150.000)

\drawline(113,0)	(109.558,0.059)
	(106.236,0.130)
	(103.032,0.212)
	(99.943,0.308)
	(96.966,0.417)
	(94.098,0.539)
	(91.336,0.676)
	(88.676,0.827)
	(86.117,0.993)
	(83.655,1.174)
	(79.009,1.585)
	(74.716,2.064)
	(70.751,2.613)
	(67.091,3.237)
	(63.711,3.939)
	(60.589,4.723)
	(57.700,5.592)
	(55.020,6.551)
	(52.526,7.603)
	(48.000,10.000)

\drawline(48,10)	(45.025,12.012)
	(41.998,14.432)
	(38.943,17.213)
	(35.886,20.309)
	(32.852,23.673)
	(29.864,27.259)
	(26.947,31.018)
	(24.127,34.906)
	(21.428,38.874)
	(18.875,42.877)
	(16.492,46.868)
	(14.305,50.799)
	(12.337,54.625)
	(10.614,58.298)
	(9.160,61.772)
	(8.000,65.000)

\drawline(8,65)	(7.070,68.055)
	(6.200,71.304)
	(5.389,74.736)
	(4.639,78.337)
	(3.949,82.094)
	(3.319,85.994)
	(2.749,90.024)
	(2.239,94.171)
	(1.789,98.422)
	(1.399,102.764)
	(1.069,107.184)
	(0.799,111.669)
	(0.589,116.207)
	(0.439,120.783)
	(0.349,125.385)
	(0.318,130.000)
	(0.349,134.615)
	(0.439,139.217)
	(0.589,143.793)
	(0.799,148.331)
	(1.069,152.816)
	(1.399,157.236)
	(1.789,161.578)
	(2.239,165.829)
	(2.749,169.976)
	(3.319,174.006)
	(3.949,177.906)
	(4.639,181.663)
	(5.389,185.264)
	(6.200,188.696)
	(7.070,191.945)
	(8.000,195.000)

\drawline(8,195)	(9.169,198.225)
	(10.646,201.690)
	(12.405,205.350)
	(14.416,209.159)
	(16.651,213.073)
	(19.083,217.045)
	(21.684,221.031)
	(24.424,224.984)
	(27.275,228.860)
	(30.211,232.612)
	(33.202,236.196)
	(36.220,239.567)
	(39.237,242.677)
	(42.225,245.484)
	(45.155,247.940)
	(48.000,250.000)

\drawline(48,250)	(52.389,252.561)
	(54.829,253.733)
	(57.465,254.839)
	(60.319,255.885)
	(63.417,256.876)
	(66.782,257.818)
	(70.437,258.717)
	(74.407,259.578)
	(78.715,260.406)
	(83.385,261.208)
	(85.863,261.600)
	(88.441,261.987)
	(91.121,262.371)
	(93.907,262.751)
	(96.801,263.128)
	(99.806,263.504)
	(102.926,263.878)
	(106.162,264.252)
	(109.520,264.625)
	(113.000,265.000)

\drawline(105.255,262.169)(113.000,265.000)(104.834,266.146)
\drawline(233,115)	(233.009,112.356)
	(233.009,109.804)
	(232.984,104.970)
	(232.923,100.480)
	(232.824,96.314)
	(232.685,92.455)
	(232.504,88.883)
	(232.279,85.582)
	(232.010,82.533)
	(231.693,79.717)
	(231.327,77.116)
	(230.441,72.486)
	(229.337,68.496)
	(228.000,65.000)

\drawline(228,65)	(226.514,61.946)
	(224.682,58.697)
	(222.541,55.292)
	(220.126,51.772)
	(217.474,48.175)
	(214.621,44.543)
	(211.603,40.915)
	(208.457,37.330)
	(205.218,33.829)
	(201.923,30.451)
	(198.608,27.236)
	(195.309,24.224)
	(192.063,21.454)
	(188.905,18.967)
	(185.872,16.803)
	(183.000,15.000)

\drawline(183,15)	(178.263,12.571)
	(175.631,11.451)
	(172.790,10.387)
	(169.713,9.374)
	(166.375,8.407)
	(162.750,7.478)
	(158.814,6.584)
	(154.539,5.719)
	(149.900,4.876)
	(147.437,4.462)
	(144.873,4.052)
	(142.205,3.644)
	(139.430,3.239)
	(136.546,2.835)
	(133.548,2.433)
	(130.433,2.031)
	(127.199,1.628)
	(123.841,1.224)
	(120.358,0.819)
	(116.745,0.411)
	(113.000,0.000)

\drawline(120.738,2.850)(113.000,0.000)(121.169,-1.127)
\put(108,135){\makebox(0,0)[lb]{\raisebox{0pt}[0pt][0pt]{\shortstack[l]{{\twlrm $0$}}}}}
\end{picture}
\vskip 2cm
\caption{The contour for the residue calculus.}
\label{fig:C4}
\end{center}
\end{figure}
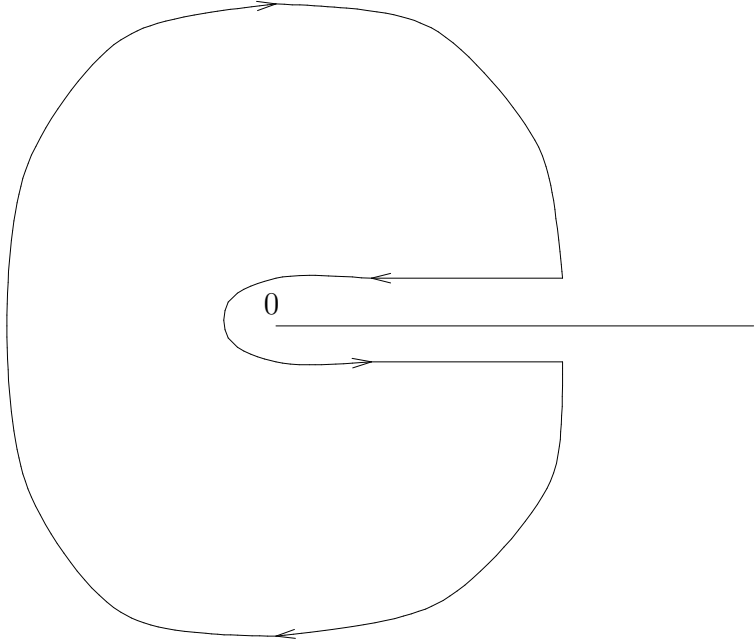

\begin{prop}For $1\le k\le n$ and $r\le i\le s$ we have 
\begin{eqnarray}
I_{ki}&=&
\frac{\sqrt{-1}^{s+1-i}}{\pi}B_i^{1/2}\left(\prod_{j=r}^sB_j\right)^{1/2}
\left(\sum_{j=r}^i(-B_j)^{k-1}\beta_jF_{ji}
+\sum_{j=i}^sB_j^{k-1}(\beta_j+\pi \sqrt{-1})G_{ji}\right),
\nonumber\\
&&\label{eqn:Iki1}\\
J_{ki}&=&2\sqrt{-1}^{s-i}B_i^{1/2}\left(\prod_{j=r}^sB_j\right)^{1/2}
\sum_{j=r}^i(-B_j)^{k-1}F_{ji}
\label{eqn:Jki1} \\
&=&
-2\sqrt{-1}^{s-i}B_i^{1/2}\left(\prod_{j=r}^sB_j\right)^{1/2}
\sum_{j=i}^s B_j^{k-1}G_{ji}.
\label{eqn:Jki2}
\end{eqnarray}
\end{prop}
\proof
Changing the integration variable to $A=e^\alpha$ we 
have 
\[
I_{ki}=\frac{\sqrt{-1}^{2r-s+1-i}}{\pi}B_i^{1/2}
\left(\prod_{j=r}^sB_j\right)^{1/2}
\times \int_0^\infty\omega_{ki},
\]
where we have set 
\[
\omega_{ki}=\frac{A^{k-1}dA}
{\prod_{j=r}^i(A+B_j)\prod_{j=i}^s(A-B_j+\sqrt{-1}\,0)}.
\]
To see \eqref{eqn:Iki1} it suffices to show that 
\[
\int_0^\infty\omega_{ki}
=(-1)^{n+1}
\left(\sum_{j=r}^i(-B_j)^{k-1}\beta_jF_{ji}
+\sum_{j=i}^sB_j^{k-1}(\beta_j+\pi \sqrt{-1})G_{ji}\right).
\]
This follows from integration of
$\omega_{ki}\log(-A)$ along the contour shown in Figure \ref{fig:C4}. 

The formula \eqref{eqn:Jki2} is a direct consequence of \eqref{eqn:Iki1}.
Counting the sum of the residues of $\omega_{ki}$ we find
\[
0=\sum_{j=r}^i(-B_j)^{k-1}F_{ji}+\sum_{j=i}^sB_j^{k-1} G_{ji}.
\]
This shows the equality of \eqref{eqn:Jki1} and \eqref{eqn:Jki2}.
\qed

Define
\begin{equation}
J_{ki}=I_{ki}+(-1)^{s+i}\bar{I}_{ki},
\label{eqn:Jki}
\end{equation}
and denote by $I_i$ and $J_i$ the column vectors 
\[
I_i={}^t(I_{1i},\cdots,I_{ni}),
\qquad
J_i={}^t(J_{1i},\cdots,J_{ni}).
\]

\begin{prop}\label{prop:id}
\begin{equation}
1=\prod_{r\le j<k\le s}2\sqrt{-1}\cosh\frac{\beta_j-\beta_k}{2}\times
\det\left(J_r,J_{r+1},\cdots,J_s\right).
\label{eqn:id}
\end{equation}
\end{prop}
\proof
To see this, we write down the known equality 
\begin{equation}
1=\br{1}=\sum_{\vep_r,\cdots,\vep_s}
\br{E^{(r)}_{\vep_r\vep_r}\cdots E^{(s)}_{\vep_s\vep_s}}
(\beta_r,\cdots,\beta_s).
\label{eqn:1}
\end{equation}
For convenience let us introduce the symbols $I_i(\pm),\bar{I}_i(\pm)$ 
by setting $I_i(+)=I_i,\bar{I}_i(+)=\bar{I}_i$ and
$I_i(-),\bar{I}_i(-)=$the empty symbol.
Then \eqref{eqn:1} can be written as
\begin{eqnarray*}
1&=&\prod_{r\le j<k\le s}2\sqrt{-1}\cosh\frac{\beta_j-\beta_k}{2}
\\
&&\times
\sum_{\vep_r,\cdots,\vep_s}
\det\left(I_r(-\vep_r),\cdots,I_s(-\vep_s),\bar{I}_s(\vep_s),
\cdots,\bar{I}_{r}(\vep_r)\right).
\end{eqnarray*}
Taking the sum over $\vep_s,\vep_{s-1},\cdots$ successively 
and keeping track of the signs, 	
we find that the last sum becomes a single determinant
$\det(J_r,\cdots,J_s)$.
\qed

\begin{prop}\label{prop:psipsi}
If $r\le m,l\le s$ then
\begin{equation}
\br{\psi_m\psi^*_l}=
D^{-1}\det\left(J_r,\cdots,J_{l-1},I_m,J_{l+1},\cdots,J_s\right)
\label{eqn:psipsi}
\end{equation}
where $D=\det(J_r,J_{r+1},\cdots,J_s)$.
\end{prop}
\proof
This can be verified in a similar way as in the proof of 
Proposition \ref{prop:id}.
As an example, let us take $r=1,s=4,m=1,l=3$:
\begin{eqnarray*}
\br{\psi_1\psi^*_3}&=&\br{\sigma_1^-\sigma_2^z\sigma_3^+1}
\\
&=&\sum_{\vep_2,\vep_4}\br{E^{(1)}_{-+}
\left(\vep_2E^{(2)}_{\vep_2\vep_2}\right)
E^{(3)}_{+-}E^{(4)}_{\vep_4\vep_4}}.
\end{eqnarray*}
Using \eqref{eqn:EE} and \eqref{eqn:id}, in 
the same notation as in Proposition \ref{prop:id}, we have
\begin{eqnarray*}
D\br{\psi_1\psi^*_3}&=&
\sum_{\vep_2,\vep_4}\vep_2
\det\Bigl(I_1(+),I_2(-\vep_2),I_3(-),I_4(-\vep_4),
\\
&&\qquad \bar{I}_4(\vep_4),\bar{I}_3(-),\bar{I}_2(\vep_2),\bar{I}_1(+)\Bigr).
\end{eqnarray*}
The sum over $\vep_4$ gives $J_4$, Summing further over $\vep_2$ we obtain
\begin{eqnarray*}
&&\det(I_1,J_4,\bar{I}_2,\bar{I}_1)-
\det(I_1,I_2,J_4,\bar{I}_1)
\\
&&=\det(I_1,-J_2,J_4,\bar{I}_1)
\\
&&=\det(J_1,J_2,I_1,J_4)
\end{eqnarray*}
where we used $J_2=I_2+\bar{I}_2$ and $J_1=I_1-\bar{I}_1$.

In general, consider the case $m<l$. 
We have 
\[
D\br{\psi_m\psi^*_l}=
\sum_{\vep_r,\cdots,\vep_s}\vep_{m+1}\cdots\vep_{l-1}
\det I(\vep),
\]
where $I(\vep)$ denotes the matrix consisting of the following array 
of column vectors:
\begin{eqnarray*}
&&I_r(-\vep_r),\cdots,I_{m-1}(-\vep_{m-1}),I_m(+),
I_{m+1}(-\vep_{m+1}),\cdots,I_{l-1}(-\vep_{l-1}),
\\
&&
I_l(-),I_{l+1}(-\vep_{l+1}),\cdots,I_s(-\vep_s),
\bar{I}_s(\vep_s),\cdots,\bar{I}_{l+1}(\vep_{l+1}),
\\
&&
\bar{I}_l(-),\bar{I}_{l-1}(\vep_{l-1}),\cdots,\bar{I}_{m+1}(\vep_{m+1}),
\bar{I}_m(+),\bar{I}_{m-1}(\vep_{m-1}),\cdots,\bar{I}_r(\vep_r).
\end{eqnarray*}
Summing over $\vep_s,\vep_{s-1},\cdots$ we see that the sum 
combines into a single determinant
\begin{eqnarray*}
&&\det(J_r,\cdots,J_{m-1},I_m,-J_{m+1},\cdots,-J_{l-1},J_{l+1},
\cdots,J_s,\bar{I}_m)
\\
&&=
\det(J_r,\cdots,J_{m-1},(-1)^{s-m}\bar{I}_m,J_{m+1},\cdots,J_{l-1},I_m,
J_{l+1},\cdots,J_s).
\end{eqnarray*}
This shows \eqref{eqn:psipsi}.
The other cases are similar.
\qed

Arguing in a similar manner, one can show 
in general the following.
\begin{prop}\label{prop:multi}
Suppose $m_1<\cdots<m_k$, $l_1<\cdots<l_k$,
and let $r\le\min(m_1,l_1)$, $\max(m_k,l_k)\le s$. 
Then 
\begin{eqnarray}
&&\br{\psi_{m_1}\cdots\psi_{m_k}\psi^*_{l_k}\cdots\psi^*_{l_1}}
\nonumber\\
&=&
D^{-1}\det\left(J_r,\cdots,J_{l_1-1},I_{m_1},J_{l_1+1},
\cdots,J_{l_k-1},I_{m_k},J_{l_k+1},
\cdots,J_s\right)
\nonumber
\end{eqnarray}
where $D=\det(J_r,J_{r+1},\cdots,J_s)$.
In the right hand side $I_{m_j}$ is placed at the $l_j$-th slot.
\end{prop}
We omit the details.

\noindent{\it Proof of Proposition \ref{prop:wick}.}\quad
Consider the matrix $X=(J_r,J_{r+1},\cdots,J_s)$, and set 
$K_m=X^{-1}I_m$. 
Then Proposition \ref{prop:multi} states that 
\[
\br{\psi_{m_1}\cdots\psi_{m_k}\psi^*_{l_k}\cdots\psi^*_{l_1}}
=
\det\left(e_1,\cdots,K_{m_1},\cdots,K_{m_k},\cdots,e_{n}\right)
\]
where the $e_j=(\delta_{ji})_{1\le i\le n}$ denote the unit vectors. 
It is clear that 
the right hand side is 
\[
\det\left( (K_{m_j})_{l_i}\right)_{1\le j,i\le k}.
\]
and that $\br{\psi_m\psi^*_l}=(K_m)_l$. 
The proposition follows from this observation.
\qed

\noindent{\it Proof of Proposition \ref{prop:two}.}\quad 
The formula \eqref{eqn:psipsi2} is already known. 
Let us show \eqref{eqn:psipsi1} by taking 
$r=m$ and $s=l$ in \eqref{eqn:psipsi}.
We wish to compute 
\[
\det(J_m,J_{m+1},\cdots,J_{l-1},I_m-\frac{1}{2}J_m).
\]
Substituting 
\begin{eqnarray*}
I_{km}-\frac{1}{2}J_{km}
&=&
\frac{\sqrt{-1}^n}{\pi}B_m^{1/2}\left(\prod_{j=m}^lB_j\right)^{1/2}
\left((-B_m)^{k-1}\beta_mF_{mm}+\sum_{j=m}^lB_j^{k-1}\beta_jG_{jm}\right)
\\
&=&
\frac{\sqrt{-1}^n}{\pi}B_m^{1/2}\left(\prod_{j=m}^lB_j\right)^{1/2}
\sum_{j=m}^lB_j^{k-1}(\beta_j-\beta_m)G_{jm}
\end{eqnarray*}
and using \eqref{eqn:Jki1}, we have the following expression
for $D\br{\psi_m\psi^*_l}$:
\[
\frac{\sqrt{-1}^n}{\pi}B_m^{1/2}\left(\prod_{j=m}^lB_j\right)^{n/2}
\sum_{j=m}^l(\beta_j-\beta_m)G_{jm}
\times \det X_jYZ.
\]
Here $X_j,Y,Z$ are the following matrices.
\begin{eqnarray*}
X_j&=&
\pmatrix{1   & \cdots & 1        & 1   \cr
        -B_m & \cdots & -B_{l-1} & B_j \cr
        \vdots&       &\vdots    &\vdots\cr
       (-B_m)^{n-1}&\cdots&(-B_{l-1})^{n-1}&B_j^{n-1}\cr
}
\\
Y&=&
\pmatrix{F_{mm}&\cdots &F_{m\,l-1}& 0 \cr
            0 & \ddots& \vdots    & 0 \cr
            0 &\cdots &F_{l-1\,l-1}& 0 \cr
            0 & \cdots&0          & 1\cr
}
\\
Z&=&
{\rm diag}\left(2\sqrt{-1}^{l-m}B_m^{1/2},\cdots,
2\sqrt{-1}B_{l-1}^{1/2},1\right).
\end{eqnarray*}
The matrix $Y$ is upper triangular with diagonal entries 
$F_{kk}$ ($m\le k\le l-1$) and $1$. 
It is therefore straightforward to compute the determinant.
Inserting the expressions for $F_{ji}$ and $G_{ji}$
in \eqref{eqn:FG} we obtain the formula
\eqref{eqn:psipsi1}.

The case of $\br{\psi_m^*\psi_l}$ can be shown similarly, using
\eqref{eqn:Jki2}.
\qed

\setcounter{equation}{0}
\def\itm#1{\begin{itemize}\item{#1}\end{itemize}}
\def\oR{\overline R}
\def\oG{\overline G}
\def\ve{\varepsilon}
\def\lb#1{\label{eqn:#1}}
\def\refeq#1{(\ref{eqn:#1})}
\def\sc[#1,#2]{{\scriptstyle{\scriptstyle#1\over\scriptstyle#2}}}
\def\cR{\check R}
\def\mt#1#2{\pmatrix{&#2\cr#1&\cr}}
\def\s{\sigma}
\def\H{{\cal H}}
\def\Ad{{\rm Ad}\,}

\def\Fa{
\begin{figure}[htb]
\begin{center}
\setlength{\unitlength}{0.0125in}
\begin{picture}(288,108)(0,-10)
\drawline(185,65)(185,25)
\drawline(183.000,33.000)(185.000,25.000)(187.000,33.000)
\drawline(145,65)(145,25)
\drawline(143.000,33.000)(145.000,25.000)(147.000,33.000)
\drawline(105,65)(105,25)
\drawline(103.000,33.000)(105.000,25.000)(107.000,33.000)
\drawline(53.000,47.000)(45.000,45.000)(53.000,43.000)
\drawline(45,45)(245,45)
\put(-20,42){\makebox(0,0)[lb]{\raisebox{0pt}[0pt][0pt]{\shortstack[l]
{{\twlrm $T(u)^{\{\varepsilon'_n\}}_{\{\varepsilon_n\}}=$}}}}}
\put(170,80){\makebox(0,0)[lb]{\raisebox{0pt}[0pt][0pt]{\shortstack[l]{{\twlrm $\varepsilon'_{n-1}$}}}}}
\put(170,0){\makebox(0,0)[lb]{\raisebox{0pt}[0pt][0pt]{\shortstack[l]{{\twlrm $\varepsilon_{n-1}$}}}}}
\put(95,80){\makebox(0,0)[lb]{\raisebox{0pt}[0pt][0pt]{\shortstack[l]{{\twlrm $\varepsilon'_{n+1}$}}}}}
\put(95,0){\makebox(0,0)[lb]{\raisebox{0pt}[0pt][0pt]{\shortstack[l]{{\twlrm $\varepsilon_{n+1}$}}}}}
\put(140,0){\makebox(0,0)[lb]{\raisebox{0pt}[0pt][0pt]{\shortstack[l]{{\twlrm $\varepsilon_n$}}}}}
\put(65,0){\makebox(0,0)[lb]{\raisebox{0pt}[0pt][0pt]{\shortstack[l]{{\twlrm $\cdots$}}}}}
\put(200,0){\makebox(0,0)[lb]{\raisebox{0pt}[0pt][0pt]{\shortstack[l]{{\twlrm $\cdots$}}}}}
\put(200,80){\makebox(0,0)[lb]{\raisebox{0pt}[0pt][0pt]{\shortstack[l]{{\twlrm $\cdots$}}}}}
\put(140,80){\makebox(0,0)[lb]{\raisebox{0pt}[0pt][0pt]{\shortstack[l]{{\twlrm $\varepsilon'_n$}}}}}
\put(65,80){\makebox(0,0)[lb]{\raisebox{0pt}[0pt][0pt]{\shortstack[l]{{\twlrm $\cdots$}}}}}
\end{picture}
\end{center}
\end{figure}}

\def\Fb{
\begin{figure}[htb]
\begin{center}
\setlength{\unitlength}{0.0125in}
\begin{picture}(172,230)(0,-10)
\drawline(100.000,98.000)(102.000,90.000)(104.000,98.000)
\drawline(102,90)(102,115)
\drawline(174.000,107.000)(172.000,115.000)(170.000,107.000)
\drawline(172,115)(172,90)
\drawline(27,215)	(24.229,214.760)
	(21.663,214.485)
	(17.104,213.819)
	(13.251,212.972)
	(10.029,211.915)
	(5.190,209.053)
	(2.000,205.000)

\drawline(2,205)	(0.729,201.877)
	(0.112,198.547)
	(0.092,195.112)
	(0.615,191.677)
	(1.623,188.344)
	(3.060,185.218)
	(4.871,182.402)
	(7.000,180.000)

\drawline(7,180)	(11.815,176.279)
	(14.444,174.816)
	(17.206,173.594)
	(20.092,172.595)
	(23.092,171.802)
	(26.194,171.196)
	(29.389,170.760)
	(32.666,170.476)
	(36.015,170.326)
	(39.426,170.293)
	(42.888,170.359)
	(46.390,170.505)
	(49.924,170.715)
	(53.478,170.970)
	(57.042,171.252)
	(60.605,171.545)
	(64.158,171.829)
	(67.690,172.087)
	(71.190,172.302)
	(74.649,172.455)
	(78.056,172.529)
	(81.401,172.506)
	(84.674,172.369)
	(87.863,172.098)
	(90.959,171.678)
	(93.952,171.089)
	(96.830,170.314)
	(99.585,169.336)
	(102.205,168.136)
	(107.000,165.000)

\drawline(107,165)	(111.828,160.018)
	(113.979,157.036)
	(115.957,153.785)
	(117.764,150.311)
	(119.403,146.657)
	(120.875,142.868)
	(122.184,138.988)
	(123.330,135.061)
	(124.317,131.133)
	(125.146,127.246)
	(125.819,123.446)
	(126.339,119.777)
	(126.708,116.284)
	(126.927,113.010)
	(127.000,110.000)

\drawline(127,110)	(126.933,106.989)
	(126.730,103.713)
	(126.385,100.216)
	(125.894,96.543)
	(125.253,92.738)
	(124.458,88.847)
	(123.503,84.913)
	(122.384,80.983)
	(121.097,77.100)
	(119.638,73.308)
	(118.001,69.654)
	(116.182,66.182)
	(114.177,62.935)
	(111.982,59.959)
	(107.000,55.000)

\drawline(107,55)	(102.328,52.063)
	(99.797,50.944)
	(97.148,50.036)
	(94.391,49.323)
	(91.533,48.789)
	(88.586,48.419)
	(85.556,48.195)
	(82.455,48.102)
	(79.290,48.124)
	(76.070,48.245)
	(72.806,48.447)
	(69.505,48.717)
	(66.178,49.036)
	(62.832,49.389)
	(59.478,49.761)
	(56.123,50.134)
	(52.778,50.493)
	(49.451,50.822)
	(46.152,51.104)
	(42.889,51.324)
	(39.672,51.465)
	(36.509,51.511)
	(33.410,51.446)
	(30.384,51.254)
	(27.439,50.919)
	(24.586,50.425)
	(21.832,49.755)
	(19.188,48.894)
	(16.662,47.825)
	(12.000,45.000)

\drawline(12,45)	(7.333,40.222)
	(5.332,37.207)
	(3.664,33.913)
	(2.413,30.442)
	(1.663,26.897)
	(1.497,23.382)
	(2.000,20.000)

\drawline(2,20)	(2.980,17.195)
	(4.427,14.595)
	(9.011,9.773)
	(12.296,7.433)
	(16.341,5.064)
	(21.217,2.606)
	(23.991,1.325)
	(27.000,0.000)

\drawline(27,215)	(31.262,214.715)
	(35.374,214.426)
	(39.340,214.130)
	(43.162,213.829)
	(46.845,213.520)
	(50.392,213.203)
	(53.807,212.878)
	(57.094,212.544)
	(60.257,212.200)
	(63.298,211.846)
	(66.222,211.480)
	(69.032,211.102)
	(71.732,210.712)
	(74.326,210.309)
	(79.209,209.460)
	(83.711,208.549)
	(87.860,207.572)
	(91.687,206.523)
	(95.219,205.396)
	(98.488,204.185)
	(101.521,202.886)
	(104.349,201.493)
	(107.000,200.000)

\drawline(107,200)	(109.761,198.186)
	(112.626,196.012)
	(115.566,193.517)
	(118.550,190.741)
	(121.550,187.724)
	(124.534,184.506)
	(127.474,181.126)
	(130.340,177.624)
	(133.101,174.040)
	(135.728,170.414)
	(138.191,166.786)
	(140.459,163.195)
	(142.505,159.681)
	(144.296,156.284)
	(145.805,153.044)
	(147.000,150.000)

\drawline(147,150)	(148.199,146.095)
	(149.226,141.748)
	(150.085,137.023)
	(150.452,134.538)
	(150.778,131.983)
	(151.063,129.365)
	(151.307,126.692)
	(151.510,123.972)
	(151.674,121.212)
	(151.798,118.422)
	(151.882,115.608)
	(151.927,112.779)
	(151.934,109.942)
	(151.902,107.106)
	(151.831,104.278)
	(151.723,101.466)
	(151.577,98.679)
	(151.394,95.924)
	(151.173,93.208)
	(150.916,90.541)
	(150.623,87.930)
	(150.293,85.382)
	(149.928,82.906)
	(149.091,78.201)
	(148.114,73.878)
	(147.000,70.000)

\drawline(147,70)	(145.836,66.773)
	(144.372,63.304)
	(142.633,59.636)
	(140.646,55.818)
	(138.437,51.894)
	(136.032,47.912)
	(133.458,43.917)
	(130.740,39.955)
	(127.906,36.073)
	(124.981,32.317)
	(121.992,28.732)
	(118.965,25.365)
	(115.926,22.262)
	(112.901,19.470)
	(109.917,17.034)
	(107.000,15.000)

\drawline(107,15)	(104.342,13.426)
	(101.508,11.963)
	(98.469,10.604)
	(95.196,9.345)
	(91.660,8.180)
	(87.831,7.104)
	(83.681,6.110)
	(79.178,5.194)
	(74.296,4.351)
	(71.703,3.954)
	(69.003,3.573)
	(66.194,3.208)
	(63.272,2.857)
	(60.232,2.520)
	(57.071,2.197)
	(53.786,1.885)
	(50.373,1.586)
	(46.829,1.298)
	(43.148,1.020)
	(39.329,0.752)
	(35.367,0.493)
	(31.259,0.243)
	(27.000,0.000)

\put(22,190){\makebox(0,0)[lb]{\raisebox{0pt}[0pt][0pt]{\shortstack[l]{{\twlrm $\times i$}}}}}
\put(17,20){\makebox(0,0)[lb]{\raisebox{0pt}[0pt][0pt]{\shortstack[l]{{\twlrm $\times -i$}}}}}
\put(57,100){\makebox(0,0)[lb]{\raisebox{0pt}[0pt][0pt]{\shortstack[l]{{\twlrm $\times S_j$}}}}}
\end{picture}
\vskip 2cm
\caption{The contour $C$.}
\label{fig:Fb}
\end{center}
\end{figure}}

\section{Inhomogenoeous Ising model}\label{app:F}
In this section, we compute the correlation functions of the inhomogenoeous
Ising model at the critical temperature. We give an explicit formula
for the vacuum expectation values $\lv\psi^*_m\psi_n\rv$ where
$\psi^*_n$, $\psi_n$ $(n\in\Z)$ are the free fermions diagonalizing
the transfer matrix $T(u)$ of the critical Ising model. The general
correlation functions are given by the Pfaffians of these matrix elements.
In \cite{AuP} the correlation functions for the critical Ising model were 
given.
We have not checked the equivalence of our result to theirs
except for some simple cases.

\subsection{Completely inhomogeneous Hamiltonian}
Consider the transfer matrix of a completely inhomogeneous six-vertex
model in the infinite volume:

\Fa
\[
=\sum_{\{\tau_n\}}\prod_n
\oR^{\varepsilon'_n\tau_{n-1}}_{\varepsilon_n\tau_n}(\beta_n+u).
\]
The horizontal line carries the spectral parameter $0$ and the vertical lines
carry the spectral parameters $\beta_n+u$. We assume that
$\beta_n=0$ if $|n|\gg 0$.
The Boltzmann weights
$\oR^{\varepsilon'_1,\varepsilon'_2}_{\varepsilon_1,\varepsilon_2}(\beta)$
are given by \refeq{BW} with $\nu=\sc[1,2]$. This is the choice in which
the six-vertex model is equivalent to the critical Ising model
(see e.g.\cite{Bax82}).

Let $S$ be the shift operator
\[
S^{\{\varepsilon'_n\}}_{\{\varepsilon_n\}}=
\prod_n\delta_{\varepsilon_{n+1}\varepsilon'_n}.
\]
Then we have
\bea
S^{-1}T(u)&=&\sum_{\{\tau_n\}}\prod_n
\cR^{\varepsilon'_n\tau_{n-1}}_{\tau_n\varepsilon_{n-1}}(\beta_n+u)\nn\\
&=&\cdots\cR_{n+1\,n}(\beta_{n+1}+u)\cR_{n\,n-1}(\beta_n+u)\cdots
\lb{TR}
\ena
where
\[
\cR(\beta)=\pmatrix{
1&&&\cr
&\bar c(\beta)&\bar b(\beta)&\cr
&\bar b(\beta)&\bar c(\beta)&\cr
&&&1\cr}
\]
and
\[
\bar b(\beta)={1-\zeta^2\over i(1+\zeta^2)},
\quad \bar c(\beta)={2\zeta\over1+\zeta^2},\quad\zeta=e^{-\beta/2}.
\]
The matrix $\cR(\beta)$ can be put in the form 
\[
\cR(\beta)=e^{\gamma X},\quad 
X=\sigma^{+}\otimes \sigma^{-}+\sigma^{-}\otimes \sigma^{+}
\]
where $\sigma^\pm=(\sigma^x\pm i\sigma^y)/2$ and 
$\gamma=\gamma(\beta)$ is related to $\beta$ by 
\bea
&&e^{-\gamma}={1+i\,\sinh{\beta\over2}\over\cosh{\beta\over2}}
={2\zeta\over1+\zeta^2}+{i(1-\zeta^2)\over1+\zeta^2},\nn\\
&&e^{-{\beta\over2}}={{1-i\sinh\gamma}\over\cosh\gamma}=\zeta.\nn
\lb{GB}
\ena
As $-i\beta$ increases from $0$ to $\pi$, $\gamma$ increases 
monotonically from $0$ to $\infty$.
We write $\gamma_n=\gamma(\beta_n)$, 
$C_n=\cosh\gamma_n$ and $S_n=\sinh\gamma_n$.
In the below, we assume that $S_n<1$.

\subsection{Jordan-Wigner transformation}
As usual, we define the Jordan-Wigner transformation
\bea
\psi^*_n=\s^+_n \prod_{m<n}\s^z_m,\nn\\
\psi_n=\s^-_n \prod_{m<n}\s^z_m.\nn
\ena
Note that the operators $\psi^*_n$ and $\psi_n$ satisfy the canonical
anti-commutation relation
\be
\lb{AR}
[\psi^*_m,\psi^*_n]_+=[\psi_m,\psi_n]_+=0,
[\psi^*_m,\psi_n]_+=\delta_{m,n}.
\en
For $m$, $n\in\Z$ such that $m>n$, we set
\[
H_{mn}=\psi_n\psi^*_m+(-1)^{m-n+1}\psi_m\psi^*_n.
\]
Note that
\be
X_{n\,n-1}=H_{n\,n-1},\quad[H_{n\,n-1},H_{m,n}]=H_{m\,n-1},
\quad[H_{n\,n-1},H_{m\,n-1}]=H_{m\,n}.
\lb{HR}
\en

We have
\begin{prop}
\be
T(0)^{-1}T(u)=1+{iu\over2}\H+O(u^2)
\lb{TrH}
\en
\end{prop}
where
\[
\H=\sum_{m>n}(-1)^{m-n}C_mS_{m-1}\cdots S_{n+1}C_nH_{mn}.
\]
\proof
Let us use $\equiv$ to mean an equality modulo $u^2$.
Using \refeq{TR}, we have
\bea
&&T(0)^{-1}T(u)-1\nn\\
&&\equiv\sum_n
\cdots\cR_{n-1\,n-2}(\beta_{n-1})^{-1}
\left(\cR_{n\,n-1}(\beta_n)^{-1}\cR_{n\,n-1}(\beta_n+u)-1\right)
\nn\\
&&\qquad \quad \times \cR_{n-1\,n-2}(\beta_{n-1})\cdots\nn\\
&&\equiv-{iu\over2}\sum_n\cdots\Ad\cR_{n-1\,n-2}(\beta_{n-1})^{-1}
C_nH_{n\,n-1}.\nn
\ena
Since $\cR_{k\,k-1}(\beta)^{-1}=e^{-\gamma_kH_{k\,k-1}}$,
the proposition follows from \refeq{HR}.\qed

If we fix $\beta_n$'s, 
the transfer matrices $T(u)$ commute with each other 
for different value of $u$. 
Therefore $\H$ also commutes with $T(u)$, and 
they can be diagonalized simultaneously.

\subsection{Diagonalization}
In order to diagonalize the Hamiltonian $\H$ we set
\bea
\phi(\theta)=\sum_nC_ne^{in\theta}\prod_{j\le n-1}(1+S_je^{-i\theta})
\prod_{j\ge n+1}(1-S_je^{i\theta})\psi_n,\lb{BG}\\
\phi^*(\theta)=\sum_nC_ne^{-in\theta}\prod_{j\le n-1}(1-S_je^{i\theta})
\prod_{j\ge n+1}(1+S_je^{-i\theta})\psi^*_n.\nn
\ena
Then we have
\begin{prop}
\be
[\H,\phi(\theta)]=-(e^{i\theta}+e^{-i\theta})\phi(\theta),\lb{P1}
\en
\be
[\H,\phi^*(\theta)]=(e^{i\theta}+e^{-i\theta})\phi^*(\theta),\lb{P2}
\en
\be
[\phi^*(\theta_1),\phi(\theta_2)]_+=\prod_j(1+S_je^{-i\theta_1})
(1-S_je^{i\theta_1})\sum_ke^{ik(\theta_1-\theta_2)}.\lb{P3}
\en
\end{prop}
\proof
Set $z=e^{i\theta}$, and write
\bea
\phi(\theta)&=&\sum_{n\in\Z}x_n\psi_n,\nn\\
x_n&=&C_nz^n\prod_{j\le n-1}(1+S_jz^{-1})
\prod_{j\ge n+1}(1-S_jz).\nn
\ena
Write also,
\bea
[\H,\psi_n]&=&-\sum_m\psi_mA_{mn}\nn\\
A_{mn}&=&\cases{C_mS_{m-1}\cdots S_{n+1}C_n&if $m>n$;\cr
0&if $m=n$;\cr
(-1)^{n-m-1}C_nS_{n-1}\cdots S_{m+1}C_m&if $m<n$.\cr}
\ena
We are to prove
\be
\sum_{n\in\Z}A_{mn}x_n=(z+z^{-1})x_m.
\lb{LE}
\en
Suppose that $\beta_n=0$, i.e., $C_n=1$ and $S_n=0$,
except for $M\le n\le N$. If $m\ge N+2$ or $m\le M-2$, then \refeq{LE}
is valid because
\[
A_{mn}=\cases{1&if $n=m\pm1$;\cr
0&otherwise,\cr}
\]
and
\[
x_n=\cases{z^n\prod_{M\le j\le N}(1+S_jz^{-1})&if $n\ge N+1$;\cr
z^n\prod_{M\le j\le N}(1-S_jz)&if $n\le M-1$.\cr}
\]
Therefore, \refeq{LE} is written in the form
\be
A^{(N,M)}x^{(N,M)}=0
\lb{AX}
\en
where $A^{(N,M)}=\Bigl(A^{(N,M)}_{mn}\Bigr)_{N+1\ge m,n\ge M-1}$
and $x^{(N,M)}=\Bigl(x^{(N,M)}_n\Bigr)_{N+1\ge n\ge M-1}$.
The matrix $A^{(N,M)}$ is of the form:
\bea
&&A^{(N,M)}=\pmatrix{
{\scriptstyle -z^{-1}}&
{\scriptstyle C_N}&\cdots\cr
{\scriptstyle C_N}&{\scriptstyle -z-z^{-1}}&\cdots\cr
{\scriptstyle -S_NC_{N-1}}&{\scriptstyle C_NC_{N-1}}&\cr
{\scriptstyle S_NS_{N-1}C_{N-2}}&
{\scriptstyle -C_NS_{N-1}C_{N-2}}&\cr
\cdot&\cdot&\cr
\cdot&\cdot&{\overline A}^{(N-1,M)}\cr
\cdot&\cdot&\cr
{\scriptstyle(-1)^{N-M}S_NS_{N-1}\cdots S_{M+1}C_M}&
{\scriptstyle(-1)^{N-M-1}C_NS_{N-1}\cdots S_{M+1}C_M}&\cr
{\scriptstyle(-1)^{N-M+1}S_NS_{N-1}\cdots S_{M+1}S_M}&
{\scriptstyle(-1)^{N-M}C_NS_{N-1}\cdots S_{M+1}S_M}&\cr}\nn\\
&&=\pmatrix{
&{\scriptstyle S_NS_{N-1}\cdots S_{M+1}C_M}&
{\scriptstyle S_NS_{N-1}\cdots S_{M+1}S_M}\cr
&{\scriptstyle C_NS_{N-1}\cdots S_{M+1}C_M}&
{\scriptstyle C_NS_{N-1}\cdots S_{M+1}S_M}\cr
&\cdot&\cdot\cr
\underline{A}^{(N,M+1)}&\cdot&\cdot\cr
&\cdot&\cdot\cr
&{\scriptstyle C_{M+2}S_{M+1}C_M}&
{\scriptstyle C_{M+2}S_{M+1}S_M}\cr
&{\scriptstyle C_{M+1}C_M}&
{\scriptstyle C_{M+1}S_M}\cr
\cdots&{\scriptstyle -z-z^{-1}}&{\scriptstyle C_M}\cr
\cdots&{\scriptstyle C_M}&{\scriptstyle -z}\cr}.\nn
\ena
Here 
\[
\overline{A}^{(N-1,M)}=\Bigl(A^{(N-1,M)}_{mn}\Bigr)_{N-1\ge m,n\ge M-1},
\quad
\underline{A}^{(N,M+1)}=\Bigl(A^{(N,M+1)}_{mn}\Bigr)_{N+1\ge m,n\ge M+1}.
\]
Similarly, the vector $x^{(N,M)}$ is of the form:
\bea
x^{(N,M)}&=&\pmatrix{(1+S_Mz^{-1})\cdots(1+S_Nz^{-1})z^{N+1}\cr
(1+S_Mz^{-1})\cdots(1+S_{N-1}z^{-1})C_Nz^N\cr
\phantom{-}\cr
\overline{x}^{(N-1,M)}\times(1-S_Nz)\cr}\nn\\
&=&\pmatrix{\underline{x}^{(N,M+1)}\times(1+S_Mz^{-1})\cr
\phantom{-}\cr
C_M(1-S_{M+1}z)\cdots(1-S_Nz)z^M\cr
(1-S_Mz)(1-S_{M+1}z)\cdots(1-S_Nz)z^{M-1}\cr}.
\ena
Here
\[
\overline{x}^{(N-1,M)}=\Bigl(x^{(N-1,M)}_n\Bigr)_{N-1\ge n\ge M-1},
\quad
\underline{x}^{(N,M+1)}=\Bigl(x^{(N,M+1)}_n\Bigr)_{N+1\ge n\ge M+1}.
\]
We prove \refeq{AX} by induction on $N-M$. If $N=M$,
we can check directly that
\[
\pmatrix{-z^{-1}&C_N&S_N\cr
C_N&-z-z^{-1}&C_N\cr
-S_N&C_N&-z}
\pmatrix{(1+S_Nz^{-1})z^{N+1}\cr C_Nz^N\cr (1-S_Nz)z^{N-1}\cr}=0.
\]
If $N>M$, noting that
\[
-S_N\cdot(1+S_Nz^{-1})z^{N+1}+C_N\cdot C_Nz^N=(1-S_Nz)z^N,
\]
we can reduce the equality $\Bigl(A^{(N,M)}x^{(N,M)}\Bigr)_n=0$
for $N-1\ge n\ge M-1$ to $\Bigl(A^{(N-1,M)}x^{(N-1,M)}\Bigr)_n=0$.
Similarly, noting that
\[
C_M\cdot C_Mz^M+S_M\cdot(1-S_Mz)z^{M-1}=(1+S_Mz^{-1})z^M,
\]
we can reduce the equality $\Bigl(A^{(N,M)}x^{(N,M)}\Bigr)_n=0$
for $N+1\ge n\ge M+1$ to $\Bigl(A^{(N,M+1)}x^{(N,M+1)}\Bigr)_n=0$.
The proof of \refeq{P2} is similar.
 
Let us prove \refeq{P3}. We have
\bea
&&[\phi^*(\theta_1),\phi(\theta_2)]_+
=\sum_kC_k^2z_2^kz_1^{-k}\nn\\
&&\times\prod_{j=-\infty}^{k-1}(1+S_jz_2^{-1})(1-S_jz_1)
\prod^{\infty}_{j=k+1}(1-S_jz_2)(1+S_jz_1^{-1})\nn
\ena
where $z_j=e^{i\theta_j}$ ($j=1,2$).
We assume that $C_n=1$ and $S_n=0$ except for $M\le n\le N$. Then we have
\bea
&&[\phi^*(\theta_1),\phi(\theta_2)]_+
=z_2^{N+1}z_1^{-N-1}{\prod_{j=M}^N
(1+S_jz_2^{-1})(1-S_jz_1)
\over1-z_2z_1^{-1}}\nn\\
&&+\sum_{k=M}^N z_2^kz_1^{-k}(1+S_k^2)
\prod_{j=M}^{k-1}(1+S_jz_2^{-1})(1-S_jz_1)
\prod^{N}_{j=k+1}(1-S_jz_2)(1+S_jz_1^{-1})\nn\\
&&+
z_2^{M-1}z_1^{-M+1}
{\prod_{j=M}^N(1-S_jz_2)(1+S_jz_1^{-1})
\over1-z_2^{-1}z_1}.\lb{AC}
\ena
Write the RHS as
\[
\Bigl(\sum_{k\in\Z}z_2^kz_1^{-k}\Bigr)
\prod_{j=M}^N
(1+S_jz_1^{-1})(1-S_jz_1)
+F^{(N,M)}(S_M,\cdots,S_N;z_1,z_2).
\]
Then, $F^{(N,M)}(S_M,\cdots,S_N;z_1,z_2)$ belongs to
$\C[S_M,\cdots,S_N,z_1,z_1^{-1},z_2,z_2^{-1}]$.

We wish to show $F^{(N,M)}=0$.
Let $G^{(N,M)}(S_M,\cdots,S_N;z_1,z_2)$ be the RHS of \refeq{AC}.
Note that $F^{(N,M)}=G^{(N,M)}$ in
$\C(z_1,z_2)[S_M,\cdots,S_N]$.
Therefore, it is enough to show $G^{(N,M)}=0$
as an element of $\C(z_1,z_2)[S_M,\cdots,S_N]$.
Note that
\bea
G^{(N,M)}(S_M,\cdots,S_N;z_1,z_2)
&=&G^{(N,M)}(-S_M,\cdots,-S_N;z_2^{-1},z_1^{-1})\nn\\
&=&z_2^{N+M}z_1^{-N-M}
G^{(N,M)}(S_N,\cdots,S_M;z_2,z_1).\nn
\ena
It is also easy to show that
\[
G^{(N,M)}(S_M,\cdots,S_N;z_1,z_2)
=G^{(N,M)}(S_{\s(M)},\cdots,S_{\s(N)};z_1,z_2)
\]
for any permutation $\s$ of $\{M,\cdots,N\}$. Now
$G^{(N,M)}(S_M,\cdots,S_N;z_1,z_2)$
is a polynomial of degree $2$ in $S_M$. Therefore, in order to show that
$G^{(N,M)}=0$, it is enough to show
\[
G^{(N,M)}(z_1,S_{M+1},\cdots,S_N;z_1,z_2)=0.
\]
This is shown by induction: we have
\[
G^{(N,M)}(z_1,S_{M+1},\cdots,S_N;z_1,z_2)
=(z_2z_1^{-1}+z_2z_1)
{\overline G}^{(N,M+1)}(S_{M+1},\cdots,S_N;z_1,z_2)
\]
where
\bea
&&\overline{G}^{(N,M+1)}(S_{M+1},\cdots,S_N;z_1,z_2)
=-z_2^{N-1}z_1^{-N+1}
\sum_{j=M+1}^N(1+S_jz_2^{-1})(1-S_jz_1)\nn\\
&&+(1-z_1z_2^{-1})
\sum_{k=M+1}^N z_2^{k-1}z_1^{-k+1}(1+S_k^2)
\nn\\
&&\qquad\quad \times \prod_{j=M+1}^{k-1}(1+S_jz_2^{-1})(1-S_jz_1)
\prod_{j=k+1}^N(1-S_jz_2)(1+S_jz_1^{-1})\nn\\
&&+\sum_{j=M+1}^N(1-S_jz_2)(1+S_jz_1^{-1}).\nn
\ena
Then we can show that
\begin{eqnarray*}
\overline{G}^{(N,M+1)}(S_{M+1},\cdots,S_N;z_1,z_2)
&=&z_2z_1^{-1}(1+S_{M+1}z_2^{-1})(1-S_{M+1}z_1)\\
&&\times \overline{G}^{(N,M+2)}(S_{M+2},\cdots,S_N;z_1,z_2).
\end{eqnarray*}
Because $\overline{G}^{(N,N)}(S_N;z_1,z_2)=0$,
we have
$\overline{G}^{(N,M+1)}(S_{M+1},\cdots,S_N;z_1,z_2)=0$
by induction.

\subsection{Correlation functions}
The vacuum vector $\rv$ satisfies
\be
\lb{RV}
\phi(\theta)\rv=0\quad\hbox{if $-{\pi\over2}\le\theta\le{\pi\over2}$},\quad
\phi^*(\theta)\rv=0\quad\hbox{if ${\pi\over2}\le\theta\le{3\pi\over2}$}.\nn
\en
Similarly, the dual vacuum $\lv$ satisfies
\be
\lb{LV}
\lv\phi(\theta)=0\quad\hbox{if ${\pi\over2}\le\theta\le{3\pi\over2}$},\quad
\lv\phi^*(\theta)=0\quad\hbox{if $-{\pi\over2}\le\theta\le{\pi\over2}$}.\nn
\en
We have also $\lv{\rm vac}\rangle=1$.
Our goal is to compute two point functions $\lv\psi^*_m\psi_n\rv$.
For this purpose we need
\begin{prop}
\label{prop:IV}
\bea
\psi_n&=&\int_0^{2\pi}\phi(\theta)A_n(\theta){d\theta\over2\pi},\nn\\
\psi^*_n&=&\int_0^{2\pi}\phi^*(\theta)A^*_n(\theta){d\theta\over2\pi}.\nn
\ena
Here 
$A_n(\theta)$ and $A^*_n(\theta)$ are given by
\bea
&&A_n(\theta)={e^{-in\theta}\over
C_n{\displaystyle\prod_{j=-\infty}^{n-1}}(1+S_je^{-i\theta})
{\displaystyle\prod_{j=n+1}^\infty}(1-S_je^{i\theta})}
\left\{{1\over1+S_ne^{-i\theta}}+{1\over1-S_ne^{i\theta}}-1\right\},\nn\\
&&A^*_n(\theta)={e^{in\theta}\over
C_n
{\displaystyle\prod_{j=-\infty}^{n-1}}(1-S_je^{i\theta})
{\displaystyle \prod_{j=n+1}^\infty}
(1+S_je^{-i\theta})}
\left\{{1\over1+S_ne^{-i\theta}}+{1\over1-S_ne^{i\theta}}-1\right\},\nn
\ena
\end{prop}
\proof
From \refeq{BG} we have
\bea
&&\int_0^{2\pi}\phi(\theta)A_n(\theta){d\theta\over2\pi}\nn\\
&&=\sum_{k\ge n}\psi_k\int_0^{2\pi}
{C_k\prod_{j=n}^{k-1}(1+S_je^{-i\theta})\over
C_n\prod_{j=n+1}^k(1-S_je^{i\theta})}
\left\{{1\over1+S_ne^{-i\theta}}+{1\over1-S_ne^{i\theta}}-1\right\}
e^{i(k-n)\theta}{d\theta\over2\pi},\nn\\
&&+\sum_{k\le n-1}\psi_k\int_0^{2\pi}
{C_k\prod_{j=k+1}^{n}(1-S_je^{i\theta})\over
C_n\prod_{j=k}^{n-1}(1+S_je^{-i\theta})}
\left\{{1\over1+S_ne^{-i\theta}}+{1\over1-S_ne^{i\theta}}-1\right\}
e^{-i(n-k)\theta}{d\theta\over2\pi}.\nn
\ena
Noting that $\int_0^{2\pi} e^{in\theta}{d\theta\over2\pi}=\delta_{n,0}$,
we can show this is equal to $\psi_n$.
The other case is similar.

The two point functions are given as follows.
We have obviously
\[
\lv\psi_m\psi_n\rv=\lv\psi^*_m\psi^*_n\rv=0.
\]
\begin{prop}
Suppose that $m<n$. We have
\begin{eqnarray}
&&\lv\psi^*_m\psi_n\rv=(-1)^{m-n+1}\lv\psi^*_n\psi_m\rv\nn\\
&&\phantom{---}\nn\\
&&=(-1)^{m-n}\lv\psi_m\psi^*_n\rv=-\lv\psi_n\psi^*_m\rv\nn\\
&&\phantom{---}\nn\\
&&={i^{m-n-1}\over\pi}(B_mB_n)^{1\over2}\sum_{j=m}^n\beta_j
{\prod_{m+1\le l\le n-1}(B_j+B_l)\over\prod_{m\le l\le n\atop l\not=j}
(B_j-B_l)}.\nn
\\
\lb{CF}
\end{eqnarray}
Here, we set $B_j=e^{\beta_j}$. In addition, we have
\[
\lv\psi^*_n\psi_n\rv=\lv\psi_n\psi^*_n\rv
={1\over2}.
\]
\end{prop}

\proof
Because of \refeq{AR}, it is enough to compute $\lv\psi^*_m\psi_n\rv$
$(m,n\in\Z)$.
Consider the anti-involution
\[
\psi_n\leftrightarrow\psi^*_n,\quad
\beta_n\leftrightarrow-\beta_n,\quad
\gamma_n\leftrightarrow-\gamma_n,\quad
\lv\leftrightarrow\rv,\quad
\phi(\theta)\leftrightarrow\phi^*(-\theta).
\]
Note that the last expression in \refeq{CF}
changes the sign by $(-1)^{m-n-1}$.
Therefore, it is enough to prove the equality for $\lv\psi^*_m\psi_n\rv$.

First, consider the case $m=n$. Using \refeq{RV},\refeq{LV} and
Proposition \ref{prop:IV}, we have
\bea
&&\lv\psi^*_n\psi_n\rv\nn\\
&=&\int_{{\pi\over2}}^{3\pi\over2}
{(1+S_ne^{-i\theta})(1-S_ne^{i\theta})\over C^2_n}
\left\{{1\over1+S_ne^{-i\theta}}+{1\over1-S_ne^{i\theta}}-1\right\}^2
{d\theta\over2\pi}\nn\\
&=&\int_{{\pi\over2}}^{3\pi\over2}{d\over d\theta}
\left(\log{1+S_ne^{-i\theta}\over1-S_ne^{i\theta}}+i\theta\right)
{d\theta\over2\pi i}\nn\\
&=&{1\over2}\phantom{-}.\nn
\ena
In general, for $m<n$, we have
\bea
&&\lv\psi^*_m\psi_n\rv\nn\\
&=&\int_{{\pi\over2}}^{3\pi\over2}
{C_m\over1+S_me^{-i\theta}}
\prod_{j=m+1}^{n-1}{1-S_je^{i\theta}\over1+S_je^{-i\theta}}
{C_n\over1+S_ne^{-i\theta}}e^{-i(n-m)\theta}
{d\theta\over2\pi}\nn
\ena
With the change of variable $z=-e^{i\theta}$, the right hand side becomes
\bea
&&-\int_{-i}^i
{C_mC_n\prod_{j=m+1}^{n-1}(1+S_jz)\over\prod_{j=m}^n(-z+S_j)}
{dz\over2\pi i}\nn\\
&=&(-1)^{n-m}{C_mC_n\over(2\pi i)^2}
\int_Cdz\log{z+i\over z-i}
{\prod_{j=m+1}^{n-1}(1+S_jz)\over\prod_{j=m}^n(z-S_j)}.\nn
\ena
Here the branch of $\log(z+i)/(z-i)$ is such that it has the value $0$ 
at $z=\infty$.
The contour $C$ is as in Figure \ref{fig:Fb}

\Fb

Taking the residues at $z=S_j$ $(m\le j\le n)$ and using
\refeq{BG} and, in particular, the equality
\begin{eqnarray*}
&&\log{S_j+i\over S_j-i}=-\beta_j-\pi i,
\\
&&\sum_{j=m}^n\frac{\prod_{l=m+1}^{n-1}(B_j+B_l)}
{\prod_{m\le l\le n\atop l\neq j}(B_j-B_l)}=0,
\end{eqnarray*}
we have \refeq{CF}.

\end{document}